\documentclass{aa}  
\usepackage{graphicx}
\usepackage{txfonts}

\usepackage[pdfencoding=auto,psdextra]{hyperref}
\hypersetup{
    colorlinks=true,
   linkcolor=blue,
    filecolor=magenta,      
    urlcolor=blue,
    citecolor=blue,
    pdftitle={Dwarf Galaxies in the MATLAS Survey: Hubble Space Telescope Observations of the Globular Cluster Systems of 74 Ultra Diffuse Galaxies},
    pdfauthor={Marleau et al.}
}
\usepackage{amssymb}
\usepackage{mathtools}
\usepackage{color}
\usepackage{siunitx}

\usepackage[dvipsnames]{xcolor}

\usepackage[flushleft]{threeparttable}

\usepackage{float}

\makeatletter
\renewcommand*\aa@pageof{, page \thepage{} of \pageref*{LastPage}}
\makeatother
\usepackage{lastpage}

\begin{document} 

\title{Dwarf Galaxies in the MATLAS Survey: Hubble Space Telescope Observations of the Globular Cluster Systems of 74 Ultra Diffuse Galaxies}

\author{ Francine R. Marleau\inst{1}, Pierre-Alain Duc\inst{2}, M\'elina Poulain\inst{3}, Oliver M\"uller\inst{4}, Sungsoon Lim\inst{5}, Patrick R. Durrell\inst{6}, Rebecca Habas\inst{7}, Rub\'en S\'anchez-Janssen\inst{8}, Sanjaya Paudel\inst{9}, J\'er\'emy Fensch\inst{10}}
   
\institute{
Universität Innsbruck, Institut für Astro- und Teilchenphysik, Technikerstraße 25/8, 6020 Innsbruck, Austria
\and
Observatoire Astronomique de Strasbourg  (ObAS), Universite de Strasbourg - CNRS, UMR 7550 Strasbourg, France
\and
Space Physics and Astronomy Research Unit, University of Oulu, P.O. Box 3000, FI-90014, Oulu, Finland
\and
Laboratoire d’astrophysique, École Polytechnique Fédérale de Lausanne (EPFL), Observatoire, 1290 Versoix, Switzerland
\and
Department of Astronomy, Yonsei University, 50 Yonsei-ro Seodaemun-gu, Seoul, 03722, Republic of Korea
\and
Department of Physics, Astronomy, Geology and Environmental Sciences, Youngstown State University, One University Plaza, Youngstown, OH 44555 USA
\and
INAF – Astronomical Observatory of Abruzzo, Via Maggini, 64100, Teramo, Italy
\and
UK Astronomy Technology Centre, Royal Observatory, Blackford Hill, Edinburgh, EH9 3HJ, UK
\and
Department of Astronomy and Center for Galaxy Evolution Research, Yonsei University, Seoul 03722, Republic of Korea
\and
Univ. Lyon, ENS de Lyon, Univ. Lyon 1, CNRS, Centre de Recherche Astrophysique de Lyon, UMR5574, 69007 Lyon, France
}
 
\abstract
{
Ultra diffuse galaxies, characterized by their low surface brightness and large physical size, constitute a subclass of dwarf galaxies that challenge our current understanding of galaxy formation and evolution. In this paper, we probe the properties of 74~UDGs, identified in the MATLAS survey, based on a comprehensive study of their globular cluster (GC) populations. We obtained high resolution HST imaging of these galaxies using the ACS $F606W$ and $F814W$ filters, allowing us to select GCs based on color and concentration index. After background subtraction and completeness correction, we calculate an overall total of 387~GCs. The number of GCs per galaxy ranges from 0 to 38, with the majority (64\%) having low counts ($0-2$ GCs). On average, the more massive UDGs host a larger number of GCs. We find that our UDGs have specific frequencies ($S_N$) ranging from 0 to 91, with a small population (9\%) with $S_N > 30$. The median $S_N$ of our sample is similar to the one for the Perseus cluster UDGs, despite the fact that our UDGs are found in lower density environments. The $S_N$ measurements for individual galaxies can extend beyond those found in Perseus, but remain below the values found for UDGs in the Virgo and Coma cluster. Based on a trending analysis of the $S_N$ values with the host galaxy properties, we find trends with host galaxy size, roundness, color, and local density. For the UDGs with sufficiently high statistics, we study 2D density maps of the GC distributions, which show a variety of appearances: symmetric, asymmetric, off-center, and elongated. The UDGs with disturbed density maps also show disturbed stellar light morphologies. We further quantify the distribution by modeling it with a S\'ersic profile, finding $R_{e,GC}/R_{e,gal} \sim 1.0$, which indicates that the GCs follow the stellar light of the host galaxy. 
}


\date{}
\titlerunning{HST observations of MATLAS UDGs}
\authorrunning{Marleau et al.}
\maketitle


\section{Introduction} 
\label{sec:intro}

Ultra diffuse galaxies (UDGs) are a sub-class of dwarf galaxies characterized by an exceptionally low surface brightness (LSB) combined with an unusually large physical size. Although a strict numerical threshold for their central surface brightness ($\mu_{0}$) and effective radius ($R_e$) was introduced by \citet{vanDokkum2015} ($\mu_{0,g} = 24-26$ mag arcsec$^{-2}$ and $R_e > 1.5$~kpc), the classification of a LSB galaxy as a UDG is somewhat arbitrary as there is no clear "break" in their physical properties \citep{Venhola2017, Lim2020, Marleau2021, Venhola2022}, making their definition somewhat variable in different studies and/or their context with other galaxies. One of the most intriguing aspects of UDGs is their exceptional size for such low luminosity galaxies. While the Milky Way has an effective radius of $R_e = 3.6$~kpc \citep{Bovy2013, vanDokkum2015} and a surface brightness $\mu_{0,g,MWdisk} \simeq 19.8$\,mag\,arcsec$^{-2}$, UDGs can exhibit effective radii that rival or even exceed the effective radius of the Milky Way galaxy, while possessing $\sim 1/1000$ of its stars. However, given that the effective radius depends on how the light is concentrated in these galaxies, alternative size measurement methods have been used to estimate the extent of UDGs, yielding values that are ten times smaller than Milky Way-like galaxies \citep{Trujillo2020, Chamba2020}. Although we can find some examples of UDGs in our local neighborhood, namely Andromeda XIX \citep{McConnachie2008},  M81-F8D1 \citep{Caldwell1998}, and Antlia II \citep{Torrealba2018}, the majority of UDGs detected until now are found at larger distances and in a wide range of environments. 

Due to the large expense in telescope time necessary to scan large regions of the sky for deep imaging surveys, the overwhelming majority of research into the photometric properties of UDGs was originally limited to select nearby galaxy groups (e.g., M101: \citealt{Merritt2016, Bennet2017}; nearby groups: \citealt{Greco2018}; three nearby isolated groups: \citealt{Roman2017b}; NGC1052: \citealt{vanDokkum2018b}; NGC5846: \citealt{Forbes2019}), and galaxy clusters (e.g., Coma: \citealt{vanDokkum2015, Yagi2016}; Virgo: \citealt{Mihos2015,Lim2020}; Perseus: \citealt{Wittmann2017, Gannon2022, Marleau2024}; Fornax: \citealt{Venhola2017, Venhola2022}; Hydra I: \citealt{LaMarca2022}; six Hubble Frontier Fields galaxy clusters: \citealt{Janssens2019}), where a high density of low surface brightness galaxies could be found. However, since then, other optical imaging surveys have contributed to filling in the gap in our knowledge of properties of UDGs across a range of lower density environments, such as the MATLAS (Mass Assembly of early-Type GaLAxies with their fine Structures) survey \citep{Marleau2021}, the SDSS Stripe 82 \citep{Barbosa2020}, isolated spiral and early-type spiral galaxies (\citealt{Crnojevic2014, Greco2018, Mueller2018, For2019, Roman2019}), and other wide-area imaging surveys (DESI Legacy Imaging Survey: \citealt{Zaritsky2019, Zaritsky2021, Zaritsky2022a}; Dark Energy Survey: \citealt{Tanoglidis2021}; GAMA survey \citealt{Prole2021}; Hyper Suprime-Cam Survey: \citealt{Greene2022}). Complementary to optical surveys, neutral hydrogen (HI) blind surveys (e.g., \citealt{Du2015} using ALFALFA+SDSS, \citealt{Leisman2017} using ALFALFA+SDSS,WIYN, \citealt{Marleau2021} using WRST+ALFALFA) have established the existence of a sub-population of HI-bearing UDGs in different environments. 

\begin{figure*}[ht!]
\centering
\includegraphics[width=\textwidth]{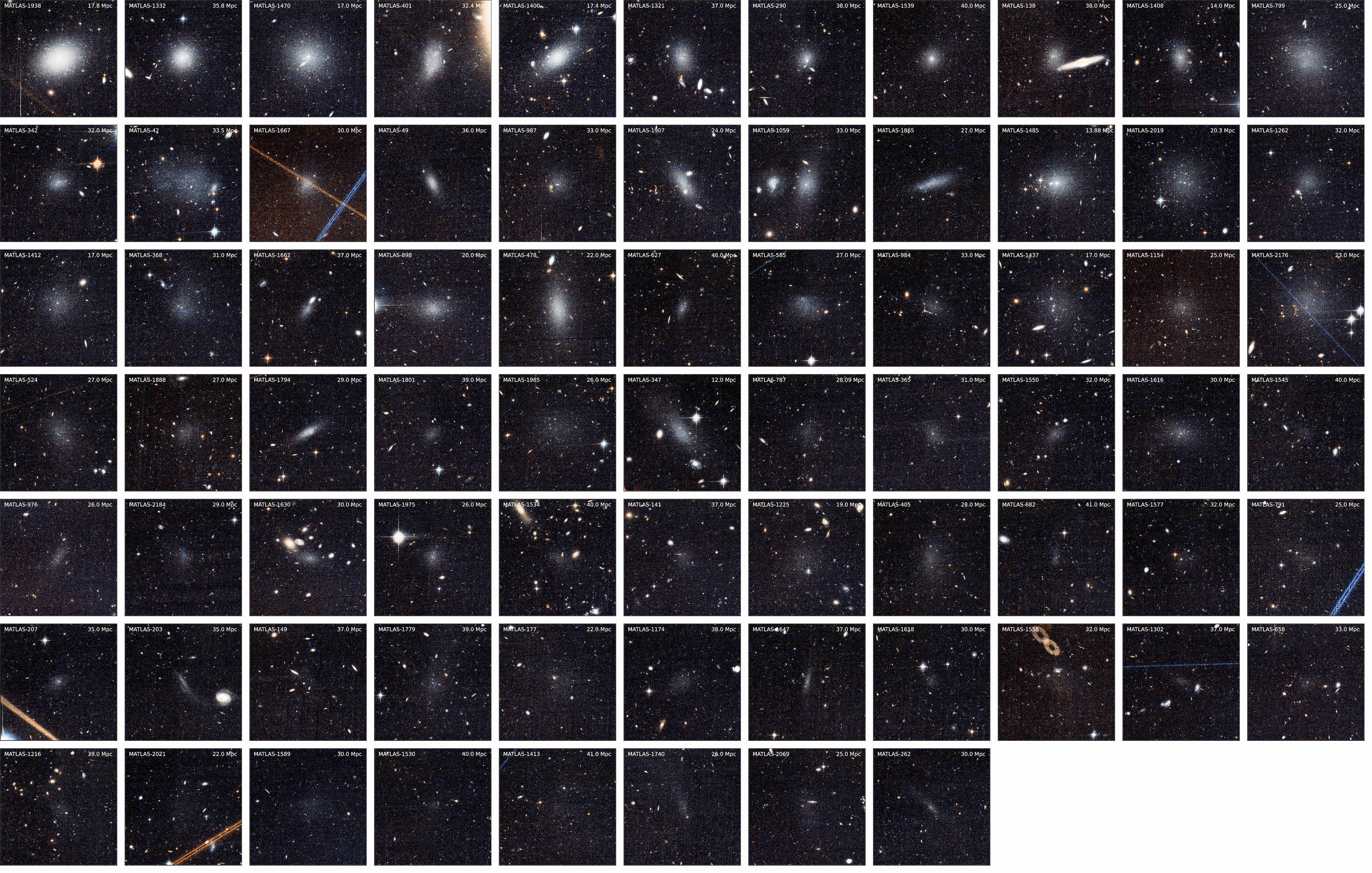}
\caption{F606W+F814W color images of the 74 MATLAS ultra diffuse galaxies, selected with a cut of $\langle \mu_{e,g} \rangle > 23.3$ mag arcsec$^{-2}$ and $R_e > 1.0$~kpc, observed with HST/ACS, ordered from {\it top, left} to {\it bottom, right} by decreasing surface brightness. Due to their low surface brightness, the galaxies in the last rows are difficult to see in the images. The names and distances of the UDGs are displayed on the images. The images are 1.3~arcminutes on a side.}
\label{fig:postagestamps}
\end{figure*}

As the number of studies identifying UDGs in a variety of environments continues to grow, a more complete determination of their formation and evolution scenario(s) has begun to emerge. One prominent formation scenario suggests that UDGs could be `failed galaxies' \citep{vanDokkum2015, Peng2016}. These galaxies would start their life like normal galaxies but would suffer from a sudden halt in star formation (SF) at early cosmic time ($z > 2$). Early quenching in these "under-developed" galaxies could be due to strong tidal forces and interactions within the intracluster medium during an early infall into a dense cluster environment \citep{Yozin2015, Tremmel2020} or an early, violent, and short-lived star formation episode \citep{Danieli2022}. This violent star forming episode would also create many GCs which would explain the high number of GCs associated with some UDGs (e.g., MATLAS-2019, \citealt{Mueller2021, Danieli2022}). Another prominent formation scenario suggests that UDGs could be `puffed-up dwarfs' \citep{Amorisco2016, Burkert2017, DiCintio2017, Chan2018, Jiang2019}. In this scenario, dwarf galaxies may have undergone tidal stripping and heating or mergers \citep{Carleton2019, Wright2021} which would have led to an increase in their effective radii. Alternatively, intrinsic processes such has high halo spins \citep{Amorisco2016, Rong2017, Amorisco2018a, Benavides2023} or strong star formation driven outflows capable of quenching star formation in the galaxies and leading to the expansion and redistribution of the remaining stars to larger radii \citep{DiCintio2017, Chan2018} could form UDGs. In this scenario, gas would remain in the UDGs and hence would explain the blue colors of some UDGs \citep{Roman2017b, Marleau2021} but not necessarily red isolated UDGs. A possible explanation for those would be that they originate as a result of backsplash orbits \citep{Benavides2021}. Finally, another possible scenario for UDG formation is that they could have formed from gas expelled from a massive galaxy after an interaction, becoming Tidal Ultra Diffuse Galaxies \citep{Duc2014, Marleau2021}. Tidal UDGs would be outliers in (located above) the mass-metallicity relation (MZR) \citep{Duc2014} and devoid of dark matter \citep{Lelli2015}. The wide range of properties of UDGs, as described below, could also indicate that UDGs form via multiple pathways \citep{Ferre-Mateu2018, Lim2018, Marleau2021, Toloba2023}. 

In terms of their stellar populations, UDGs display a range in ages ($\sim 3$ to $11$~Gyr), metallicities ([Z/H] $\sim -1.5$ to $-0.4$), and alpha-element abundances ([$\alpha$/Fe] $\gtrsim 0.3$~dex) \citep{Gu2018, Ferre-Mateu2018, Ruiz-Lara2018, Chilingarian2019, Fensch2019a, Martin-Navarro2019, Mueller2020, Rong2020, Villaume2022, Heesters2023}. Their stellar populations are in general similar to those of dwarf elliptical galaxies, but some have stellar populations similar to GCs (old age) and dwarf spheroidal galaxies (low metallicity). UDGs found in high density environments appear to have formed earlier and faster (according to their high [$\alpha$/Fe] values) than those found in low-density environments (which tend to have lower [$\alpha$/Fe] values and be younger). The analysis of UDGs using SED fitting has revealed that GC-poor UDGs are more metal-rich, on average younger and more elongated than their GC-rich counterparts \citep{Buzzo2022, Buzzo2024}. In these works, the GC richness classification is based directly on GC numbers and the GC-rich UDGs are defined, fairly subjectively, as those with 20 or more GCs. This definition can vary between different works. The GC-poor UDGs have properties similar to those of dwarfs and are consistent with the dwarf MZR, which is consistent with the `puffed-up dwarf' scenario. On the other hand, GC-rich UDGs are located below the dwarf MZR and therefore are more consistent with early-quenching and hence the `failed galaxies' scenario. However, spectral analysis does not show this trend and indicates no GC dependence \citep{Ferre-Mateu2023}. In agreement with the stripped/tidal formation scenario, some UDGs are found to be positive outliers in the MZR \citep{Buzzo2024}. 

With respect to their globular cluster population, some studies find that UDGs tend to host two to three times as many GCs as dwarf galaxies of the same stellar mass \citep{Lim2018, Lim2020} whereas others find no significant difference in GC numbers between the two types of galaxies \citep{Amorisco2018b, Prole2019a, Marleau2021}. In general, the UDGs have a wide range of GC counts, from GC-poor to GC-rich (e.g., MATLAS-2019) \citep{Lim2018, Lim2020, Gannon2021, Marleau2021, Mueller2021}. Some studies have shown that UDGs in denser environments can have higher specific frequencies \citep{Peng2008,Mistani2016,Lim2018}. Other GC properties, such as their color, luminosity function (LF), and radial distribution (measured in term of the ratio of the GC half-number radius $R_{e,GC}$ to the UDG effective radius $R_e$) seems to match those of dwarf galaxies \citep{Lim2024}. In terms of their radial distribution, UDGs are found to have $R_{e,GC}/R_{e,gal} \sim 0.7-1.7$ \citep{Peng2016, Mueller2021, Montes2021, Danieli2022, Janssens2022}, compared to dwarfs in local groups and in the Virgo cluster with $R_{e,GC}/R_{e,gal} \sim 1.06^{+0.21}_{-0.18}$ and $1.25^{+0.24}_{-0.18}$, respectively \citep{Carlsten2022}. The total mass locked up in GCs compared to the stars are somewhat higher for the UDGs, with $M_{GC}/M_\ast$ up to $\sim 15\%$, compared to Virgo dwarfs with $M_{GC}/M_\ast$ up to $\sim 5\%$. The GC colours are dominantly blue ($0.5 < (m_{F606W}-F_{814W})_0 < 0.9$~mag), metal-poor ([Fe/H] $\sim -1.8$ to $-1.2$), and old ($\sim 7.8$ to $11.6$~Gyr) \citep{Fensch2019a, Mueller2020, Mueller2021}. The UDG GC luminosity function (GCLF) is similar to what is found for GCs in dwarf galaxies \citep{Beasley2016a, Peng2016, vanDokkum2017, Amorisco2018b, Prole2019a, Lim2020, Somalwar2020, Mueller2021, Saifollahi2021}, with the exception of NGC\,1052-DF2 and -DF4 which appear to have GCs that are too bright \citep{vanDokkum2018b}. Some studies find that HI-rich field UDGs appear to host remarkably few GCs, with many having no candidate GCs \citep{Jones2023}. For the UDGs in the Perseus cluster, GC-rich UDGs appear to have higher velocity dispersions than GC-poor UDGs, resulting in GC-rich UDGs having higher halo masses than GC-poor UDGs \citep{Gannon2022}.

\begin{figure*}[ht!]
\includegraphics[width=\textwidth]{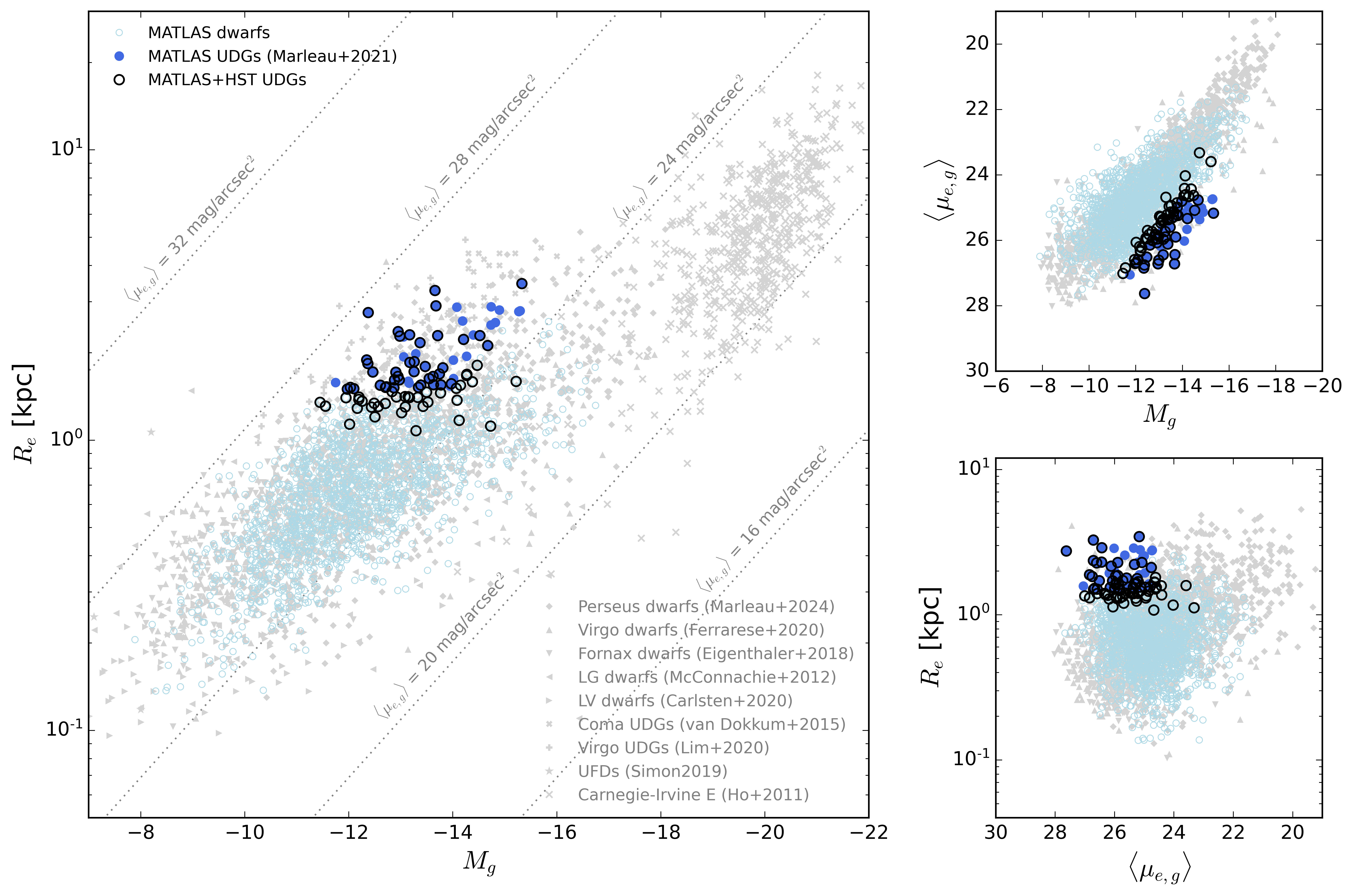}
\caption{The 74 MATLAS UDGs, selected with a cut of $\langle \mu_{e,g} \rangle > 23.3$ mag arcsec$^{-2}$ and $R_e > 1.0$~kpc observed with HST ({\it black empty circles}), overlayed on the MATLAS dwarfs of \citet{Habas2020} ({\it empty pale blue circles}) and UDGs of \citet{Marleau2021} ({\it filled blue circles}). Also shown are the dwarfs identified in the Euclid Early Release Observations of the Perseus cluster ({\it grey diamonds}; \citealt{Marleau2024}), the Next Generation Virgo Cluster Survey ({\it grey triangles}; \citealt{Ferrarese2020}) and the Next Generation Fornax Survey ({\it inverted grey triangles}; \citealt{Eigenthaler2018}) images, as well as Local Group dwarfs ({\it rotated left grey triangles}; \citealt{McConnachie2012}) and Local Volume ({\it rotated right grey triangles}; \citealt{Carlsten2020}). The UDGs found in the Coma ({\it grey cross symbols}; \citealt{vanDokkum2015}) and Virgo cluster ({\it grey plus symbols}; \citealt{Lim2020}) are shown separately. For comparison, we have also included ultra faint dwarfs (UFDs; {\it grey stars}; \citealt{Simon2019}) and massive ellipticals from the Carnegie–Irvine catalogue ({\it grey large cross symbols}; \citealt{Ho2011}).
\label{fig:scalingrelation}}
\end{figure*}

The number of GCs can provide a test to discriminate between different UDG formation scenarios \citep{Saifollahi2022}. For example, if UDGs are more massive galaxies that essentially faded due to a lack of subsequent star formation (the `failed galaxies' model), the number of GCs per stellar mass, $N_{GC}/M_\ast$, is expected to be larger than that of traditional dwarfs. If UDGs, however, have always been less massive dwarf galaxies (the `puffed-up dwarf' model), the predictions vary depending on the physical mechanisms driving the larger radii. If the UDG formed via tidal interactions, $N_{GC}/M_\ast$ is expected to be comparable to that of other dwarf galaxies, since the LSB is due to the dynamical expansion of the galaxy rather than a lower star formation efficiency, assuming that no GCs are lost during the interaction. For high spins, the expected $N_{GC}/M_\ast$ value is unclear; the stars are more spread out and the star formation efficiency (likely) drops, it is uncertain how this impacts the GC formation. Similarly, the impact of mergers is not well constrained; $N_{GC}/M_\ast$ might be lower than in galaxies with no early-time major mergers if GCs form as a consequence of the merger, but if GCs form only at high redshift, then $N_{GC}/M_\ast$ might be larger in UDGs. Interestingly, if a UDG was formed as a result of stellar feedback from intense star formation, the $N_{GC}$ is not expected to change, but the radial distribution, traced by measuring $R_{e,GC}/R_{e,gal}$, is likely to. The spatial distribution of GC systems can also provide a way to decipher the galaxy mass and assembly history of the host galaxy, as cosmological simulations are now suggesting \citep{Kruijssen2015, Renaud2017, Li2019, ReinaCampos2022}.

In this paper, we report on the analysis of data taken with the Hubble Space Telescope (HST) of the GC systems of a sample of 74 MATLAS (Mass Assembly of early-Type GaLAxies with their fine Structures) UDGs, selected with a cut of $\langle \mu_{e,g} \rangle > 23.3$ mag arcsec$^{-2}$ and $R_e > 1.0$~kpc, tracing the GCLF two magnitudes below its peak. The MATLAS survey \citep{Duc2015}, reaching a local surface brightness depth of $\mu_g \sim 28.5-29$ mag arcsec$^{-2}$, contains a sample of 2210 dwarf galaxies \citep{Habas2020}. Of these, a total of 11\% (248 dwarfs) satisfy the selection cut above. In Section\,\ref{sec:data}, we describe the observing strategy and data reduction. The detection of GCs, the counts, background subtraction and completeness correction are described in Section\,\ref{sec:analysis}, along with the calculation of the specific frequency and halo mass. The spatial distribution is explored in detail in Section\,\ref{sec:radial}. In Section\,\ref{sec:pca}, a quantitative determination of the key parameters driving the GC count is obtained. Finally, in Section\,\ref{sec:conclusions}, we summarize our results.

\section{Data} 
\label{sec:data}

\subsection{Observations} 
\label{sec:obs}

Our HST images were each obtained through a single orbit in the Cycle 28 and 29 SNAPSHOT programs GO-16257 and GO-16711 (PI: Marleau). We present the observations of 74 MATLAS UDGs selected with a cut at $\langle \mu_{e,g} \rangle > 23.3$ mag arcsec$^{-2}$ and $R_e > 1.0$~kpc, including the data from the Cycle 27 program GO-16082 for MATLAS-2019 \citep{Mueller2021}, with the HST Advanced Camera for Surveys (ACS) in both the $F606W$ and $F814W$ filters (see Figure~\ref{fig:postagestamps}). The galaxies were placed at the center of one of the charge coupled devices (CCDs) to optimize a uniform coverage map of the galaxy and maximize a suitable background sample. Two dithered images (separated by 0.5$\arcsec$) of 412\,s each (total exposure time 824\,s) were taken in each filter. We used the final, reduced CTE-corrected (Charge Transfer Efficiency corrected) {\em .drc.fits} images produced by the standard pipeline that have a spatial sampling of 0.05 arcsec/pixel and a point spread function (PSF) full width half maximum (FWHM) of $\sim 0.1$~arcsec. The list of our HST targets is given in Appendix~\ref{AppendixA}, Table~\ref{tab:sample}.

Magnitudes were computed using the VEGAmag zeropoints provided by the ACS online documentation\footnote{https://www.stsci.edu/hst/instrumentation/acs/data-analysis/zeropoints}. Aperture corrections were applied according to the ACS documentation\footnote{https://www.stsci.edu/hst/instrumentation/acs/data-analysis/aperture-corrections.}. The magnitudes were corrected for extinction using {\em python ned\_extinction\_calc}\footnote{https://github.com/mmechtley/ned\_extinction\_calc}. 

\begin{figure*}
\centerline{
\includegraphics[width=0.5\textwidth]{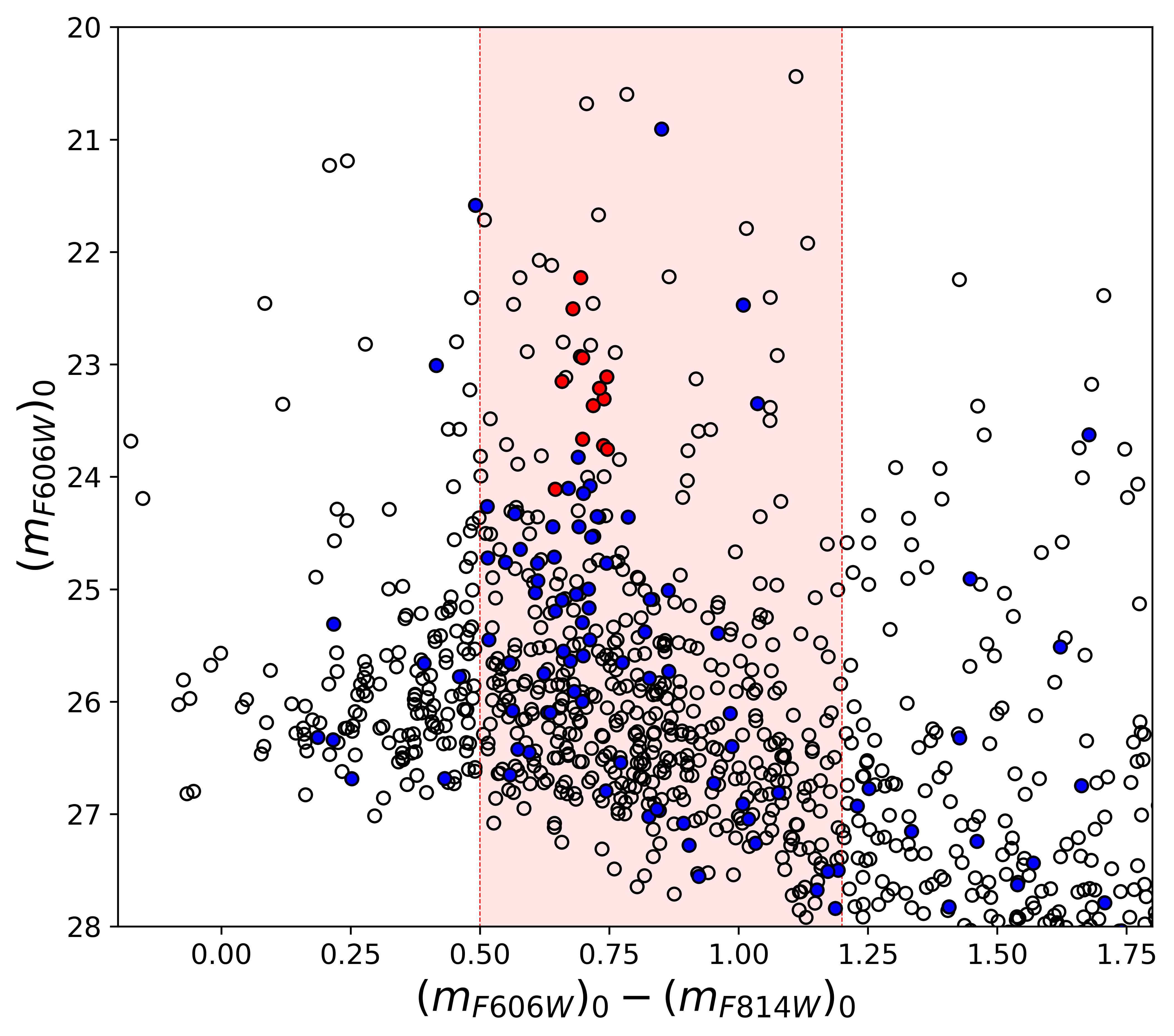}
\includegraphics[width=0.5\textwidth]{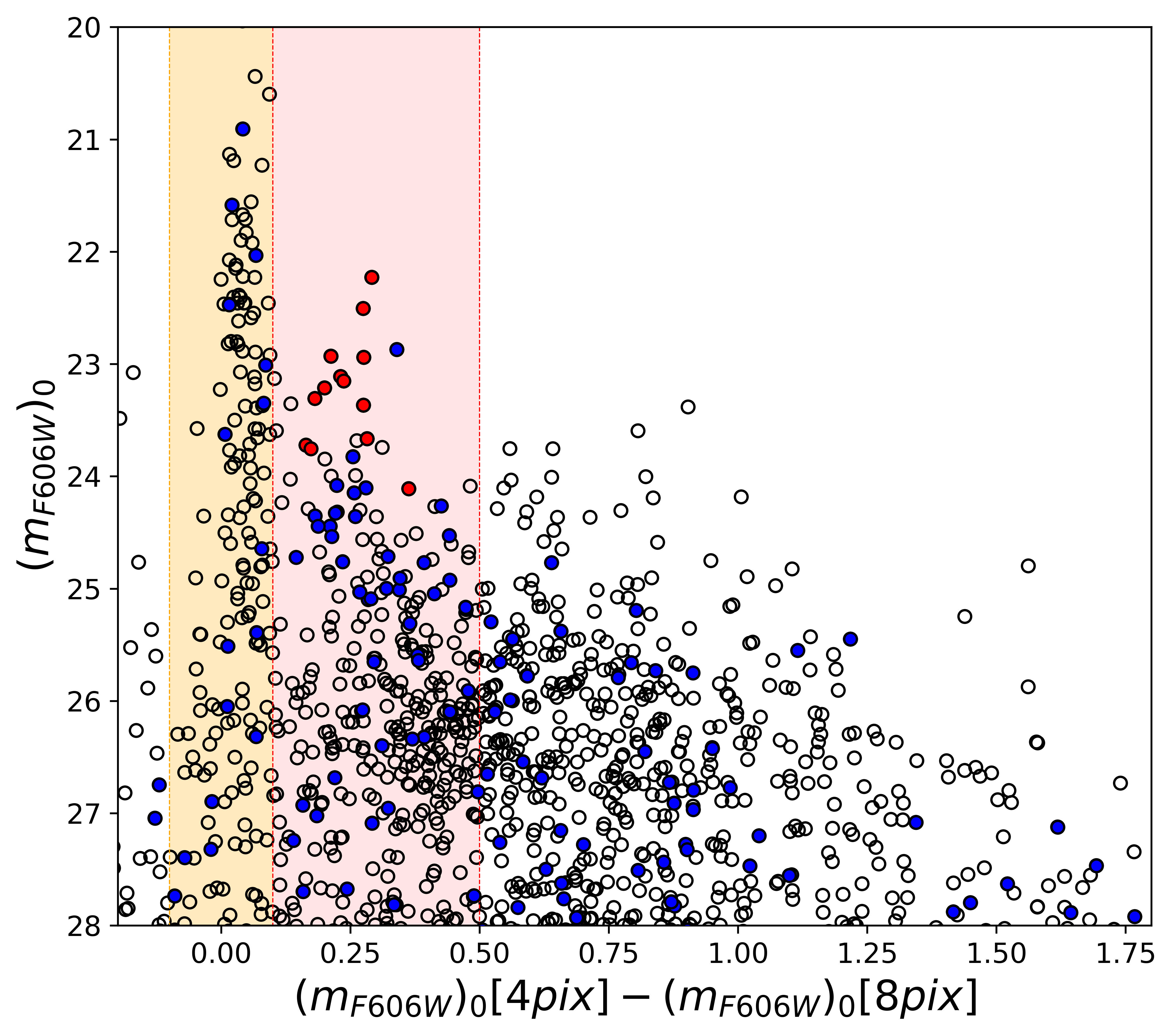}
}
\caption{Color ({\it left}) and concentration index ({\it right}) cut of all GC detections ({\it black open circles}) and the GC detections within $2 R_e$ ({\it blue circles}), based on the spectroscopically confirmed GCs in MATLAS-2019 ({\it red circles}: \citealt{Mueller2021}). The {\it light red region} highlights the selection cut we used for color, $0.5 < (m_{F606W}-F_{814W})_0 < 1.2$, and for the concentration index, $0.1 < \Delta m_{4-8} < 0.5$, for galaxies with a distance $\leq 25$~Mpc. For galaxies at larger distances, i.e., $> 25$~Mpc, the lower concentration index cut was extended to include the unresolved sources, highlighted by the {\it light orange} and {\it light red regions}, corresponding to a cut of $-0.1 < \Delta m_{4-8} < 0.5$.
\label{fig:Cindexcut}}
\end{figure*}

\subsection{Sample properties} 
\label{sec:properties}

The photometric and structural properties of our HST targets were taken from \citet{Poulain2021}. They have assumed or confirmed distances ranging from 12.3 to 45.8 Mpc, g-band absolute magnitude between $-15.32$ and $-11.44$, effective radius, $R_e$, in the range of 1.07 to 3.46 kpc, and central[average] surface brightness ranging from 22.54[23.3] to 26.84[27.6] mag arcsec$^{-2}$ in the g-band \citep{Poulain2021, Poulain2022, Heesters2023}. Of the 74 UDGs in our HST sample, a total of 38 fall under the selection criteria from \citet{vanDokkum2015}, i.e., $\mu_{0,g} = 24-26$ mag arcsec$^{-2}$ and $R_e > 1.5$~kpc (hereafter referred to as "UDGvD15"; \citealt{Marleau2021}). The remaining 36 are slightly below this selection cut in surface brightness and effective radius, as can be seen in Figure\,\ref{fig:scalingrelation}. The full UDG sample contains 40 nucleated and 34 non-nucleated UDGs. 

\begin{figure}[ht!]
\centerline{
\includegraphics[width=\linewidth]{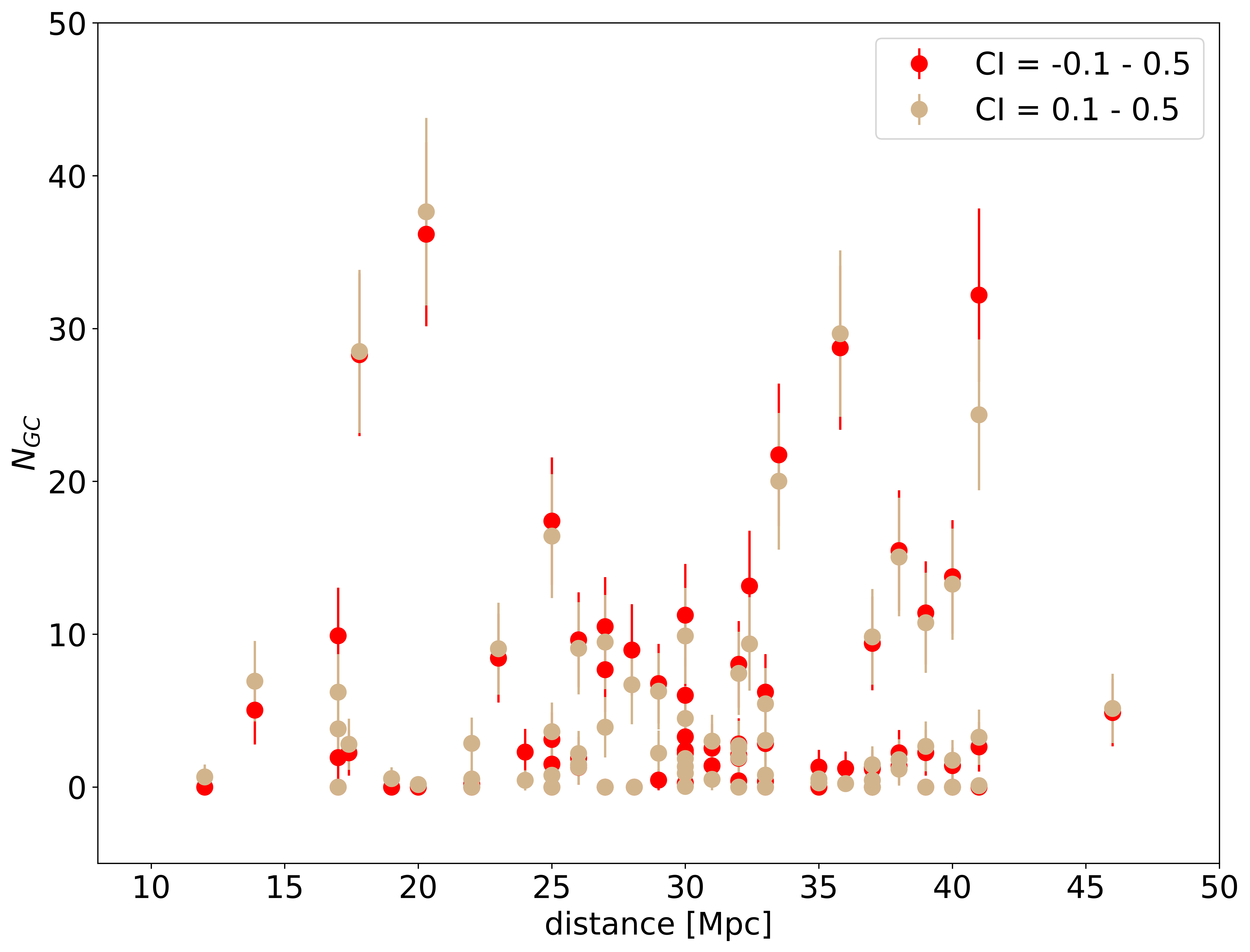}
}
\caption{Comparison between the GC counts (background and completeness corrected) measured excluding unresolved GCs ({\it tan circles}) and including unresolved GCs ({\it red circles}) as a function of distance. Below $\sim 25$~Mpc, the GC counts are dominated by the marginally resolved sources (in {\it tan circles}) whereas above $\sim 25$~Mpc, the counts become dominated by the unresolved sources (in {\it red circles}). Based on this comparison, we included unresolved GCs only at distances greater than 25~Mpc.
\label{fig:Cindexdist}}
\end{figure}

\section{Analysis}
\label{sec:analysis}

\subsection{Detection}

To optimize the detection of GCs both near and far from the center of the UDG, the smooth light profile of the galaxy was first modeled and subtracted using the GALFIT algorithm \citep{Peng2002}. The detection of GCs was then performed on the galaxy-subtracted $F814W$ image using {\em sep.extract}, the {\em Source Extractor} implementation in {\em python} \citep{Bertin1996}, using a 3 sigma threshold and a minimum of 5 adjacent pixels. The detection was done on the full $F814W$ image. The following cuts in size and ellipticity were applied: 1) a semi-major axis $a < 30$~pc, (transformed in pixels using the distance of the UDG; e.g., for MATLAS-2019 with $d=20.3$~Mpc, this corresponds to 6~pix), and 2) an ellipticity of $(1-b/a) < 0.37$. The upper limit in size for the GCs was chosen based on the visual extent of the compact GCs around M31 with half-light radii $\sim 3-5$\,pc (with maximum sizes of $\sim 8$\,arcsec or $\sim 30$\,pc at the assumed distance to M31 of 784\,kpc; \citealt{Huxor2014}).

Photometry was performed with {\em sep.sum\_circular} with a fixed aperture of 3\,pix in radius (background annulus $8-23$\,pix) in both the $F814W$ and $F606W$ images using the $F814W$ image detections. GC candidates were then selected on the basis of their color, magnitude and concentration index, $\Delta m_{4-8}$, defined as the difference between 4-pixel and 8-pixel diameter aperture magnitudes in the F606W filter normalized so that point sources have a mean value of zero \citep{Durrell2014}. 

\begin{figure}[ht!]
\centerline{
\includegraphics[width=0.5\textwidth]{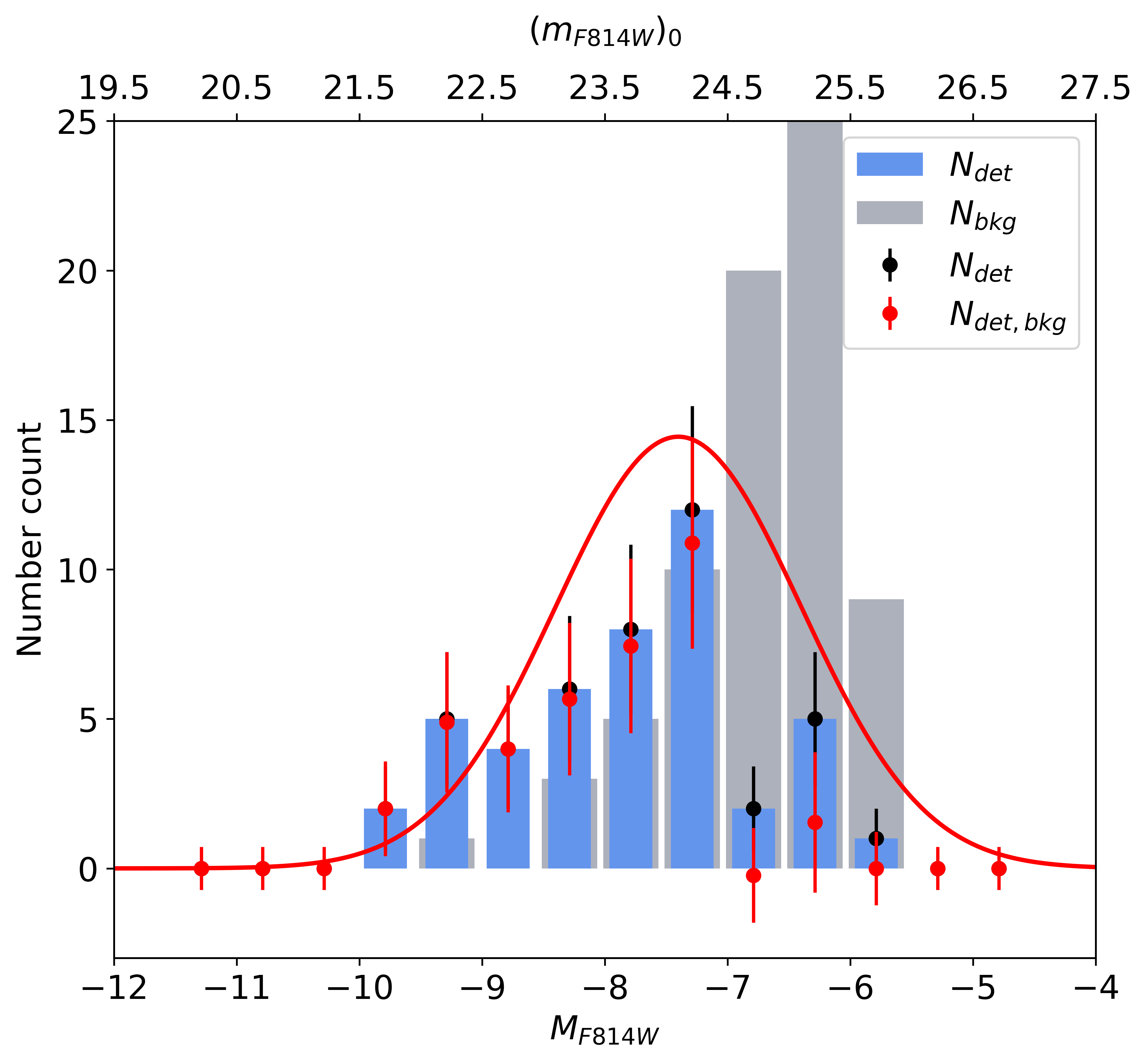}
}
\caption{GCLF as a function of absolute magnitude $M_{F814W}$ ({\it bottom}) and apparent magnitude $(m_{F814W})_0$ ({\it top}) of the detected GCs within $2 R_e$ (blue), and the detections outside $2 R_e$ (grey) for the UDG MATLAS-2019, which has the largest number of GCs in our sample. The black points are for the raw count within $2 R_e$ and the red points show the GC count after background subtraction. The {\it red solid line} shows the Gaussian fit with peak absolute magnitude $M_V = -7.3$~mag and a dispersion of 1~mag.
\label{fig:NGCbkgcomp}}
\end{figure}

\begin{figure}[ht!]
\includegraphics[width=\linewidth]{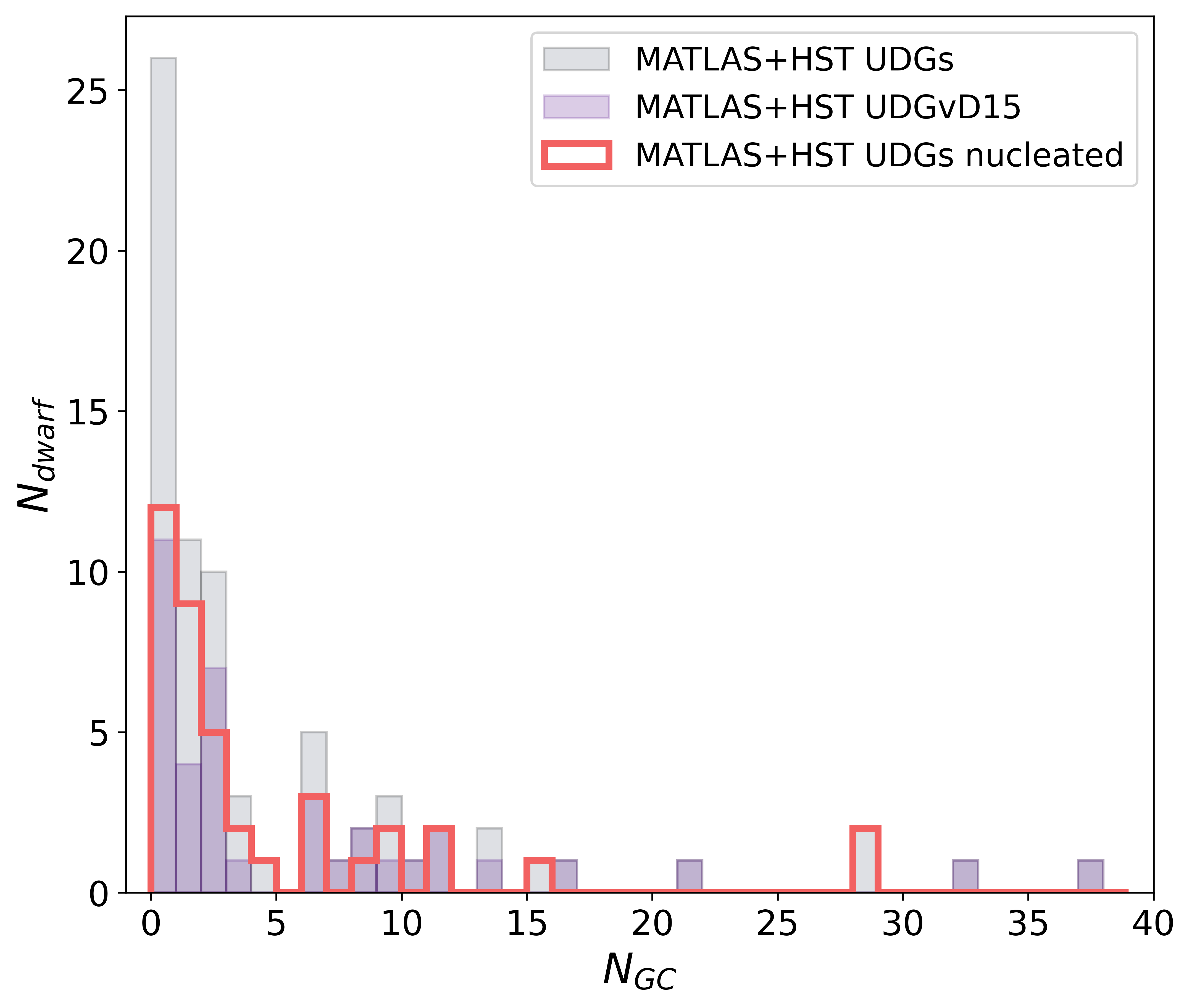}
\caption{Distribution of the number of GCs within $2 R_e$ for the 74 UDG sample ({\it light grey}), the 36 UDGvD15 as defined by \citet{vanDokkum2015} ({\it pale indigo}), and the 40 nucleated UDGs ({\it bright coral}). The majority of the sample (47 UDGs or 64\%) are measured to have $N_{GC} < 3$, while the remaining 36\% is spread across a wider range of values ($N_{GC}=3-38$). A total of 21 UDGs (28\%) have $N_{GC}$ values consistent with zero, with the tail of the distribution extending to our most GC-rich galaxy, MATLAS-2019, with $N_{GC} = 38$. The percentage of nucleated UDGs with a large number of GCs (14 UDGs or 35\% have $N_{GC} \geq 3$) is similar to the one for the whole sample.
\label{fig:NGCbkgcompUDG}}
\end{figure}

The selection cut limits are shown in Figure~\ref{fig:Cindexcut} and were selected based on the measurements obtained for the spectroscopically confirmed GCs in MATLAS-2019 (\citealt{Mueller2021}) and are consistent with the $(m_{F606W}-m_{F814W})_0$ colors of old (12~Gyr) GCs with metallicities $-2.2 < $[Fe/H] $< 0.0$ \citep{harris2023}, while also allowing for photometric uncertainties in our fainter GC candidates. They consist of a color cut in color of $0.5 < (m_{F606W}-m_{F814W})_0 < 1.2$, a magnitude cut of $-10.4 < (m_{F606W})_0 < -5.1$ (assuming a peak at $M_V=-7.3$; \citealt{Miller2007, Peng2016, Lim2018}), and a concentration cut of $0.1 < \Delta m_{4-8} < 0.5$ for UDGs at $d < 25$~Mpc and $-0.1 < \Delta m_{4-8} < 0.5$ for those at larger distances. We did not extend our selection cut any further to the blue to avoid including young GC candidates, which do not offer reliable information about the properties of the galaxy halos, and to exclude star forming clumps. Note that the central compact source, which can either be a nuclear star cluster (NSC) or GC candidate, were included in this catalogue without a color cut. 

As our sample of UDGs span a large range of distances, the concentration index was adjusted to include the unresolved GCs at distances greater than 25~Mpc (see Figure~\ref{fig:Cindexdist}). The GC counts displayed in Figure~\ref{fig:Cindexdist} have been background and completeness corrected (see Section\,\ref{sec:bkgcompcorr} below). The comparison between the GC counts measured excluding unresolved GCs and including unresolved GCs as a function of distance shows that below $\sim 25$~Mpc, the GC counts are dominated by the marginally resolved sources whereas above $\sim 25$~Mpc, the counts become dominated by the unresolved sources. At the distance of $\sim 25$\,Mpc, the resolution limit of our HST observations corresponds to a radial extent of $\sim$\,6\,pc.

Following the arguments in \citet{Mueller2021} and \citet{Lim2024}, we adopted a radius of $2 R_e$ to determine the association of GCs and host UDGs, where $R_e$ was measured for each UDG. This was uniformly applied to all UDGs, to allow a direct comparison of the GC distribution in units of effective radius. The number of GCs within that radius is subsequently referred to as $N_{det}$. The GC candidates beyond $2 R_e$ were used to determine the background contamination ($N_{bkg}$). We identified a total of 676 GC candidates associated with all 74 UDGs within this radius.

\begin{figure}[ht!]
\centerline{
\includegraphics[width=\linewidth]{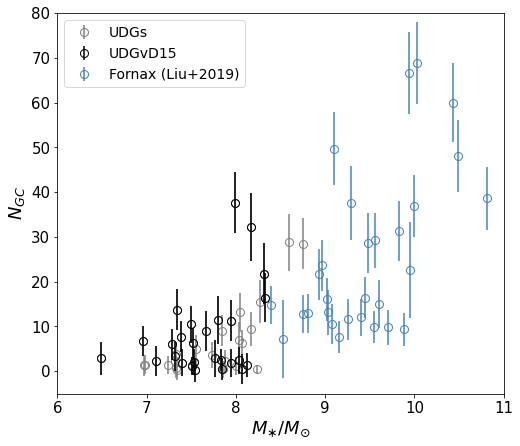}
}
\caption{The total number of GCs for the UDG sample ({\it black open circles}) and those belonging to the UDGvD15 selection criteria ({\it indigo open circles}) as a function of the galaxy stellar mass. The number of GCs is shown to steadily increase with total stellar mass for the full sample of UDGs. The trend continues as the stellar mass increase to massive early-type galaxies, as can be seen using the GC counts obtained from the HST ACS Fornax cluster survey \citep{Liu2019} ({\it blue open circles}). The galaxies in the UDGvD15 sample extend higher at each stellar mass than the galaxies in the non-UDGvD15 sample with the exception of the two most massive bins which contain no galaxies from the UDGvD15 sample. 
\label{fig:NGCmstar}}
\end{figure}

\subsection{Background and Completeness Correction} 
\label{sec:bkgcompcorr}

Following the detection of the GCs, a background correction was applied. The number of sources per unit area was computed for the region outside $2 R_e$ within the ACS FOV for each galaxy. This number was then scaled to the circular area of radius $2 R_e$ and removed from $N_{det}$, resulting in a background corrected $N_{det,bkg}$. As an example, Figure~\ref{fig:NGCbkgcomp} shows the distribution of the background corrected $N_{det,bkg}$ for MATLAS-2019. To check the validity of our background estimates and compute the uncertainties on the background, the background correction was also computed by i) only including the GC candidates detected in the CCD without the UDG, and ii) by only including the GC candidates detected outside $2 R_e$ in the CCD containing the UDG. We find that the median difference between these two latter methods for all UDGs in our sample is only $\sim 0.04$ counts ($\sim 1$\% of the background value). The standard deviation of all three methods was then used to estimate the error on the background counts.

\begin{figure}[ht!]
\centerline{
\includegraphics[width=\linewidth]{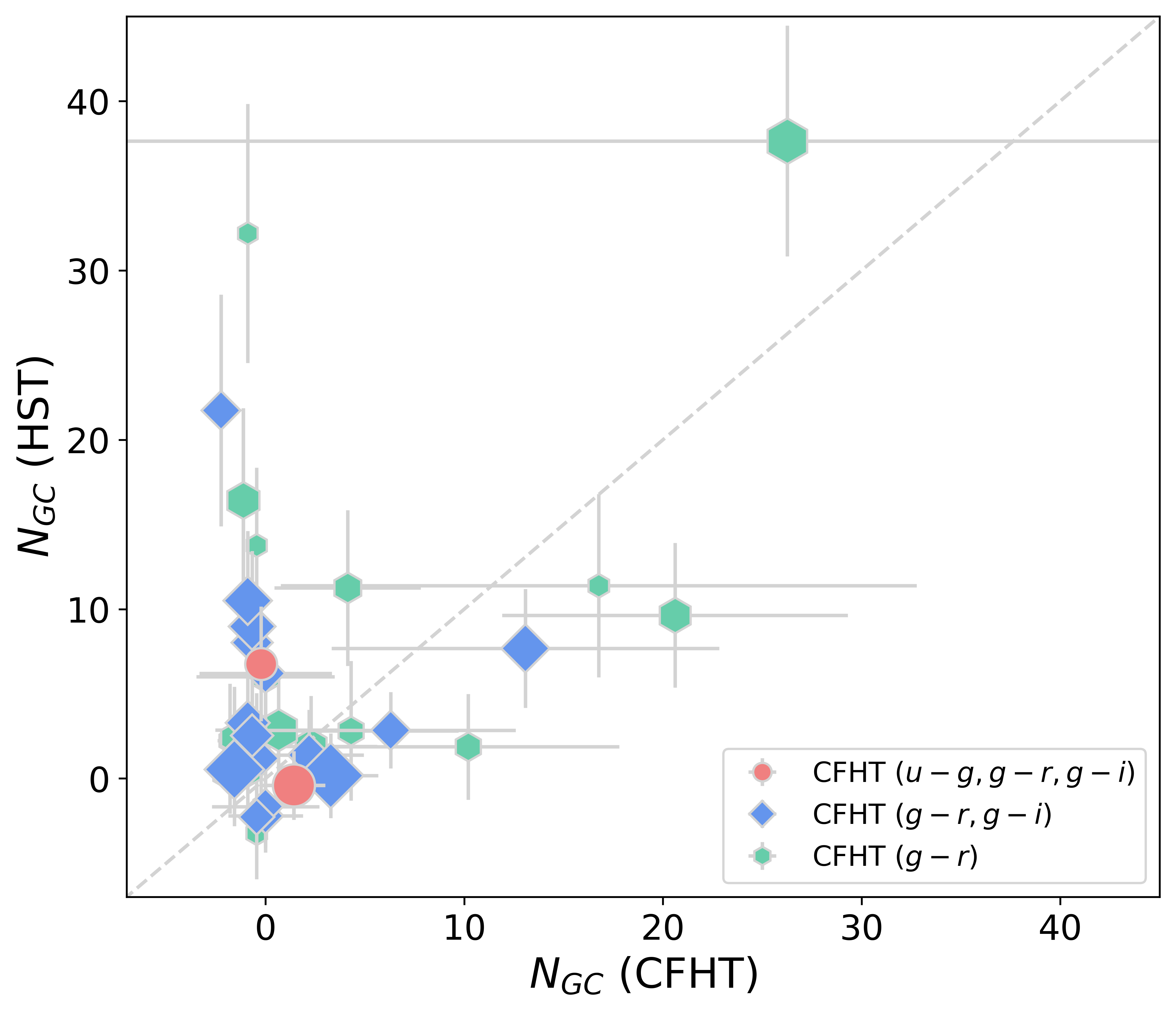}
}
\caption{Comparison between the GC counts measured in the MATLAS CFHT images and those measured from the HST images after background and completeness corrections have been applied. The {\it light coral}, {\it light blue} and {\it light green circles} are for GC detection in the CFHT images using three ($u-g,g-r,g-i$), two ($g-r,g-i$), and one ($g-r$) colors, respectively. The size of the symbol decreases with increasing distance. We do not find any trend between how many colors were used to identify the GCs in the CFHT images and the agreement with the HST GC numbers. As expected, the scatter is large but there is a clear agreement in the general trend, i.e., the UDGs with the largest number of GC candidates in the ground-based images remain the same UDGs that have the largest GC candidates in the higher resolution HST images. The large scatter at the low CFHT $N_{GC}$ values is most likely an effect of distance as the GC selection in the CFHT data was based on a fixed apparent magnitude cut. 
\label{fig:CFHT-HST}}
\end{figure}

\begin{figure}[ht!]
\includegraphics[width=\linewidth]{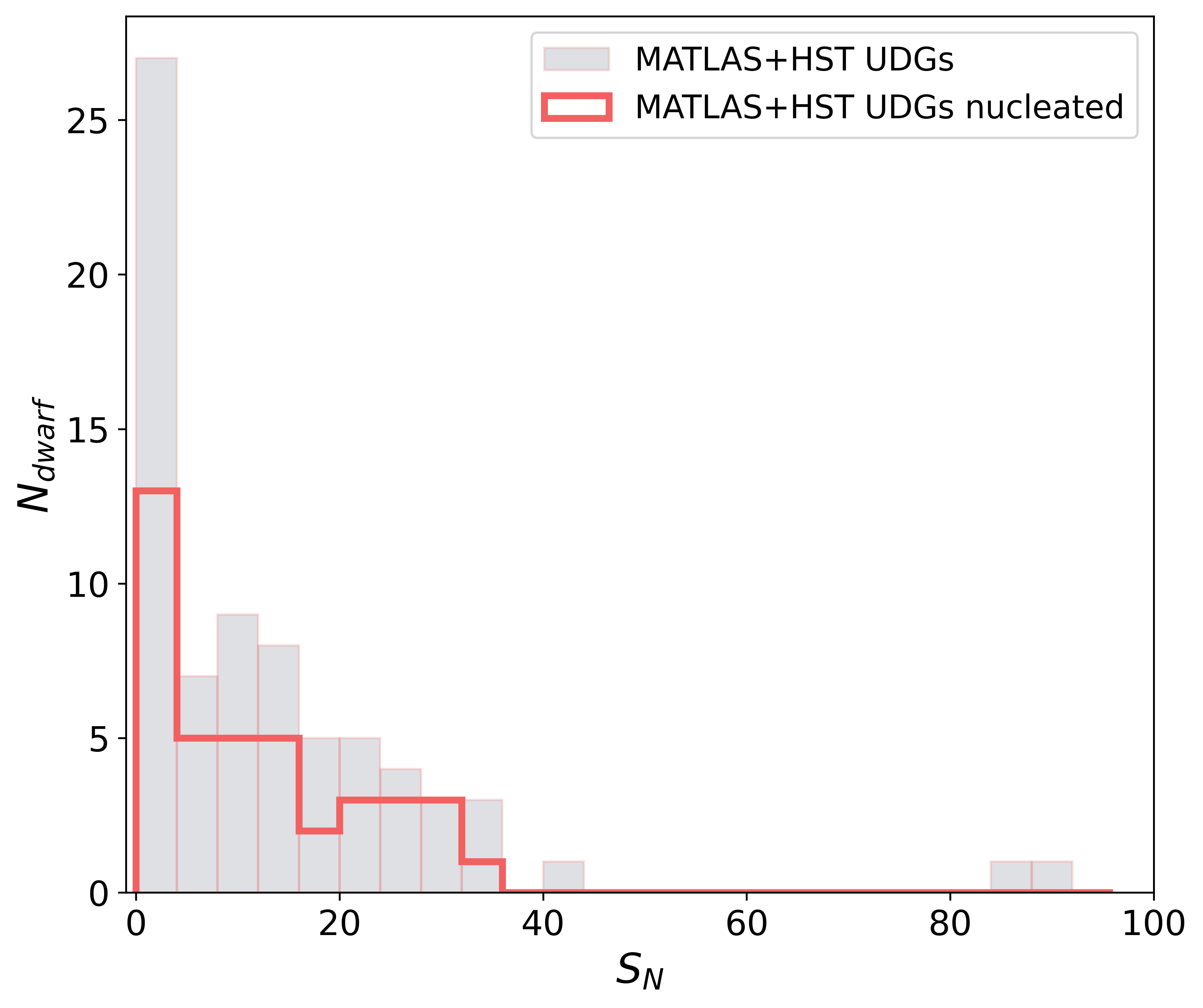}
\caption{Distribution of the specific frequency for the 74 UDG sample ({\it light grey}) and the 40 nucleated UDGs ({\it bright coral}). The nucleated UDGs are all found to have $S_N < 40$, consistent with the majority of the UDGs. The highest $S_N$ values are attributed to UDGs that are not identified as nucleated.
\label{fig:SNnuc}}
\end{figure}

Finally, a completeness correction was applied by assuming a GCLF represented by a Gaussian with peak absolute magnitude $M_V = -7.3$\,mag and a dispersion of 1 mag \citep{Miller2007, Peng2016, Lim2018}, and taking into account the completeness limit of each HST images. A correcting factor was then applied to the background corrected $N_{det,bkg}$ based on the part of the GCLF that is fainter than the observed limit, resulting in the quantity $N_{GC}$. 

After background subtraction and completeness correction, we calculate a total of 387 GCs associated with all 74 UDGs. The final $N_{GC}$ values for each of our sample galaxies are listed in Table~\ref{tab:sample} and their distribution shown in Figure~\ref{fig:NGCbkgcompUDG}. We find that 21 UDGs (28\%) have $N_{GC}$ values consistent with zero, with the tail of the distribution extending to our most GC-rich galaxy, MATLAS-2019, with $N_{GC} = 38$. The majority of the sample (47 UDGs or 64\%) are measured to have $N_{GC} < 3$, while the remaining 36\% host a GC system with $N_{GC}=3-38$. Our analysis reveals that the percentage of nucleated UDGs with a large number of GCs (14 UDGs or 35\% have $N_{GC} \geq 3$) is similar to the one for the whole sample. \footnote{A slightly more elevated fraction (16/38 or 42\%) of the most GC-rich UDGs with $N_{GC}=3-38$ are found to be from the UDGvD15 sample.}

MATLAS-2019 was the most GC-rich galaxy in the MATLAS dwarfs sample, based on CFHT and HST imaging \citep{Marleau2021,Mueller2021}, and remains the most GC-rich galaxy based on the new sample of galaxies with HST data. With our larger GC population, we confirm the result previously reported in \citet{Mueller2021}, that MATLAS-2019 does not have an anomalous GCLF, as has been suggested for NGC\,1052-DF2 and -DF4, which appear to have GCs that are too bright \citep{vanDokkum2018b}. As we re-analysed MATLAS-2019, we compare our results with those of \citet{Mueller2021} to check for consistency. We calculate for this UDG the background and completeness corrected count to be $N_{GC}=38\pm7$, consistent with $N_{GC}=36\pm6$ from \citep{Mueller2021}, using a similar method.

\begin{figure*}
\centerline{
\includegraphics[width=\textwidth]{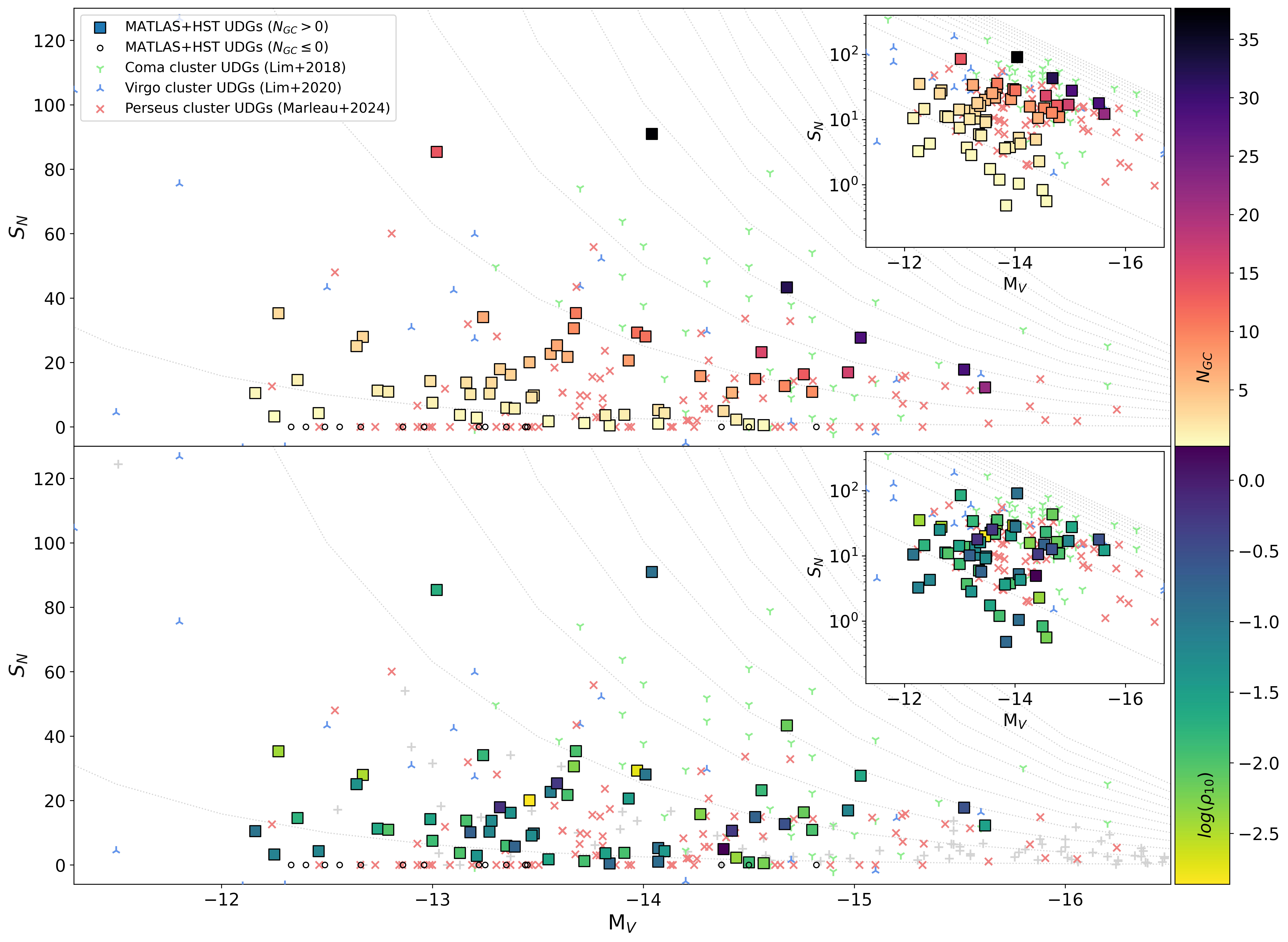}
}
\caption{{\it Top}: The specific frequency $S_N$ as a function of absolute magnitude for the 74 UDGs in our sample. The UDGs with $N_{GC} > 0$ are shown as {\it squares} with colors that scale with the value of the background and completeness corrected GC count, $N_{GC}$. The UDGs with $N_{GC} \leq 0$ are shown as {\it small black circles}. {\it Bottom}: Same as top but now showing the local density parameter, $log(\rho_{10})$, in the color bar. In both plots, the speciﬁc frequency of the Coma cluster UDGs (\citealt{Lim2018}; {\it light green symbols}), Virgo cluster UDGs (\citealt{Lim2020}; {\it light blue symbols}), and Perseus cluster UDGs (\citealt{Marleau2024}; {\it light red symbols}) are also plotted. The {\it light grey dashed lines} correspond to $S_N$ with a constant $N_{GC}$ of 1, 10, 20, ..., 90, 100. The error bars are not shown since they are quite large and can be found in Table~\ref{tab:sample}. The inset plots display the $S_N$ values using a logarithmic scale.
\label{fig:NgcSn}}
\end{figure*}

We examine how $N_{GC}$ varies as a function of the host galaxy stellar mass for our UDG sample. The results are displayed in Figure~\ref{fig:NGCmstar} along with the GC counts of massive early-type galaxies obtained from the HST ACS Fornax cluster survey \citep{Liu2019}. For dwarf galaxies with stellar masses below $M_* \sim 2 \times 10^8 M_{\odot}$, the number of GCs per UDG range from zero to 13, while the most massive UDGs with $M_* > 2 \times 10^8 M_{\odot}$ (where a jump in $N_{GC}$ is observed) host on average a larger number of GCs ($13-38$ GCs). The galaxies that extend higher at each stellar mass in terms of GC count are mostly those belonging to the UDGvD15 sub-sample, except in the two more massive bins which contain no galaxies from that sub-sample.

The comparison between the previously measured GC counts around the MATLAS dwarfs in the CFHT images \citep{Marleau2021} and those measured from the HST images after background and completeness corrections are shown in Figure~\ref{fig:CFHT-HST}. The GC detection in the ground-based CFHT images was done using either three ($u-g,g-r,g-i$), two ($g-r,g-i$), or one ($g-r$) colors, depending on the bands available. The GC color regions and magnitude limits used for the GC selection in the CFHT data are described in \citet{Marleau2021}. In particular, a g-band magnitude limit of 24.5, 24.0 and 23.5 was used for fields with three and more filters, two filters, and two filters when the seeing is worse than 1.1 arcsec, respectively.

For $N_{GC} > 2$, there is a clear agreement in the general trend at increasing GC richness. However, for $N_{GC} \leq 2$, the scatter is large, mostly for low (near zero) CFHT $N_{GC}$ values. The large scatter at the low CFHT $N_{GC}$ values is most likely an effect of distance as the GC selection in the CFHT data was based on a fixed apparent magnitude cut. Therefore, targets at larger distances of $\sim$\,40\,Mpc (smaller symbols in Figure~\ref{fig:CFHT-HST}), have a $\sim$\,1.5\,mag brighter GC selection limit in the CFHT data than those at smaller distances of $\sim$\,20\,Mpc (larger symbols in the same figure). Moreover, we expect to detect more GCs with increased spatial resolution within the higher surface brightness regions of the galaxy. This effect has been reported in the context NSCs, in which case a higher fraction of NSCs were identified in HST images as compared to lower resolution ground-based images (\citealt{Cote2006}; Poulain et al. 2024, in prep.). 

We do not find any trend between how many colors were used to identify the GCs in the CFHT images and the agreement with the GC numbers measured with the HST data. This seems to indicate that a single color is sufficient to identify the majority of GCs we have identified in the HST data, based on the methodology described above. However, at faint magnitudes ($\gtrsim$\,24\,mag in the g-band), Milky Way halo main sequence stars can become a major contaminant when using a single color selection \citep{Ferrarese2012}.

\begin{figure}[ht!]
\includegraphics[width=\linewidth]{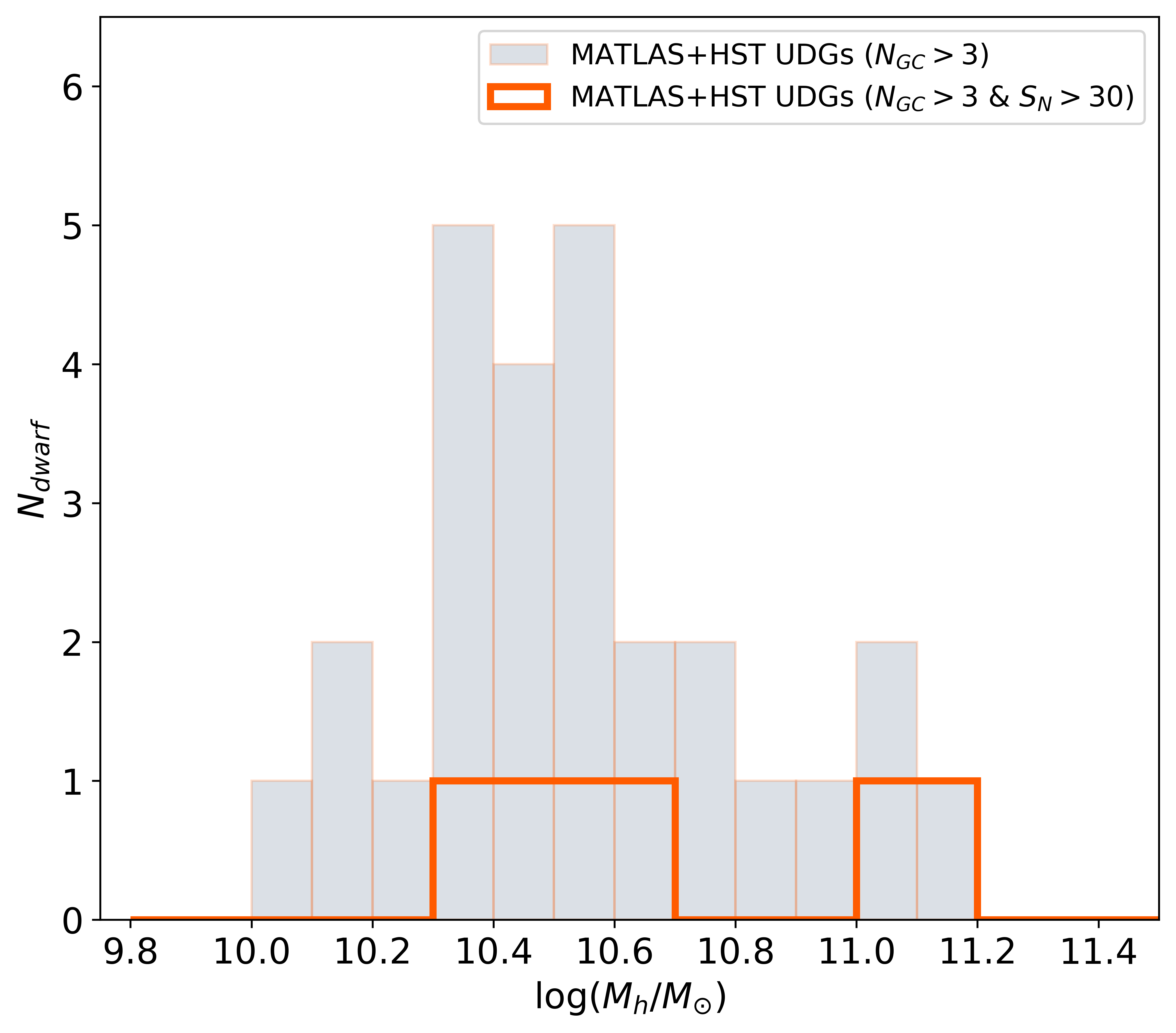}
\includegraphics[width=\linewidth]{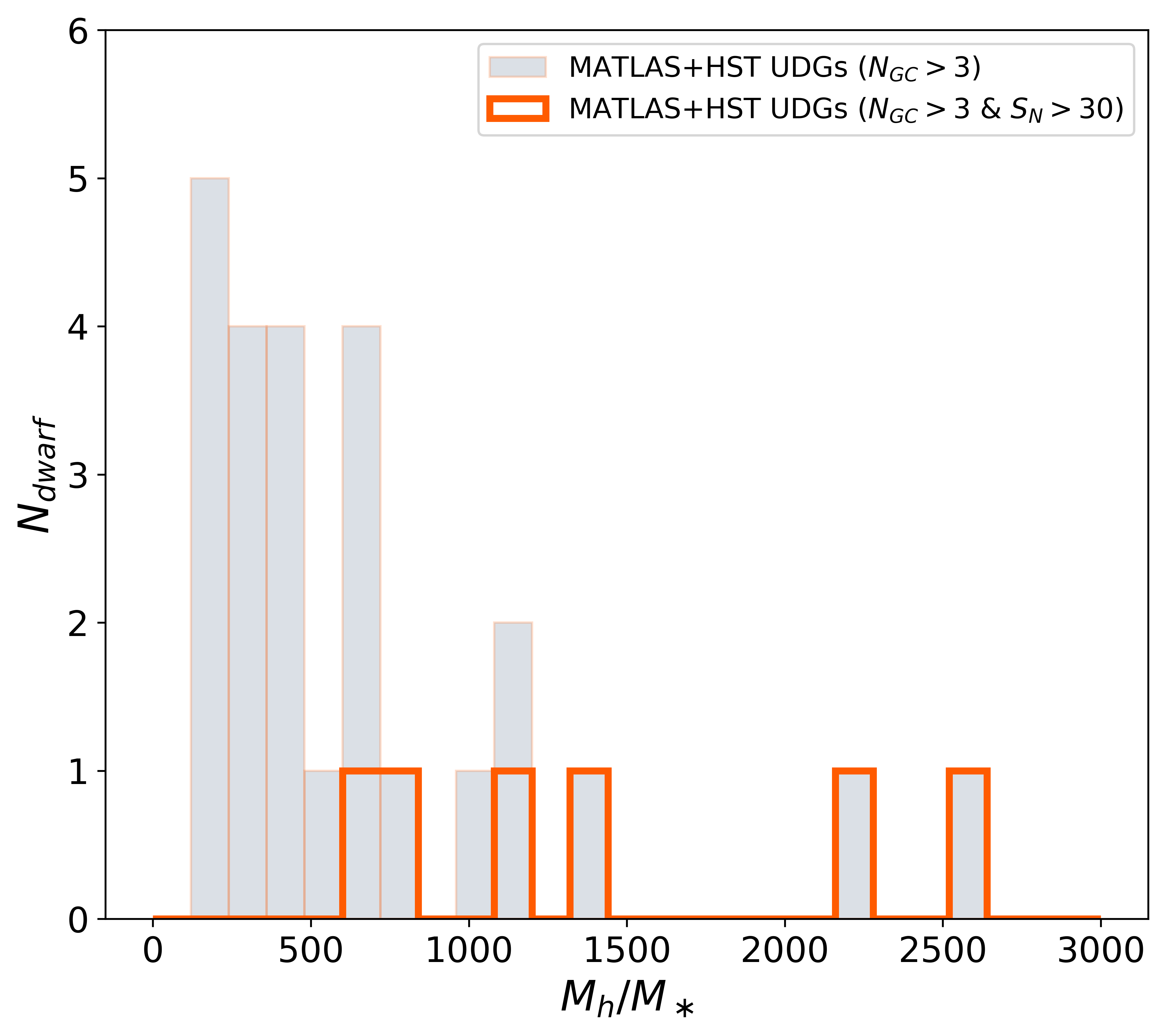}
\caption{{\it Top:} Distribution of the halo mass for the 27 out of the 74 UDGs with $N_{GC}>3$ ({\it gray}) and 6 out of the 27 UDGs with $N_{GC}>3$ and $S_N>30$ ({\it orange}). The UDGs with high $S_N$ are found to have a range of halo masses, with some at the high end of the distribution. {\it Bottom}: Same as above but showing the distributions of halo-to-stellar mass ratios. The stellar masses are from \citet{Habas2020}. The UDGs with the largest $S_N$ values are all found to have a halo-to-stellar mass ratios above $\sim$\,500.}
\label{fig:halotostarmass}
\end{figure}

\begin{figure*}
\centerline{
\includegraphics[width=0.52\textwidth]{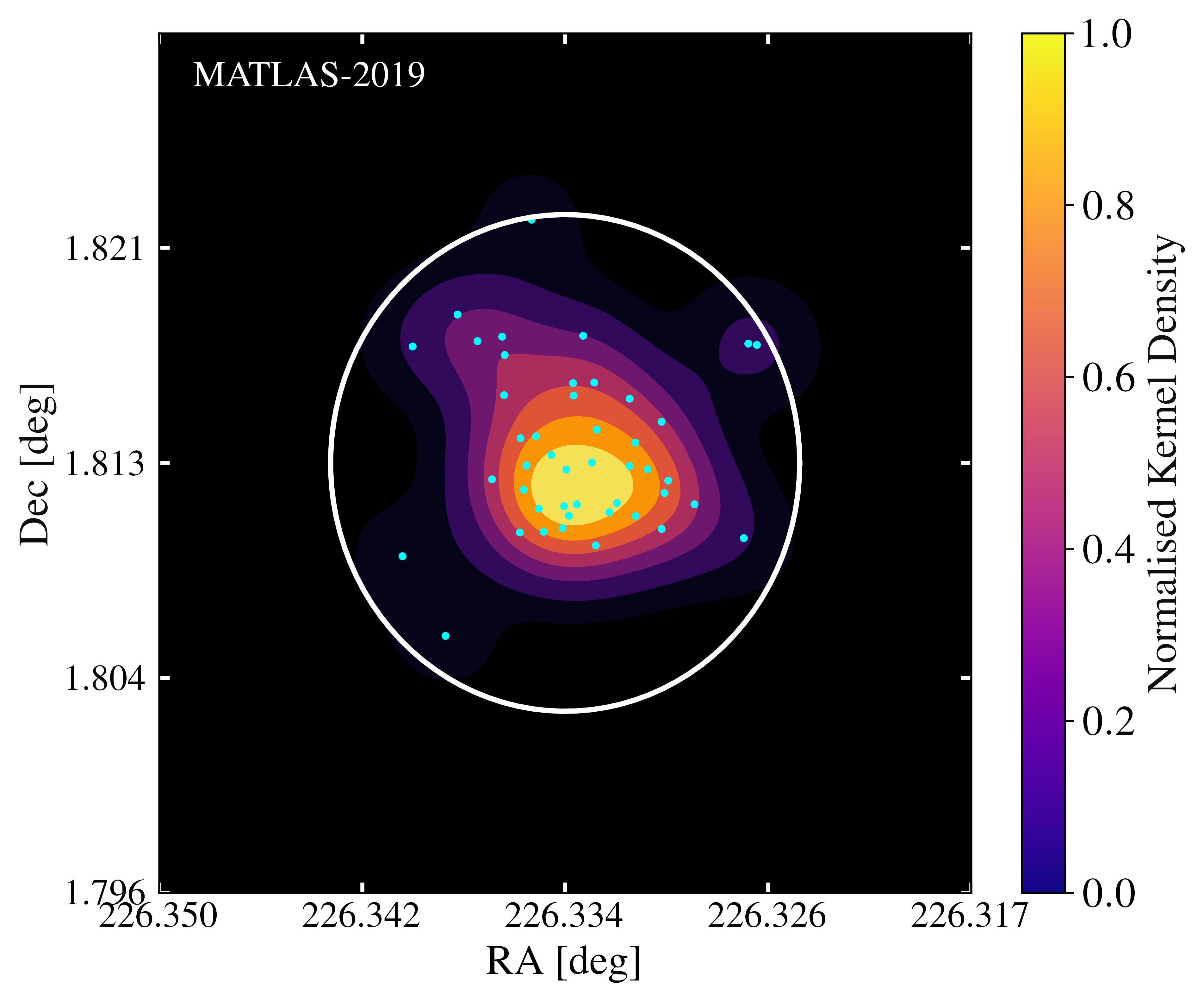}
\includegraphics[width=0.48\textwidth]{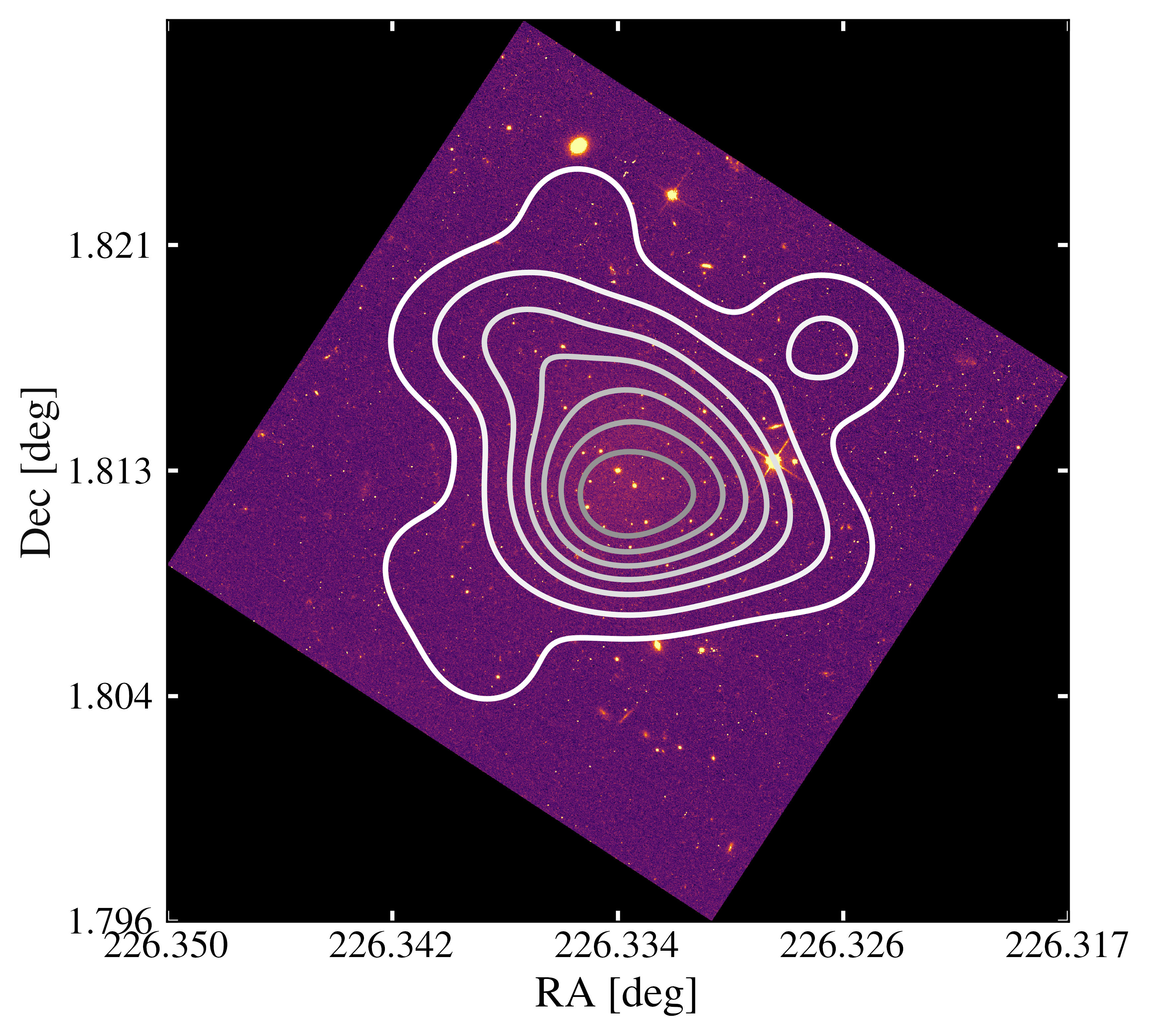}
}
\caption{{\it Left:} The projected distribution ({\it cyan dots}) and density map ({\it colored map}) of the GC candidates detected within 2$R_e$ ({\it large white circle}) for one of the 12 UDGs with $N_{GC}>10$, MATLAS-2019. {\it Right:} The density map ({\it white contours}) of the GC candidates is overlayed on the HST $F814W$ image of the UDG. The images are 2.5$R_e$ on a side.
\label{fig:densitymap}}
\end{figure*}

\subsection{Specific Frequency}

The specific frequency, $S_N$, for GCs is a measure used to quantify the abundance of GCs in a galaxy relative to its total luminosity. It provides insights into the efficiency of GC formation within a galaxy and the conditions and processes that prevailed during the early stages of galaxy formation and evolution. It is a useful tool for understanding the impact of galaxy interactions and mergers on cluster formation and investigating the interplay between GCs and the overall stellar content of a galaxy. Prior studies of the GC population of UDGs in the Coma and Virgo clusters \citep{Lim2018,Lim2020} have claimed that the $S_N$ of these UDGs varies dramatically, with the mean $S_N$ being higher for UDGs than for classical dwarf galaxies.

The specific frequency $S_N$ \citep{Harris1981} was computed using the formula:
\begin{equation}
S_N=N_{GC}\cdot10^{0.4(M_V+15)}
\end{equation}

where $N_{GC}$ is the total, background and completeness corrected number of GCs and $M_V$ is the host galaxy absolute magnitude in the $V$-band. As the number of GCs is a function of the brightness of the host galaxy, we show in Figure~\ref{fig:NgcSn} the computed values of $S_N$, as a function of the host galaxy absolute magnitude in the $V$-band, $M_V$, taken from \citet{Poulain2021}. The $N_{GC}$ values associated with each UDG are displayed in the top diagram of Figure~\ref{fig:NgcSn} using the color map. In particular, there are 7 UDGs (9\%) with $S_N > 30$ (MATLAS-405, 585, 658, 1413, 1534, 2019, 2184) and these galaxies are the ones making up the extended tail component of the GC count distribution displayed in Figure~\ref{fig:NGCbkgcompUDG}. \footnote{All of the UDGs in our sample with $S_N>30$ would be defined as UDGs even when using the stricter selection criteria defined by \citet{vanDokkum2015}.}

The specific frequency of nucleated UDGs are of special interest, because the NSC could be formed by merging GCs \citep{Fahrion2021,Fahrion2022}. However, we find no statistically significant difference between the GC populations of the nucleated and non-nucleated UDGs in our sample. As shown in Figure~\ref{fig:SNnuc}, the nucleated UDGs all have $S_N<40$, which is consistent with the majority of the MATLAS UDGs.

In Figure~\ref{fig:NgcSn}, we compare the $S_N$ versus $V$-band magnitude relation for our UDG sample with the values measured for the UDGs in the Virgo cluster \citep{Lim2020}, the Perseus cluster \citep{Marleau2024} and the Coma cluster \citep{Lim2018}. We calculate that the median value of our sample with specific frequencies greater than zero, $S_N=12.7$, is similar to the median value of 10.7 for UDGs in the Perseus cluster but less than the ones for the Virgo and Coma clusters ($S_N=42.4$ and $S_N=30.6$, respectively). Furthermore, the range of $S_N$ values (with $S_N>0$) of the MATLAS UDGs ($S_N=0.5-91.0$) extends beyond the Perseus $S_N$ values ($S_N=1.0-60.1$) but stays below the highest $S_N$ values measured for the Virgo and Coma UDGs ($S_N=187.1$ and $S_N=348.3$, respectively). The comparison of the range and median values of the samples are given in Table~\ref{tab:specfreq}.

\begin{table}
    \caption{Comparison of the specific frequencies of the UDGs in MATLAS with the ones in the Virgo, Perseus and Coma clusters.}
    \label{tab:specfreq}
    \centering
    \begin{tabular}{lrrr}
    \hline
    \hline
    Survey        & Min       & Max & Median \\
                  & ($S_N>0$) &     & ($S_N>0$)\\
    \hline
    \noalign{\smallskip}
     MATLAS+HST UDGs                                & 0.5 & 91.0 & 12.7 \\
     MATLAS+HST nuc-UDGs\tablefootmark{*}           & 0.6 & 35.3 & 12.7 \\
     Virgo UDGs (Lim+2020)                          & 1.5 & 187.1 & 42.4 \\
     Perseus UDGs (Marleau+2024)                    & 1.0 & 60.1 & 10.7 \\
     Coma UDGs (Lim+2018)                           & 2.1 & 348.3 & 30.6 \\
    \hline
    \end{tabular}
\tablefoot{
\tablefoottext{*}{Considering only the nucleated UDGs.}
}    
\end{table}

Previous studies have shown that low-mass galaxies in denser environments can have higher $S_N$ \citep{Peng2008,Mistani2016,Lim2018}. This may be due to the formation of more GCs relative to the galaxy luminosity, resulting from an increase in interactions and/or mergers at high redshift. It is also possible that in high density peaks at high redshift (proto-galaxy cluster environments), the SF rate density was higher, and consequently, so was the cluster formation efficiency. To explore this effect for our sample of UDGs, we use the local density parameter $\rho_{10}$ described in \citet{Habas2020} for the MATLAS sample. The values of $log(\rho_{10})$ for our UDGs are displayed in the bottom diagram of Figure~\ref{fig:NgcSn} using the color map and are in the range $-2.86 < log(\rho_{10}) < 0.24$. Based on the definition in \citet{Cappellari2011} of $log(\rho_{10}) > -0.4$ for galaxies in the Virgo cluster, we calculate that 69 out of the 74 UDGs (93\%) fall in the lower density side of this cut. The five UDGs (MATLAS-1437, 1412, 1400, 1470, 1485) that are located in Virgo-like densities ($ -0.29 < log(\rho_{10}) < 0.24$) are from fields on the outskirts of the Virgo cluster. The UDGs with a high $S_N$ at the same given galaxy luminosity as the UDGs in the Virgo and Perseus clusters have low $\rho_{10}$ values.

\subsection{Halo Mass}

The total number of GCs in a given galaxy has been shown to correlate over six orders of magnitude with the modelled virial mass of the host dark matter halo (\citealt{Harris2013,Beasley2016b,Burkert2020}). The scaling relation only appears to flatten for halos with virial masses smaller than $10^{10}$\,M$_{\odot}$ or $N_{GC} \lesssim 3$ (\citealt{Burkert2020}, see their Figure 1), although the predictive power of the relation is reduced below $N_{GC} \sim 15$ due to the increased scatter. There is observational evidence that the correlation holds at these low masses \citep{Zaritsky2022b} but more studies are needed to understand the physical mechanism that drives this trend.

According to \citet{Harris2017}, the virial mass $M_{h}$ of a galaxy is connected to the total mass of the GC system $ M_{GC,tot}$ via the following formula:
\begin{equation}
    M_{GC,tot}/ M_{h} = 2.9\times10^{-5}
\end{equation}

As in \citet{Marleau2021}, we assume a mean mass of a GC to be $1\times10^5$\,M$_{\odot}$ for dwarf galaxies \citep{Harris2017} and therefore multiply that number by $N_{GC}$ to compute $ M_{GC,tot}$. Therefore:
\begin{equation}
    M_{h} = 3.45\times10^9 \, N_{GC,tot}
\end{equation}

The distribution of the halo masses for the UDGs in our sample are displayed in Figure~\ref{fig:halotostarmass}. Based on the higher level of reliability of the scaling relation between halo mass and GC count for $N_{GC}>3$, only the 27 UDGs with $N_{GC}>3$ are used to produce the distribution shown in the figure. The values range from $log(M_{h}) \sim 10.05\pm0.43$ to $11.11\pm0.08$\,$M_{\odot}$, with a median value of $log(M_{h}) \sim 10.54$~$M_{\odot}$. Out of these UDGs, the 7 with the highest specific frequency, defined as $S_N>30$ (a cut that includes $\sim$\,25\% of the UDGs), are found to have a range of halo masses, with some at the high end of the distribution. The distributions of halo-to-stellar mass ratios for the 25 out of the 74 UDGs with both $N_{GC}>3$ and stellar mass estimates from \citet{Habas2020} are also shown in Figure~\ref{fig:halotostarmass}. The values range from $M_{h} \sim 176\pm142$ to $2542\pm1430$\,$M_\ast$ with a median value at $\sim$\,472\,$M_\ast$, in agreement with the sample of 59 UDGs of \citet{Marleau2021}. Therefore, our UDGs with $N_{GC}>3$ are consistent with having significant dark matter halos, although it is important to note that the uncertainties in the halo-to-stellar mass ratios can be substantial, as shown in the values quoted above. The UDGs with the largest $S_N$ values are all found to have a halo-to-stellar mass ratios above $\sim$\,500.

\section{Spatial distribution} 
\label{sec:radial}

The spatial distribution of GCs, i.e., the arrangement and pattern of the GCs relative to their host galaxies, holds valuable insights into the formation history, dynamics, and interactions of both the GCs and their host galaxies \citep{Kruijssen2015, Renaud2017, Li2019, ReinaCampos2022}. Of particular interest are the radial distribution, (a-)symmetries, and spatial anisotropy of the GC populations. By separating the GCs in the core and outer regions, the radial distribution may provide clues about the dynamical properties and formation mechanisms of the host and GC population; GCs in the outer regions can also provide information about the underlying dark matter distribution of the host galaxy. Asymmetric distributions, meanwhile, are indicators of recent interactions, mergers, or gravitational influences from nearby companions. Finally, the orientation and elongation of the spatial distribution is expected to trace the rotation and overall shape of the host galaxy, past tidal interactions, and/or indicate preferential directions from which GCs were acquired or formed. 

\subsection{Density maps}
\label{sec:densitymaps}

Given the large number of GCs ($N_{GC} > 10$) for 12 UDGs in our HST sample, we were able to examine the two-dimensional distribution of GCs associated with the following galaxies: MATLAS-1938, 2019 and 799 (all with $d <=25$~Mpc) and MATLAS-42, 585, 1413, 1779, 1616, 138, 401, 1332, and 1534 (all with $d > 25$~Mpc). An example of the projected distribution and density map for MATLAS-2019 is shown in Figure~\ref{fig:densitymap}. For all 12 UDGs, the density maps are displayed in Appendix~\ref{AppendixB}, Figure~\ref{fig:densitymapall}. The density maps for the 12 UDGs can be classified into the following types:\\

\noindent {\it Symmetric}: MATLAS-1938 and 1413 appear mostly symmetric in their GC and stellar light distributions. MATLAS-1413 has an extremely low surface brightness ($\mu_0$~=~26.21~mag/arcsec$^2$) which makes it difficult to see in the figure. \\

\noindent {\it Elongated}: MATLAS-799, 1534, 1616 and 2019 have slightly elongated GC distributions. These match well the elongation of the stellar light distribution. MATLAS-1534 has a close (low-mass) neighbour with extended LSB features stretching in its direction and therefore is possibly part of an interacting system. MATLAS-2019 had a slight elongation in the GC distribution towards the massive early-type galaxy NGC~5838. \\

\noindent {\it Off-center}: MATLAS-138 and 1332 have symmetric distribution that are off-center. Both are classified as nucleated dwarf elliptical. In the case of MATLAS-138, the offset in the distribution is possibly due to the bright galaxy contaminant located to the North-West (NW) of the galaxy, in the opposite direction to the offset. MATLAS-1332 has a bright nucleus and shows a small offset in the GC distribution towards the North. \\

\noindent {\it Asymmetric and off-center}: MATLAS-42, 401, 585 and 1779 show an asymmetric and off-center GC distribution as compared to the stellar light. All galaxies have a disturbed stellar light morphology and except for MATLAS-585, are classified as dwarf irregulars. In all cases, the GC distribution follows the disturbed stellar distribution. MATLAS-42 is detected in HI \citep{Marleau2021} and has a nearby dwarf companion (MATLAS-41) situated NW of the galaxy. The large field-of-view of the CFHT observations allows us to see that MATLAS-1779 displays tidal features extending in the NE and SW directions, i.e.\ in the direction of the massive early-type galaxy NGC~5493. The GC density clump located in the NE of the UDG follows this tidal feature. In general, a disturbed morphology does not appear to correlate with a high $S_N$ value. \\

It is important to point out that the completeness limits of the detected GCs could introduce some biases in the above results, in particular if fainter (less massive) GCs have an intrinsically wider spatial distribution than brighter (more massive) GCs, as expected from tidal friction \citep{Tremaine1975}. In addition, for UDGs at $d < 25$~Mpc, the concentration cut might have missed some unresolved compact (with half-light radii $\sim 2$\,pc) GCs which may also cause a bias as compact GCs are usually more spatially concentrated than the more extended clusters found in the outer halos of massive galaxies \citep{Huxor2014}.  

\begin{figure}[ht!]
\includegraphics[width=0.84\linewidth]{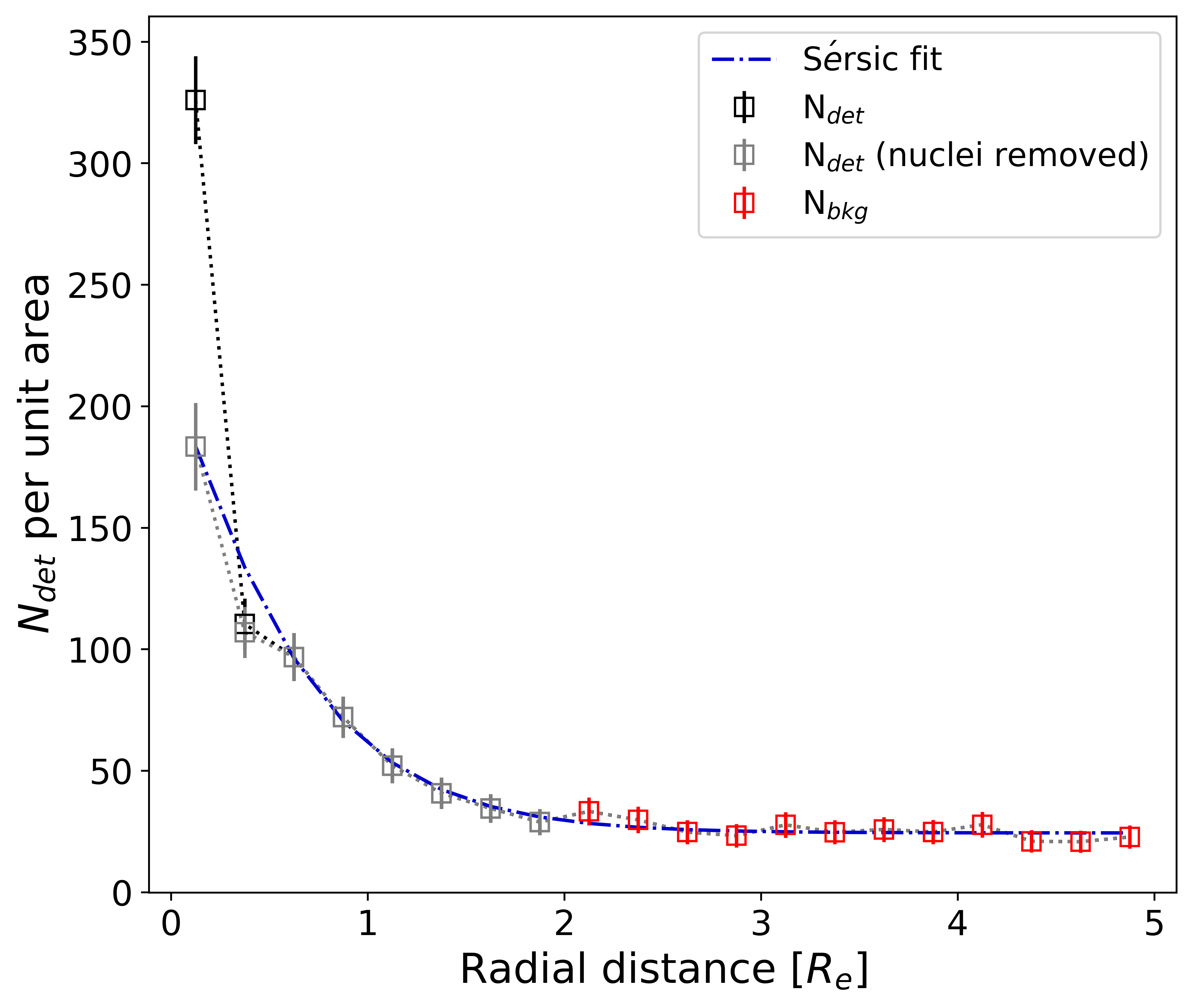}
\includegraphics[width=0.84\linewidth]{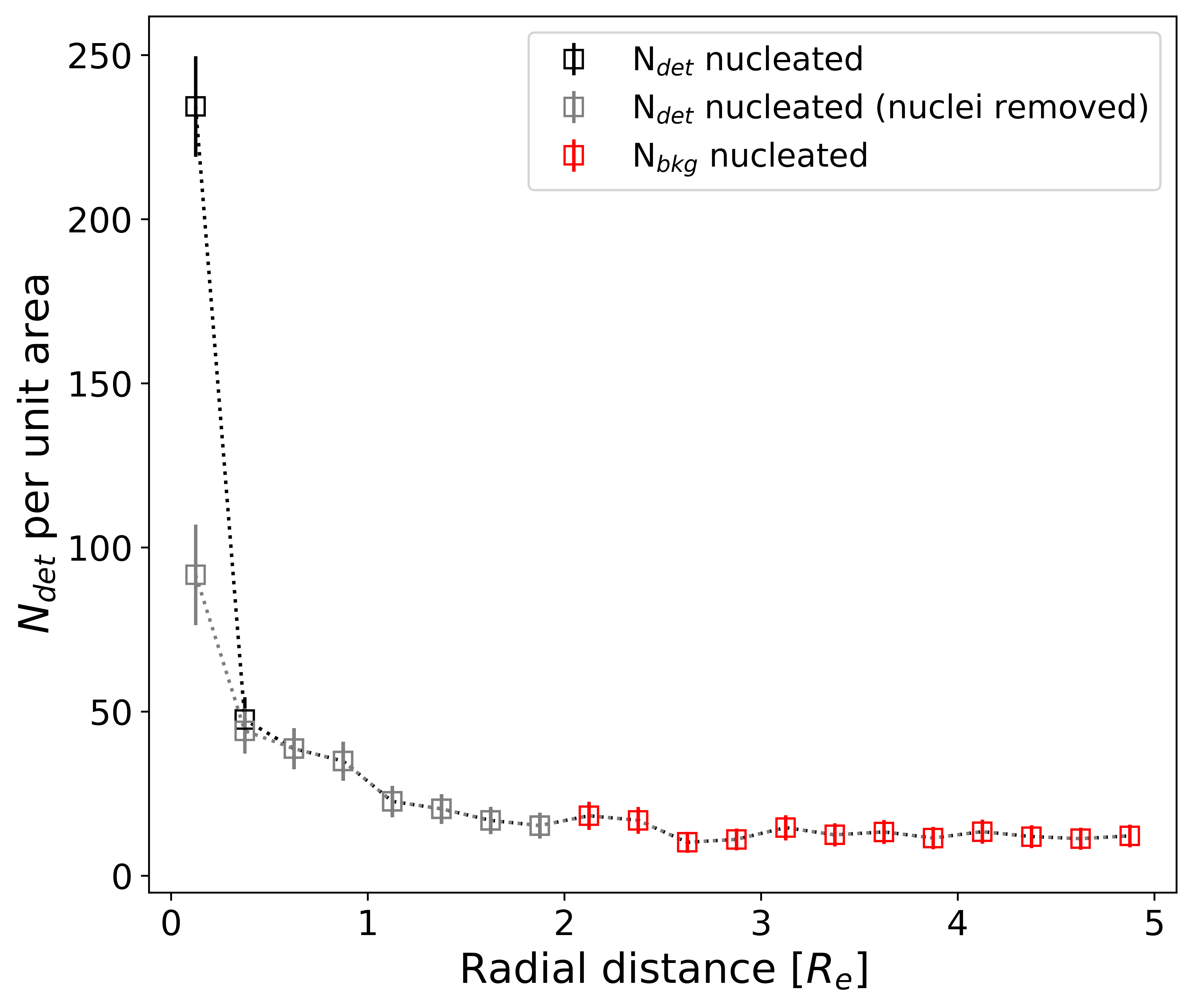}
\includegraphics[width=0.84\linewidth]{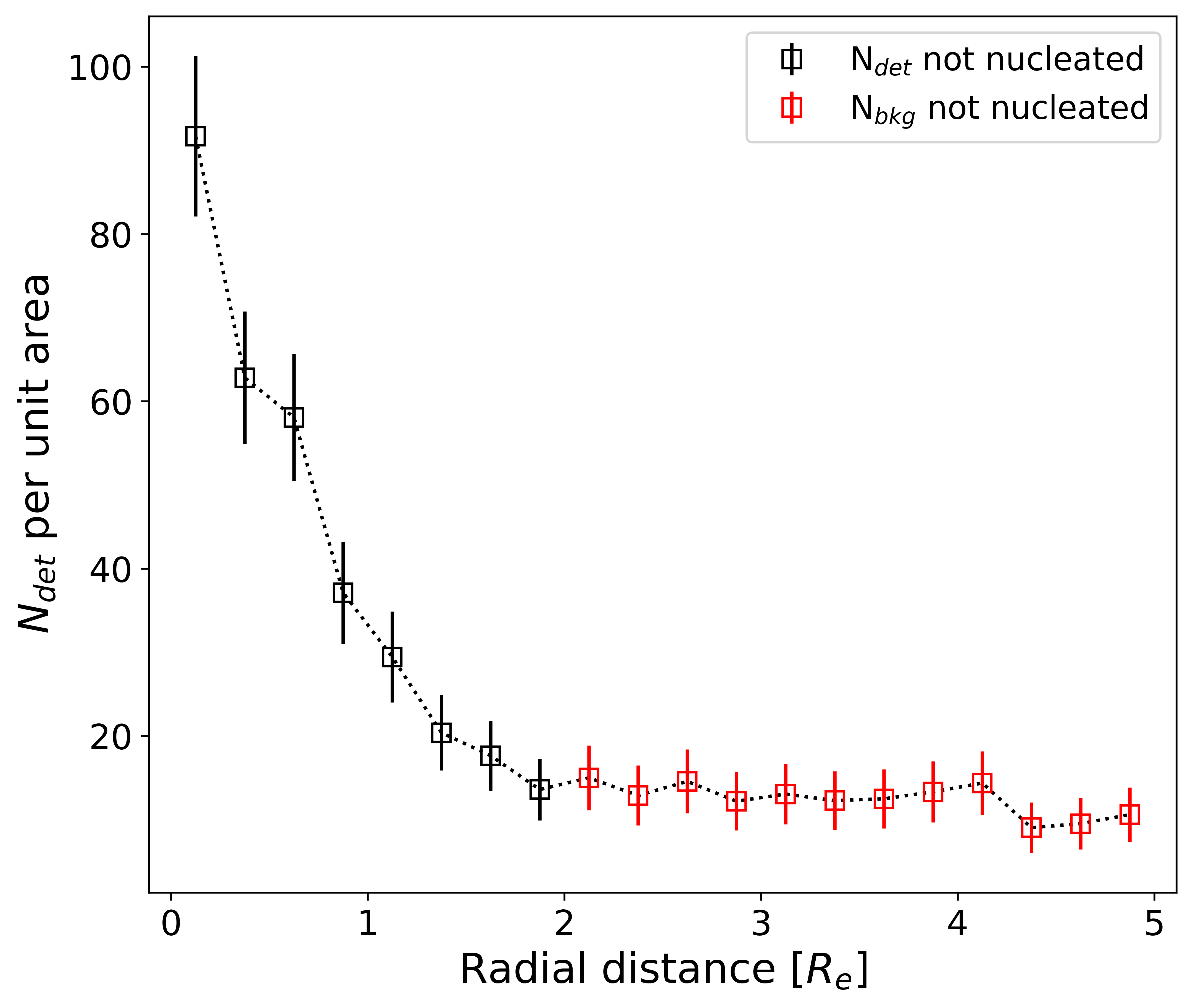}
\caption{{\it Top}: Radial distribution of the stacked GCs for all UDGs in our HST sample. The {\it black squares} are for the GC counts within 2$R_e$ while the {\it red squares} are the counts outside (background). The counts are also shown with the detection corresponding to the candidate nucleus (or nuclei, in the case of MATLAS-138 and 987) in each galaxy removed ({\it grey points and dotted line}). The S\'ersic fit is shown with the {\it dash-dot blue line}. {\it Middle}: Same as {\it Top} but for the nucleated UDGs only. {\it Bottom}: Same as {\it Top} but for the non-nucleated UDGs only.
\label{fig:raddistall}}
\end{figure}

\begin{figure}[ht!]
\includegraphics[width=0.84\linewidth]{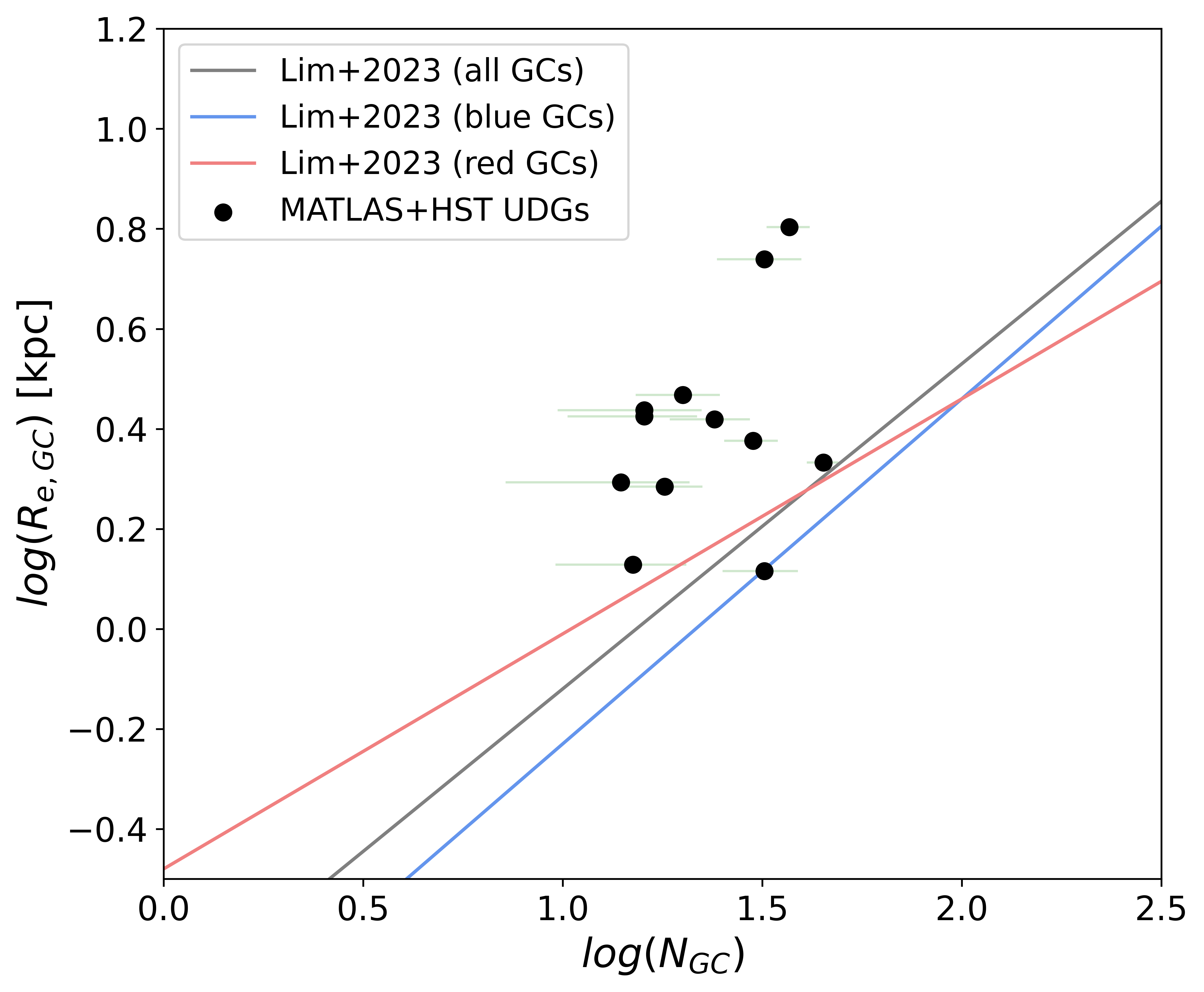}
\includegraphics[width=0.84\linewidth]{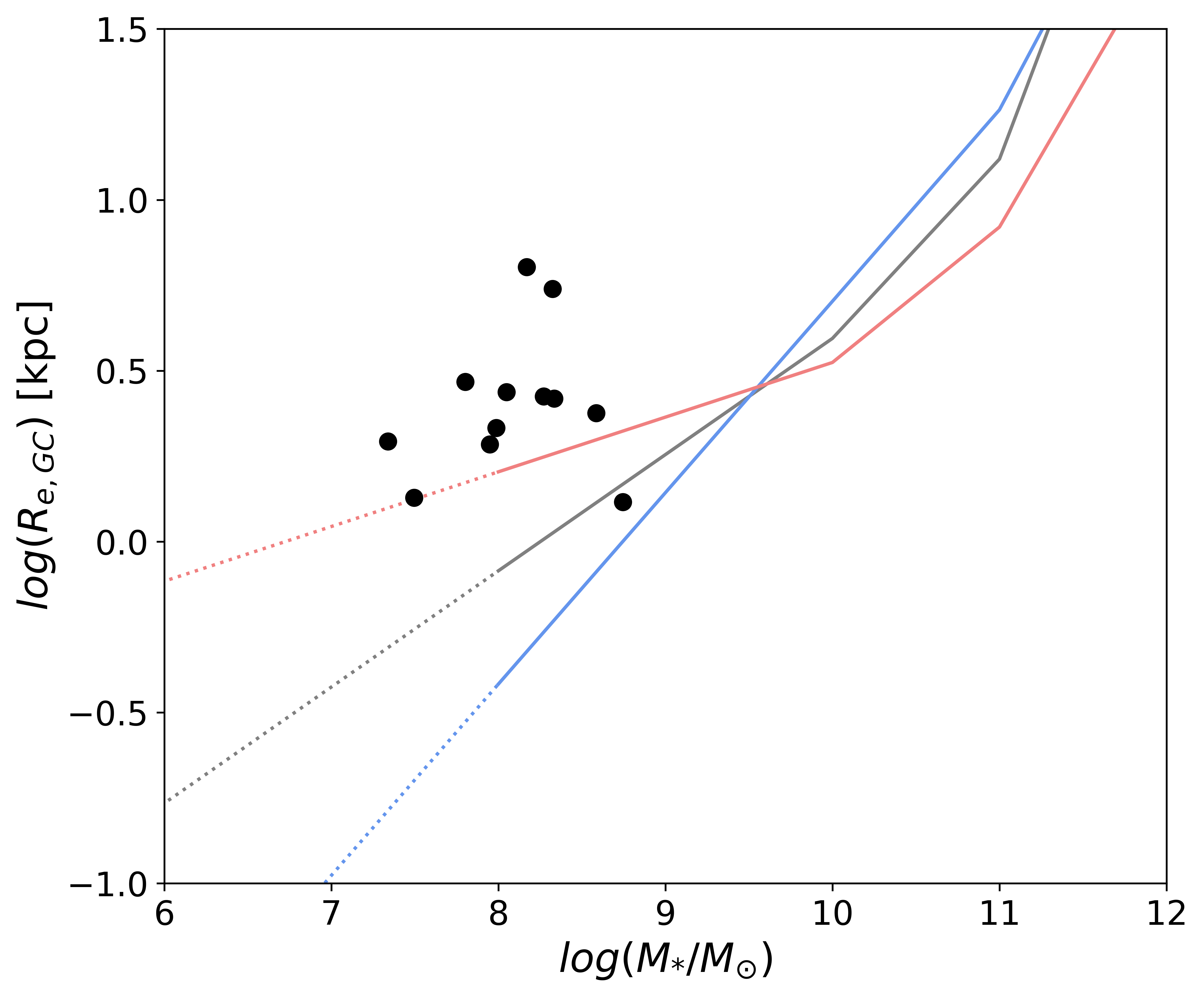}
\includegraphics[width=0.84\linewidth]{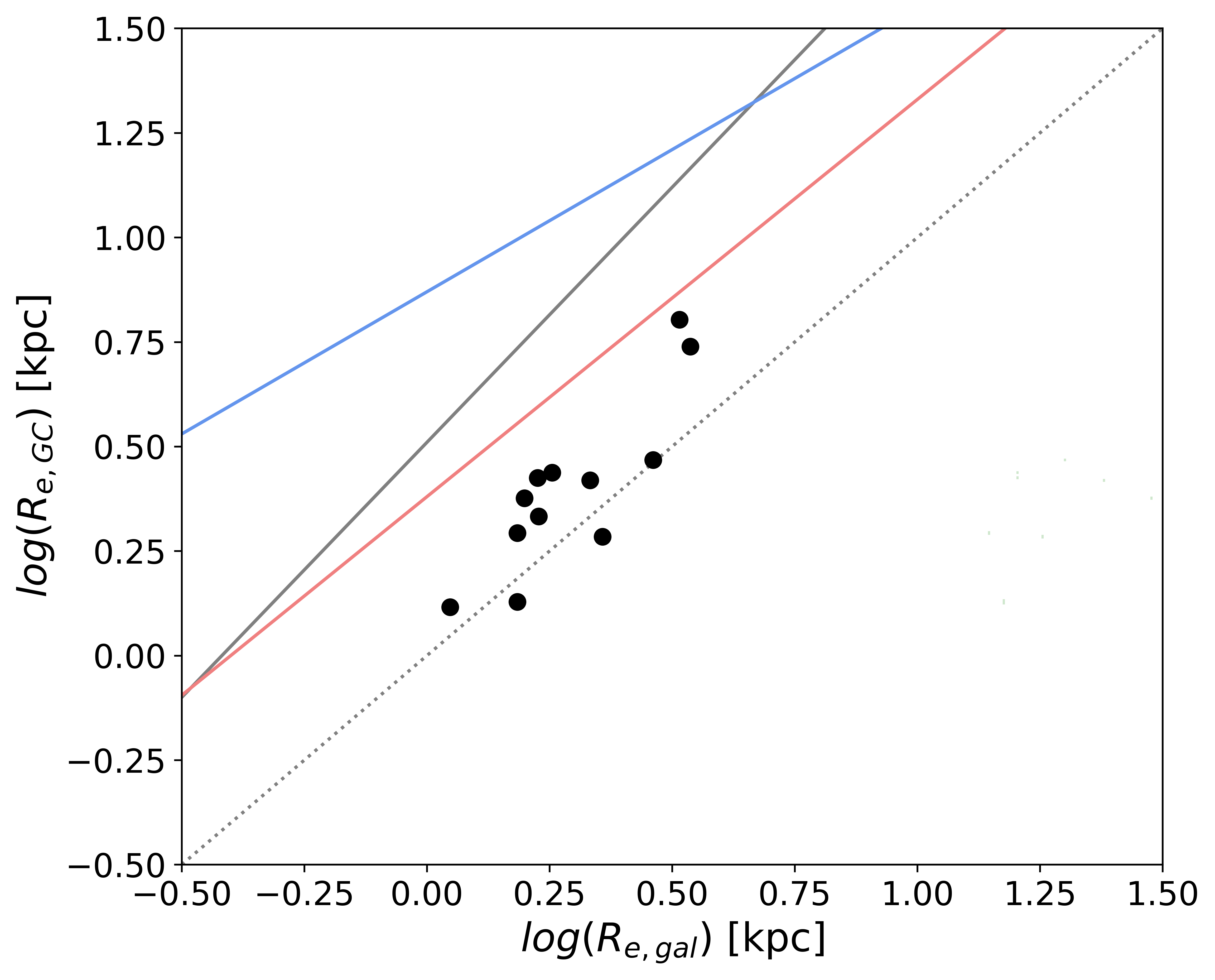}
\caption{{\it Top}: The effective radii of GC systems plotted against the total number of GCs for the 12 UDGs with $N_{GC}>10$ ({\it black filled circles}). {\it Middle}: The effective radii of GC systems plotted against the stellar masses of their host galaxies for the same 12 UDGs. Both plots show no obvious trends. {\it Bottom}: The effective radii of GC systems plotted against the effective radii of their host galaxies for the same 12 UDGs. We find that the effective radius of the GC system matches the effective radius of the galaxy ({\it grey dotted line}), indicating that the radial distribution of GCs follows the galaxy light. For all plots, the best-fit relations of \citet{Lim2024} are shown by the {\it grey line} (all GCs), the {\it blue line} (blue GCs), and the {\it red line} (red GCs). The error bars are shown for both axes in all plots but are only visible when larger than the symbols.
\label{fig:raddist}}
\end{figure}

\begin{figure*}
\centerline{
\includegraphics[width=0.33\textwidth]{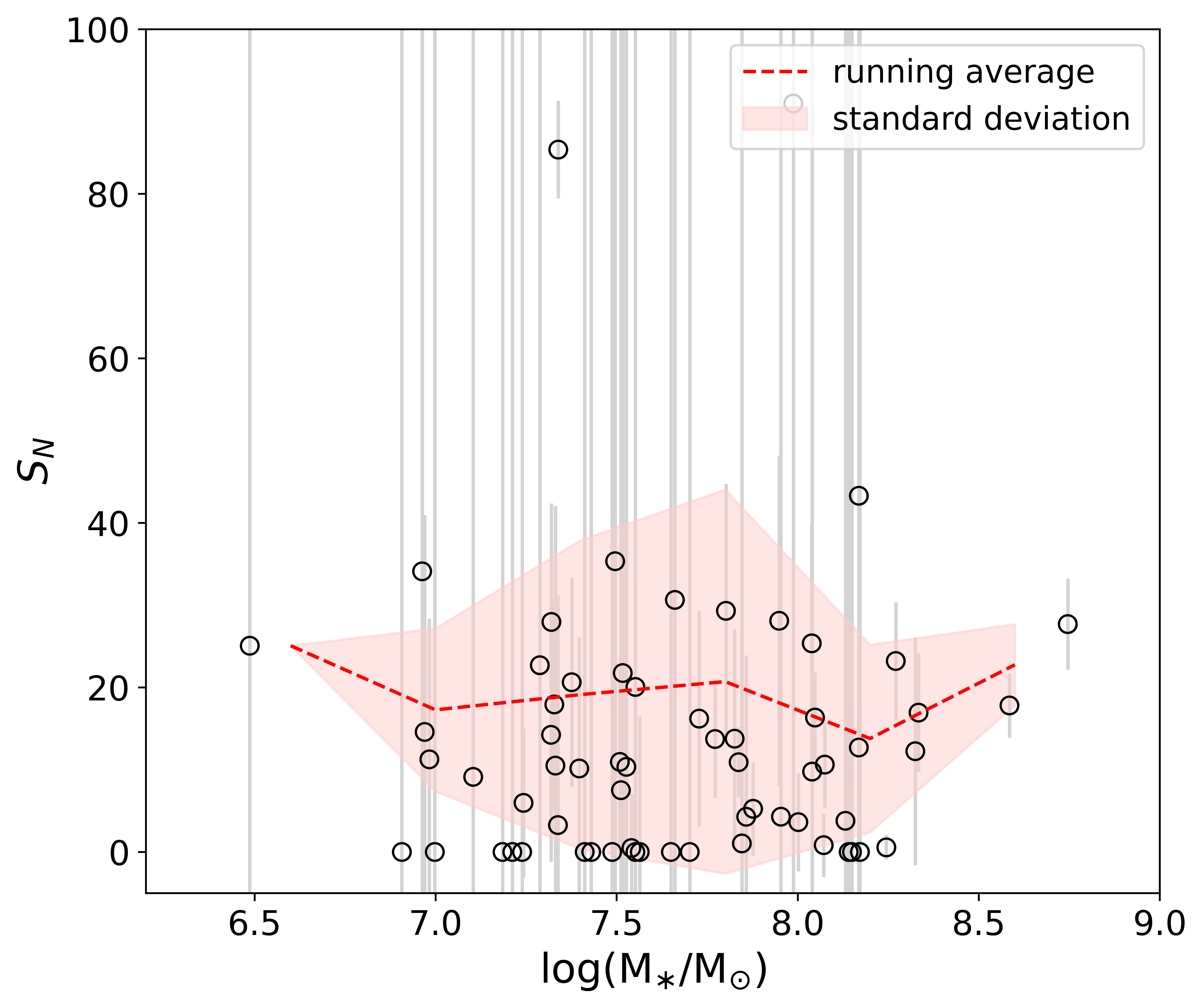}
\includegraphics[width=0.33\textwidth]{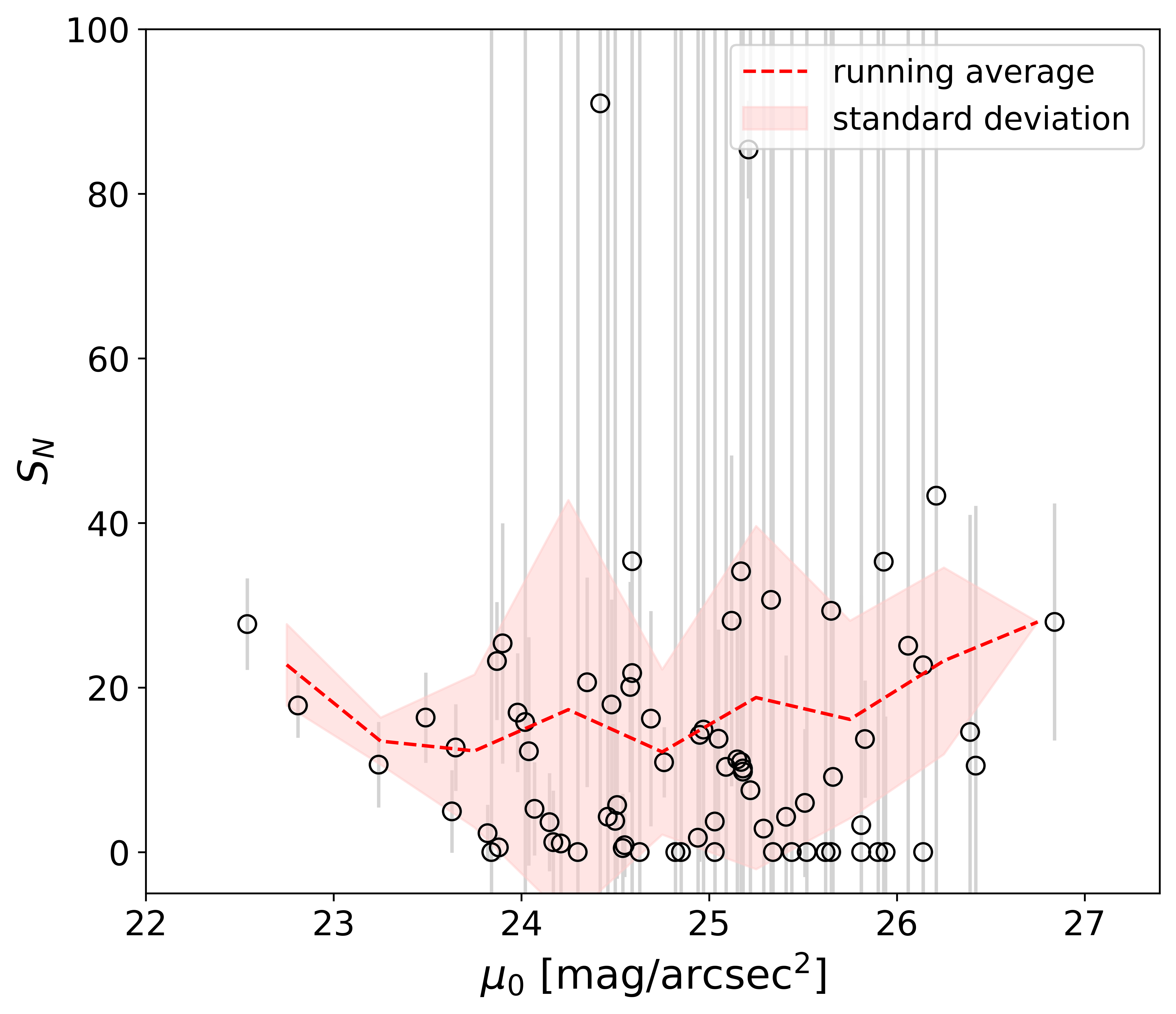}
\includegraphics[width=0.33\textwidth]{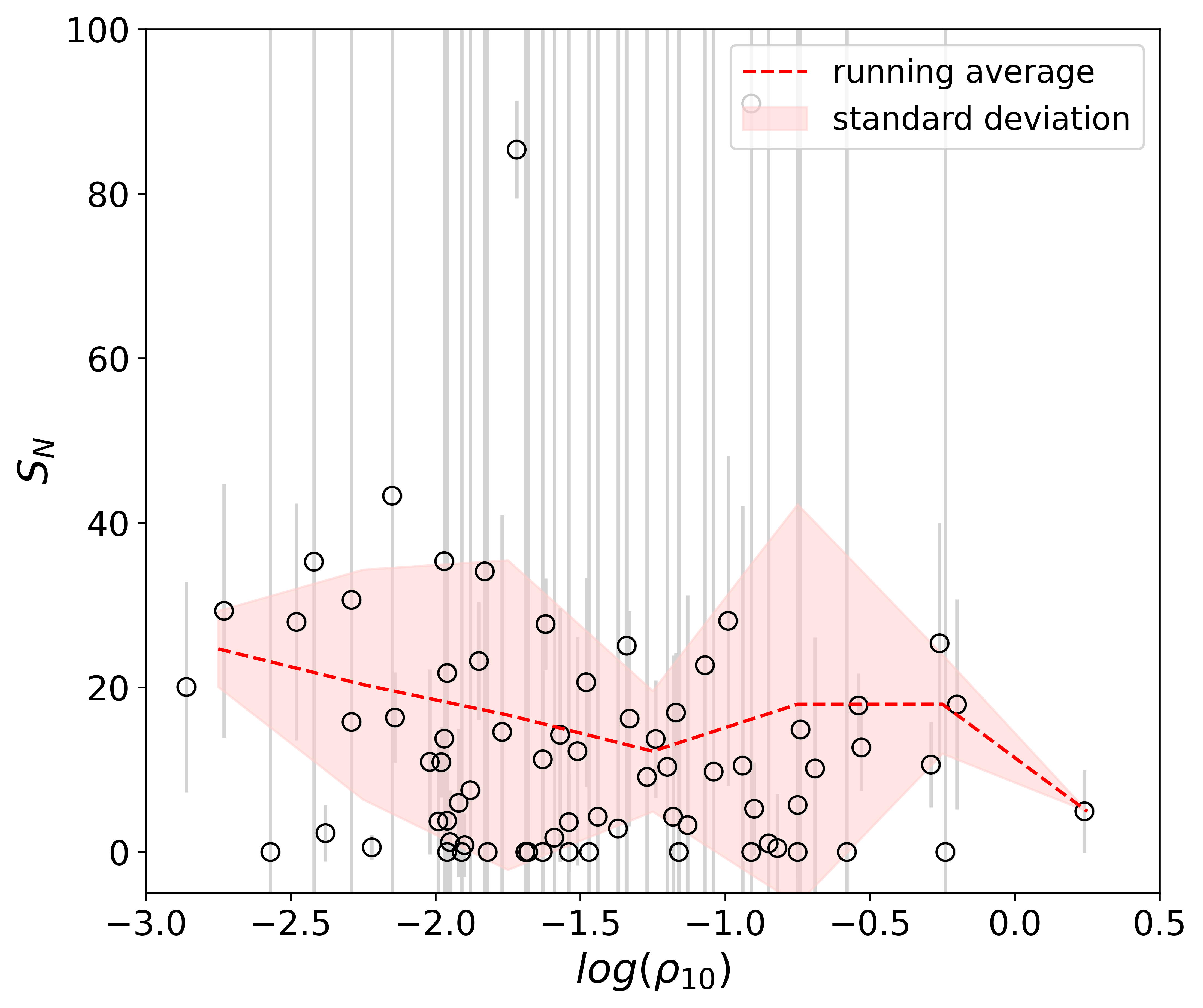}}
\centerline{
\includegraphics[width=0.33\textwidth]{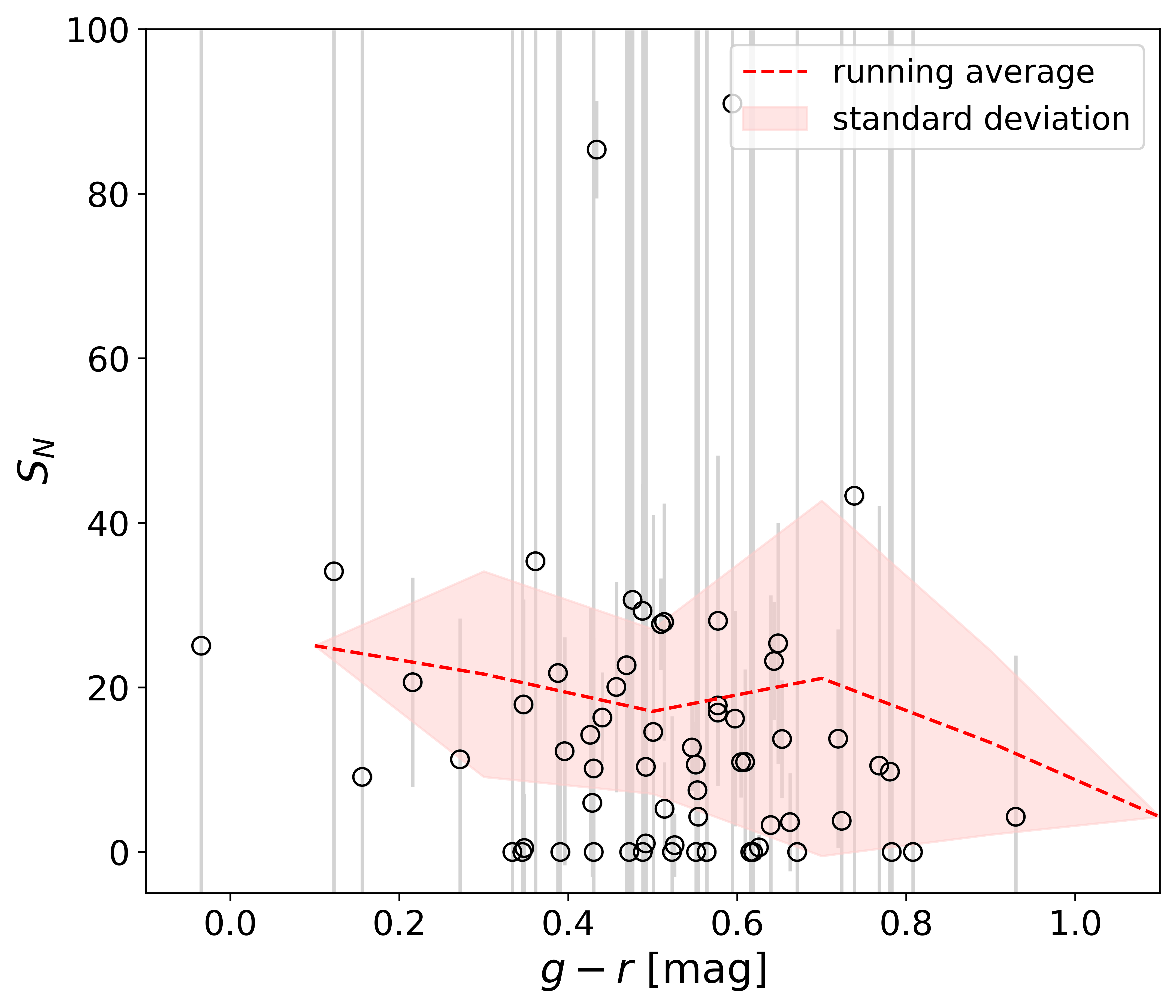}
\includegraphics[width=0.33\textwidth]{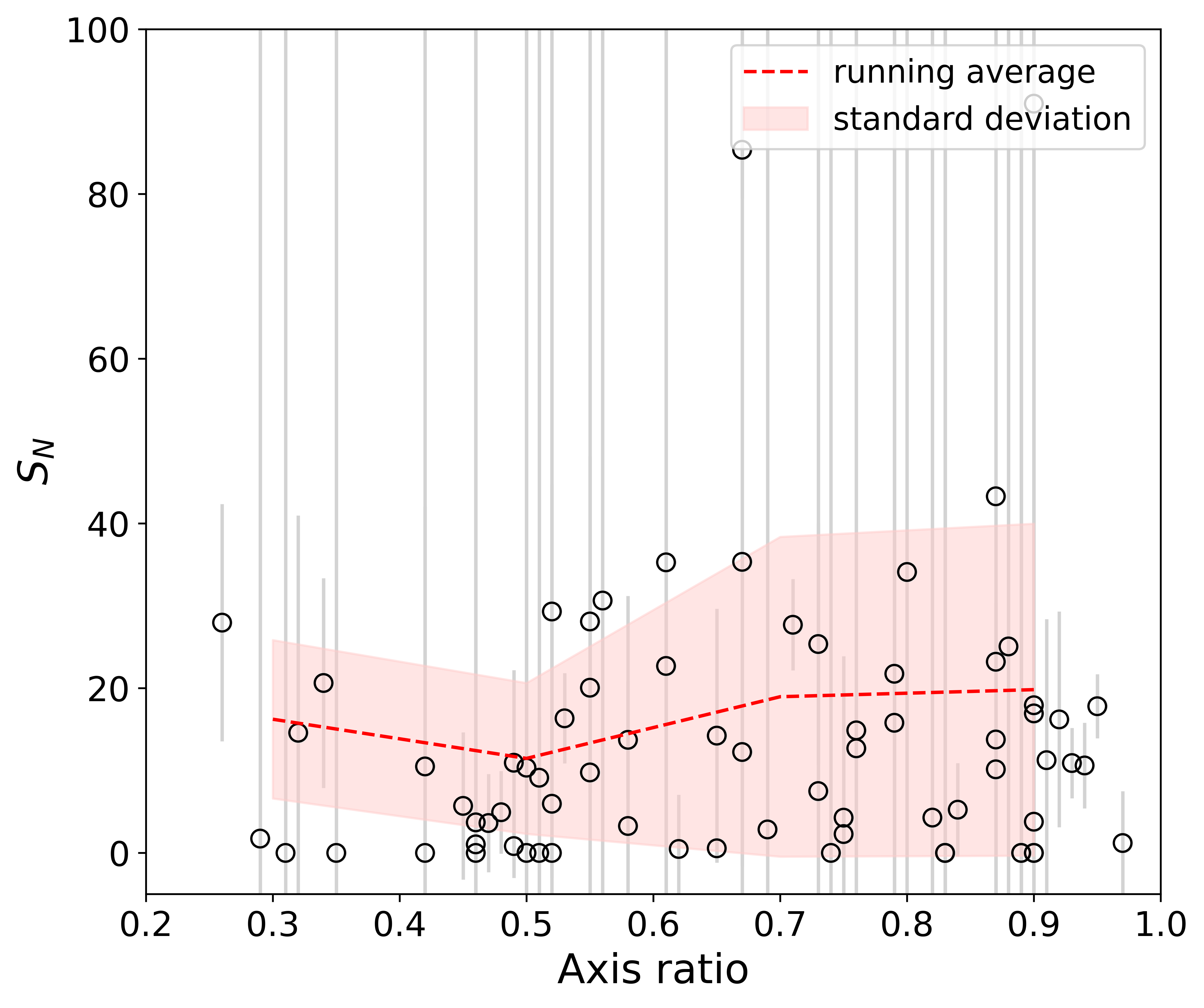}
\includegraphics[width=0.33\textwidth]{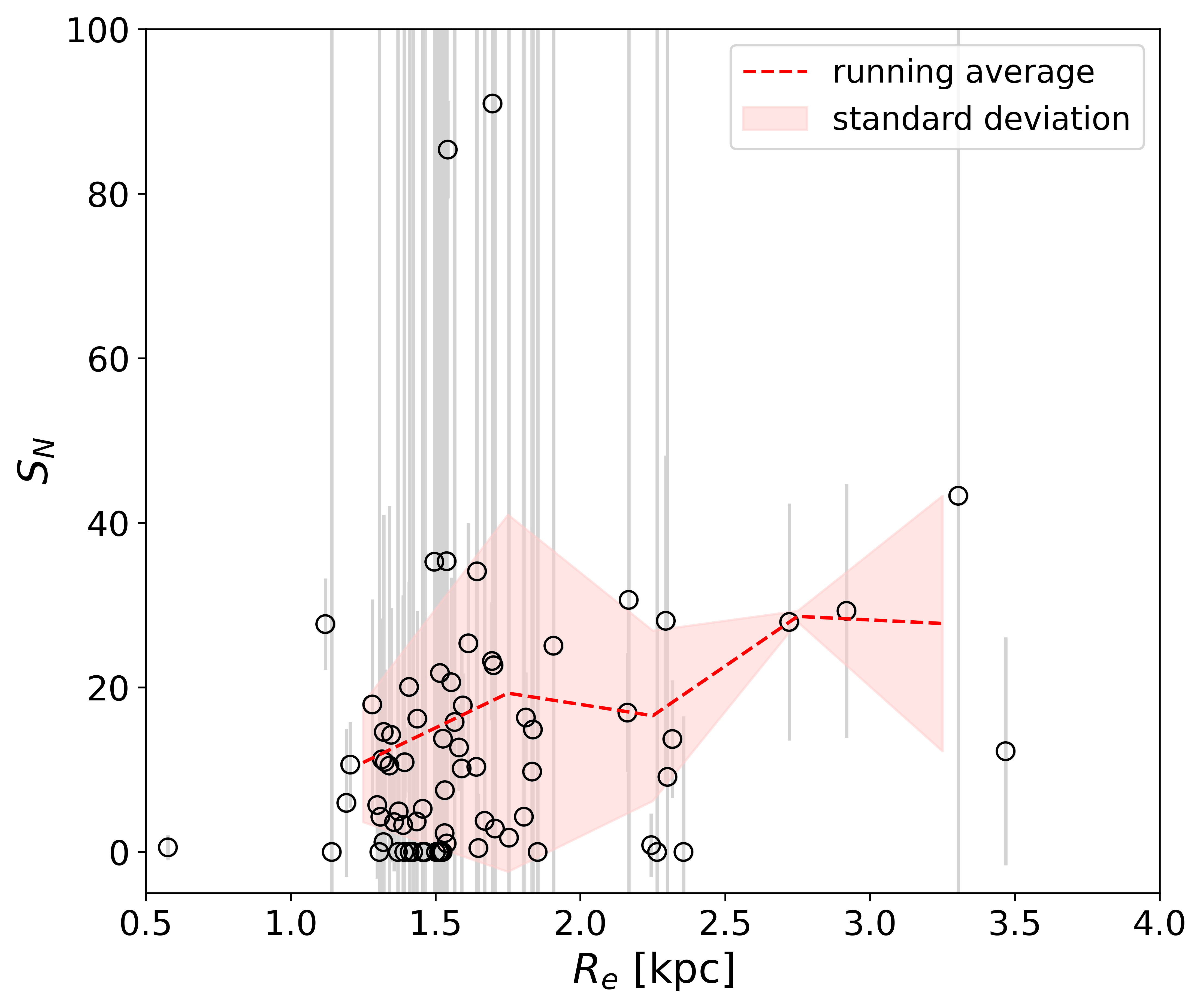}}
\caption{{\it Top}: From left to right, dependence of specific frequency $S_N$ on host galaxy stellar mass, central surface brightness ($\mu_{0}$), and local density ($log(\rho_{10})$). There are no obvious trends, except with $\rho_{10}$, which shows an increase in specific frequency as the local density becomes less dense. {\it Bottom:} From left to right, dependence of $S_N$ on host galaxy $g-r$ color, axis ratio, and effective radius ($R_e$). The bottom dependencies show the most prominent trends, with an increase in GC counts with bluer colors, increased roundness, and larger size. For all plots, the running average and standard deviation, computed using only non-zero values of $N_{GC}$, is shown as the {\it red dashed line} and {\it light red region}, respectively.
\label{fig:dependenceSn}}
\end{figure*}

\subsection{Radial profiles}

We characterized the radial profile of GCs in a UDG by counting the number of GCs per unit area in increasing annuli of multiples of 0.25\,$R_{e,gal}$, the effective radius of the galaxy. In order to quantify the radial extent of the GC distribution, the resulting counts were then fitted with a modified S\'ersic function of the form:

\begin{equation}
I(R) = I_e \, exp(-(\frac{R}{R_e})^{1/n} - 1 ) \, + \, I_b
\label{equ:sersic}
\end{equation}

The fitted S\'ersic function provides us with an estimate for both the size ($R_{e,GC}$) and shape (S\'ersic index $n$) of the bright GC system. Because we have to use the uncorrected counts, $N_{det}$, to measure the radial profile of the galaxy, the counts include some contamination so we include a constant background term, $I_b$, in our fits. 

Using this method, we characterized the stacked radial profile for all GCs detected in the 74 galaxies in our sample (see Figure~\ref{fig:raddistall}). For the 40 nucleated UDGs, the detection corresponding to the candidate nucleus (or nuclei, in the case of MATLAS-138 and 987) in each galaxy were removed. For the stacked data, we obtain a S\'ersic index of 0.3 and $R_{e,GC} = 1.0$~$R_{e,gal}$. Based on the error bars associated with the data points, the small dip at small galactocentric radii (at the bin value 0.375~$R_e$) appears to be real and could be related to the dynamical friction infall of massive GCs to create a nucleus/NSC \citep{Tremaine1975}. However, when plotting the stacked radial profile separately for the 40 nucleated and 34 non-nucleated UDGs in our sample, we find that the deficit is seen in both sub-samples, albeit within larger error bars. For the non-nucleated UDGs, the dip is followed by an overdensity of GCs in the next bin. Although the dynamical friction infall of GCs could still be a possible reason for the deficit for the nucleated UDGs, another mechanism must be producing the dip for the non-nucleated UDGs. The presence of a dark matter core and the stalling of GCs at the core radius could potentially be responsible for the observed dip in the radial distribution of GCs. This effect is due to the fact that GCs lose orbital energy through dynamical friction and can stall at the radius where the gravitational potential changes slope, which is typically at the core radius in cored dark matter profiles \citep{Hernandez1998, Goerdt2006, Read2006, Cole2012, Kaur2018}.

For the 12 UDGs with a large number of GCs ($N_{GC} > 10$), we were also able to characterize their individual GC radial profiles. The results are displayed in Appendix~\ref{AppendixC}, Figure~\ref{fig:raddistap}. Of these 12 UDGs, 5 galaxies are nucleated (MATLAS-138, 1332, 1616, 1779, 1938). For the UDG MATLAS-1938, which has a bright nucleus and a large number of centrally located GCs, the central (first) bin was not considered in order for the fit to converge. For the UDGs MATLAS-1534, 1616, and 1779, the third, second and second+third bin were not considered, respectively, in order for the fit to converge.

The deficit seen in the stacked radial profile of the 74 UDGs at galactocentric radius of $\sim$\,0.4\,$R_e$ is also seen in the individual radial profiles of 6 UDGs (MATLAS-138, 799, 1332, 1413, 154, 1616). The galaxies MATLAS-42, 401, 585 and 1779 show an even stronger deficit towards the center of their radial profiles. As can be seen in Figure~\ref{fig:densitymapall}, the GC density maps of these galaxies are both off-center and asymmetric, which leads to this deficit. These four galaxies have a disturbed stellar light morphology. The UDG MATLAS-42 is classified as a dwarf irregular and is detected in HI. It appears to have more than one stellar light component: a second component extends along the South-East (SE) direction and is well traced by the GC density extension in that direction. As mentioned above, this UDG has a close nearby dwarf companion, MATLAS-41, which is located NW of the UDG. The UDG MATLAS-401 is also classified as a dwarf irregular and its stellar light appears extended in the same direction as the GC distribution. When comparing the colors of its GCs with the color distribution of the GCs in all of the UDGs, we do not find a significant bluer population of GCs in this galaxy. MATLAS-585 is classified as a dwarf elliptical but clearly displays a disturbed stellar light morphology and its GC distribution is offset from the center. As discussed above, the UDG MATLAS-1779 is a nucleated dwarf elliptical which shows extended tidal features in the direction of the massive early-type galaxy NGC~5493. The stretch in the GC distribution along the NE and SW directions seen in the GC density map corroborates the GC deficit annulus region see in the radial profile of this UDG and the fact that this galaxy has suffered from tidal interactions.

\subsection{Dependence on host}

We analyzed the dependence of the effective radius of the GC system on three host parameters: the host total GC count, the host stellar mass, and the host effective radius. The results are displayed in Figure~\ref{fig:raddist}. The relations found for 118 early-type galaxies drawn from the Next Generation Virgo Cluster Survey (NGVS), MATLAS, and ACSVCS surveys from \citet{Lim2024} are shown for comparison.

The effective radii of the GC systems are measured to be $0.8 < R_{e,GC} < 1.9 R_{e,gal}$ (or $1.3 < R_{e,GC} < 6.3$~kpc, given the distance to each galaxy), with a median value of $1.2 R_{e,gal}$ ($2.5$~kpc). The effective radii of the GC systems as a function of the host GC count and host stellar mass are above the relations of \citet{Lim2024} for more massive galaxies ($\log(M_\ast)>8.3$). Furthermore, we do not find any obvious trend with host total GC count or host stellar mass.

When comparing the effective radius of the GC spatial distribution and the galaxy light profile, we find a simple one-to-one correspondence between them. This means that in our UDG sample, the GCs follow the galaxy light. The relation of \citet{Lim2024} comparing the effective radii of the GC systems to the effective radii of their host galaxies, which includes an environmental effect, sits mostly above our measurements.

Although the scatter is large, the measured S\'ersic indices mostly fall in the range $0.01 < n < 0.5$ with a median value of $n \sim 0.26$, albeit with large measurement errors and sometimes imperfect fits. In comparison, the GC systems of \citet{Lim2024} have S\'ersic indices in the range $1 < n < 4$, reaching larger S\'ersic index values, but this is to be expected as the galaxies in their sample contain more massive hosts and contain galaxies located in higher density environments.

\section{Key host properties governing $S_N$} 
\label{sec:pca}

The GC counts have been said to correlate with many host UDG properties, such as its color, roundness and the local environment it inhabits. We explored the dependence of the GC count per unit galaxy luminosity, i.e.\ the specific frequency $S_N$, on the host UDG properties in order to determine the key variable(s) governing this UDG property. 

Figure~\ref{fig:dependenceSn} displays the behavior of the specific frequency $S_N$ with host galaxy stellar mass, central surface brightness ($\mu_{0}$), local density ($log(\rho_{10})$), CFHT $g-r$ color, axis ratio, and effective radius ($R_e$). As we are using the GC counts per unit host galaxy luminosity, we expect to find no trend with host galaxy stellar mass, as shown in the first plot. There is no clear trend with host galaxy central surface brightness and only weak trends with local density, color and axis ratio. We note that the color plot does not contain the same amount of data points as the other plots due to the fact that 10 out of the 74 UDGs do not have CFHT $g-r$ colors. The rise towards blue colors is driven mainly by two blue UDGs (MATLAS-2021 and 2184). Contrary to expectations, we find that the $S_N$ values of the UDGs in the lowest density environments appear to be on average slightly higher than those in the highest density environments.

The most prominent trend is seen with host galaxy size. The UDGs with effective radius greater than 2.5~kpc are MATLAS-42, 262, 1413 and 1779. Three of those (MATLAS-42, 262 and 1779 with $R_e =$~3.46, 2.75 and 2.90~kpc, respectively) show signs of disturbed stellar light morphology (MATLAS-42) or tidal features (MATLAS-262 and 1779). This may hint at a connection between a UDG specific frequency and its recent merger and/or tidal interaction history which is accompanied by an increase in the measured effective radius of the UDG.  

Although some trends are hinted from our current UDG dataset, more detailed studies of both the host UDGs and their GC populations are needed in order to fully probe and understand the key variable(s) governing $S_N$.

\section{Conclusions} 
\label{sec:conclusions}

Globular clusters are key to constructing a complete picture of the formation and evolution of UDGs. Their presence and number reveal important details about the dark matter halos and masses of their host galaxy, and studying their properties can give us insights into past events that shaped these galaxies, helping us uncover their formation histories. Making use of the ACS instrument on the Hubble Space Telescope, we have obtained high resolution images in two filters, $F606W$ and $F814W$, of 74 UDGs, selected from the MATLAS dwarf sample with cuts in surface brightness and effective radius of $\langle \mu_{e,g} \rangle > 23.3$ mag arcsec$^{-2}$ and $R_e > 1.0$~kpc, respectively. With one of the largest UDG samples observed with HST, we have been able to study the link between GC counts and UDG properties as well as construct two-dimensional GC density maps. Some of our key findings are summarized below:

\begin{itemize}

\item The number of GCs associated with each UDG in our sample ranges from $0 - 38$($\pm7$). The distribution of GC counts is characterized by a prominent peak at extremely small GCs (64\% have $N_{GC} < 3$), while the remaining galaxies populate an extended tail out to the the maximum value. More massive UDGs host a larger number of GCs, on average; the UDGs with stellar masses below $M_* \sim 2 \times 10^8 M_{\odot}$ have GC counts between zero and 13, while the most massive UDGs with $M_* > 2 \times 10^8 M_{\odot}$ have counts ranging from 13 -- 38.
\\
\item The distributions of estimated halo masses based on GC counts for the MATLAS UDGs with $N_{GC} > 3$ range from $M_{h} \sim 176\pm142$ to $2542\pm1430$\,$M_\ast$ with a median value at $\sim$~472\,$M_\ast$. Therefore, our UDGs are consistent with having significant dark matter halos, although the uncertainties in the halo-to-stellar mass ratios can be substantial and therefore require further confirmation.
\\
\item We also calculate the specific frequency of GCs per galaxy, which allows for a more accurate comparison between galaxies of different stellar masses or luminosities. We find $S_N$ values between 0.5 and 91.0 (for $S_N>0$), with a small population (9\%) at high specific frequencies (arbitrarily defined as $S_N > 30$). We note that the median value of our sample with specific frequencies greater than zero, $S_N=12.7$, is similar to the median value of 10.7 for UDGs in the Perseus cluster but less than the ones for the Virgo and Coma clusters ($S_N=42.4$ and $S_N=30.6$, respectively). Furthermore, the range of $S_N$ values (for $S_N>0$) of the MATLAS UDGs extends beyond the Perseus $S_N$ values ($S_N=1.0-60.1$) but stays below the highest $S_N$ values measured for the Virgo and Coma UDGs ($S_N=187.1$ and $S_N=348.3$, respectively). Contrary to expectations, we find that the $S_N$ values of the UDGs in the lowest density environments appear to be on average slightly higher than those in the highest density environments.
\\
\item The GC populations of nucleated UDGs are of special interest, because the NSC could be formed by merging GCs. However, we find no statistically significant difference between the GC populations of the nucleated and non-nucleated UDGs in our sample. For both populations, the percentage of galaxies with $N_{GC} \geq 3$ is approximately 35\%. The nucleated UDGs all have $S_N<40$, which is also consistent with the majority of the MATLAS UDGs.
\\
\item We also examined density maps for the systems with large enough $N_{GC}$ counts to be meaningful. The 12 UDGs with $N_{GC}>10$ show a variety of distributions: symmetric, elongated, off-center, as well as asymmetric and off-center. Both symmetric and elongated GC distributions match well with the stellar distribution. In general, a disturbed morphology does not appear to correlate with a high $S_N$ value. Two UDGs show an off-center (with respect to the stellar light) symmetric density map. In one case, this is likely due to the bright galaxy contaminant located in the opposite direction to the offset. The UDGs with asymmetric and off-center GC distributions show a disturbed stellar light morphology, with one galaxy exhibiting extended tidal features. The UDGs with disturbed GC density maps also show disturbed stellar light morphology, but no correlation is found with the specific frequency of the system.
\\
\item The spatial distribution of the GCs around the host galaxy is measured for the first time for UDGs. It is derived by fitting a S\'ersic profile and characterized by the effective radius of the GC system. For the 12 UDGs with GC counts greater than 10, we find that they are spatially distributed with $R_{e,GC}/R_{e,gal}$ values between $0.8-1.9$. For these UDGs, the GC radial distribution appears to follow the galaxy light profile. The stacked spatial distribution of the GCs belonging to all UDGs also gives a one-to-one relation with $R_{e,GC}/R_{e,gal}=1.0$. For the same sample of 12 UDGs, the GC radial distribution does not seem to correlate with the GC count or the stellar mass of the galaxy.
\\
\item A common feature of the individual radial profiles is a deficit of GCs at small galactocentric radii ($\sim$~0.4~$R_e$), followed by an overdensity of GCs in the adjacent higher radial bin. As some of them are nucleated UDGs, this could be due to the dynamical friction infall of massive GCs to create a nucleus/NSC. However, as the stacked profile of the non-nucleated UDGs also show this deficit, the presence of a dark matter core and the stalling of GCs at the core radius could potentially be responsible for the observed dip in the radial distribution of GCs.
\\
\item Based on a trending analysis of the UDG specific frequency $S_N$ with the host galaxy properties, we find a relation with host galaxy size, roundness, and $g-r$ color. These findings in particular hint at a connection between the UDG specific frequency and its recent merger and/or tidal interaction history (accompanied by an increase in the size of the UDG). We also find a trend of higher $S_N$ values with decreasing local density. This may be a sign that the UDGs in the lowest density environments have not undergone enough major mergers to disrupt their GC populations. However, these possible trends are hampered by large uncertainties due to low statistics, and therefore more data are needed to confirm them.

\end{itemize}

The imaging data of the Euclid Wide Survey (covering $\sim$\,14\,000\,deg$^{2}$; \citealt{Scaramella-EP1}) will soon give us access to a much larger statistical sample of UDGs with which we will be able to study in more details the results obtained with the 74 MATLAS UDGs. It will provide high resolution imaging data for UDGs across a wide range of environments, from clusters to MATLAS-like environments, and even extending to lower local densities. A detailed study of the GC populations of the dwarf and ultra diffuse galaxies identified in the Euclid Early Release Observations of the Perseus cluster \citep{Marleau2024} is currently underway (Saifollahi et al. 2024, in prep.) that will already more than double the number of UDGs with a detailed study of their GC systems.

\section{Acknowledgements}

We would like to thank the referee for their valuable comments and suggestions, which have greatly improved the quality of this manuscript. This research is based on observations from the NASA/ESA Hubble Space Telescope obtained [from the Data Archive] at the Space Telescope Science Institute, which is operated by the Association of Universities for Research in Astronomy, Incorporated, under NASA contract NAS5-26555. Support for Program number GO-16257 and GO-16711 was provided through a grant from the STScI under NASA contract NAS5-26555. O.M.\ is grateful to the Swiss National Science Foundation for ﬁnancial support. P.R.D.\ gratefully acknowledges support from grant HST-GO-16257.002-A. S.P.\ acknowledges support from the Mid-career Researcher Program (No.\ RS-2023-00208957). S.L.\ acknowledges the support from the Sejong Science Fellowship Program by the National Research Foundation of Korea (NRF) grant funded by the Korea government (MSIT) (No.\ NRF-2021R1C1C2006790). M.P.\ is supported by the Academy of Finland grant No.\ 347089. R.H.\ acknowledges funding from the Italian INAF Large Grant 12-2022. This research was supported by the International Space Science Institute (ISSI) in Bern, through ISSI International Team project No.\ 534. The work presented in this paper made use of GALFIT \citep{Peng2002}, sep \citep{Barbary2016}, photutils \citep{Bradley2020}, astropy \citep{Astropy2013}, as well as Source Extractor \citep{Bertin1996}.

\bibliographystyle{aa}
\bibliography{bibliography}

\begin{appendix}
\onecolumn

\section{Tables}
\label{AppendixA}

Table~\ref{tab:sample} summarizes the number of GCs identified in the HST images for each of the 74 UDGs in our sample, and includes the calculated value of the specific frequency ($S_N$) and the halo mass ($M_h$). The distance and absolute magnitude ($M_V$) of the UDG is also provided. 

\newpage

\begin{table}[H]
\renewcommand{\arraystretch}{0.9}
\setlength{\tabcolsep}{1.5pt}
\small
\caption{GC counts for the 74 MATLAS UDGs in our sample.}
\centering
\begin{tabular}{l r r r r r r r}
\hline\hline
Name  & dist & $M_V$ & N$_{det}$  & N$_{det,bg}$  & N$_{GC}$  & S$_N$ & log(M$_{h})$ \\
     & [Mpc] & [mag]  &           &              &                    &      &         \\
\hline      \\[-2mm]
MATLAS-42  & 33.5  &$ -15.62 \pm 0.01 $ &$ 32 \pm 5 $ &$ 19.54 \pm 6.15 $ &$ 21.74 \pm 6.84 $ &$ 12.25 \pm 3.86 $ &$ 10.87 \pm 0.14 $ \\
MATLAS-49  & 36.0  &$ -13.82 \pm 0.01 $ &$ 3 \pm 1 $ &$ 1.06 \pm 1.74 $ &$ 1.22 \pm 2.00 $ &$ 3.63 \pm 5.96 $ &$ 9.62 \pm 0.71 $ \\
MATLAS-138  & 38.0  &$ -14.56 \pm 0.01 $ &$ 16 \pm 4 $ &$ 13.03 \pm 4.02 $ &$ 15.48 \pm 4.77 $ &$ 23.21 \pm 7.16 $ &$ 10.73 \pm 0.13 $ \\
MATLAS-141  & 37.0  &$ -13.00 \pm 0.02 $ &$ 3 \pm 1 $ &$ 1.02 \pm 1.73 $ &$ 1.19 \pm 2.03 $ &$ 7.51 \pm 12.74 $ &$ 9.61 \pm 0.74 $ \\
MATLAS-149  & 37.0  &$ -12.66 \pm 0.02 $ &$ 2 \pm 1 $ &$ -1.89 \pm 1.60 $ &$ -2.21 \pm 1.88 $ & --  & --  \\
MATLAS-177  & 22.0  &$ -13.35 \pm 0.01 $ &$ 4 \pm 2 $ &$ -0.40 \pm 2.00 $ &$ -0.41 \pm 2.03 $ & --  & --  \\
MATLAS-203  & 35.0  &$ -12.49 \pm 0.02 $ &$ 2 \pm 1 $ &$ -0.12 \pm 1.42 $ &$ -0.14 \pm 1.61 $ & --  & --  \\
MATLAS-207  & 35.0  &$ -13.35 \pm 0.02 $ &$ 3 \pm 1 $ &$ 1.15 \pm 1.73 $ &$ 1.30 \pm 1.97 $ &$ 5.97 \pm 9.00 $ &$ 9.65 \pm 0.65 $ \\
MATLAS-262  & 30.0  &$ -12.67 \pm 0.05 $ &$ 8 \pm 2 $ &$ 3.07 \pm 3.03 $ &$ 3.28 \pm 3.24 $ &$ 27.96 \pm 27.6 $ &$ 10.05 \pm 0.43 $ \\
MATLAS-290  & 38.0  &$ -14.44 \pm 0.01 $ &$ 3 \pm 1 $ &$ 1.15 \pm 1.73 $ &$ 1.37 \pm 2.06 $ &$ 2.29 \pm 3.45 $ &$ 9.67 \pm 0.65 $ \\
MATLAS-342  & 32.0  &$ -14.27 \pm 0.02 $ &$ 11 \pm 3 $ &$ 7.35 \pm 3.32 $ &$ 8.03 \pm 3.62 $ &$ 15.80 \pm 7.13 $ &$ 10.44 \pm 0.20 $ \\
MATLAS-347  & 12.0  &$ -13.13 \pm 0.02 $ &$ 3 \pm 1 $ &$ 0.66 \pm 1.73 $ &$ 0.66 \pm 1.73 $ &$ 3.71 \pm 9.71 $ &$ 9.36 \pm 1.13 $ \\
MATLAS-365  & 31.0  &$ -13.16 \pm 0.10 $ &$5 \pm 2 $ &$ 2.35 \pm 2.25 $ &$ 2.54 \pm 2.44 $ &$ 13.77 \pm 13.28 $ &$ 9.94 \pm 0.42 $ \\
MATLAS-368  & 31.0  &$ -13.91 \pm 0.02 $ &$ 6 \pm 2 $ &$ 1.29 \pm 2.46 $ &$ 1.39 \pm 2.66 $ &$ 3.79 \pm 7.23 $ &$ 9.68 \pm 0.83 $ \\
MATLAS-401  & 32.4  &$ -14.76 \pm 0.004 $ &$ 16 \pm 4 $ &$ 11.98 \pm 4.02 $ &$ 13.15 \pm 4.41 $ &$ 16.34 \pm 5.48 $ &$ 10.66 \pm 0.15 $ \\
MATLAS-405  & 28.0  &$ -13.67 \pm 0.02 $ &$ 18 \pm 4 $ &$ 8.54 \pm 4.24 $ &$ 8.97 \pm 4.46 $ &$ 30.64 \pm 15.24 $ &$ 10.49 \pm 0.22 $ \\
MATLAS-478  & 22.0  &$ -14.50 \pm 0.01 $ &$ 10 \pm 3 $ &$ 0.52 \pm 3.29 $ &$ 0.53 \pm 3.34 $ &$ 0.83 \pm 5.29 $ &$ 9.26 \pm 2.76 $ \\
MATLAS-524  & 27.0  &$ -14.50 \pm 0.004 $ &$ 3 \pm 1 $ &$ -0.73 \pm 1.90 $ &$ -0.76 \pm 1.98 $ & --  & --  \\
MATLAS-585  & 27.0  &$ -13.68 \pm 0.01 $ &$ 15 \pm 3 $ &$ 10.08 \pm 3.96 $ &$ 10.50 \pm 4.12 $ &$ 35.34 \pm 13.87 $ &$ 10.56 \pm 0.17 $ \\
MATLAS-627  & 46.0  &$ -13.46 \pm 0.02 $ &$ 5 \pm 2 $ &$ 3.52 \pm 2.24 $ &$ 4.88 \pm 3.11 $ &$ 20.06 \pm 12.79 $ &$ 10.23 \pm 0.28 $ \\
MATLAS-658  & 33.0  &$ -12.27 \pm 0.17 $ &$ 4 \pm 2 $ &$ 2.58 \pm 2.03 $ &$ 2.85 \pm 2.25 $ &$ 35.28 \pm 28.3 $ &$ 9.99 \pm 0.34 $ \\
MATLAS-682  & 41.0  &$ -12.96 \pm 0.02 $ &$ 1 \pm 1 $ &$ -0.03 \pm 1.02 $ &$ -0.03 \pm 1.28 $ & -- & -- \\
MATLAS-787  & 28.09  &$ -12.86 \pm 0.01 $ &$ 3 \pm 1 $ &$ -0.89 \pm 1.82 $ &$ -0.94 \pm 1.91 $ & --  & --  \\
MATLAS-791  & 25.0  &$ -12.56 \pm 0.03 $ &$ 3 \pm 1 $ &$ -1.09 \pm 1.74 $ &$ -1.12 \pm 1.79 $ & --  & --  \\
MATLAS-799  & 25.0  &$ -14.97 \pm 0.01 $ &$ 24 \pm 4 $ &$ 15.96 \pm 5.30 $ &$ 16.42 \pm 5.45 $ &$ 16.94 \pm 5.63 $ &$ 10.75 \pm 0.14 $ \\
MATLAS-898  & 20.0  &$ -13.84 \pm 0.02 $ &$ 6 \pm 2 $ &$ 0.17 \pm 2.48 $ &$ 0.17 \pm 2.50 $ &$ 0.48 \pm 7.25 $ &$ 8.76 \pm 6.52 $ \\
MATLAS-976  & 26.0  &$ -12.79 \pm 0.03 $ &$ 2 \pm 1 $ &$ 1.38 \pm 1.42 $ &$ 1.43 \pm 1.47 $ &$ 10.95 \pm 11.24 $ &$ 9.69 \pm 0.45 $ \\
MATLAS-984  & 33.0  &$ -13.64 \pm 0.02 $ &$ 9 \pm 3 $ &$ 5.62 \pm 3.01 $ &$ 6.21 \pm 3.33 $ &$ 21.75 \pm 11.67 $ &$ 10.33 \pm 0.23 $ \\
MATLAS-987  & 33.0  &$ -13.72 \pm 0.02 $ &$ 3 \pm 1 $ &$ 0.33 \pm 1.75 $ &$ 0.37 \pm 1.94 $ &$ 1.20 \pm 6.28 $ &$ 9.11 \pm 2.27 $ \\
MATLAS-1059  & 33.0  &$ -14.82 \pm 0.01 $ &$ 6 \pm 2 $ &$ -1.50 \pm 2.45 $ &$ -1.66 \pm 2.71 $ & --  & --  \\
MATLAS-1154  & 25.0  &$ -13.37 \pm 0.01 $ &$ 8 \pm 2 $ &$ 3.53 \pm 2.85 $ &$ 3.63 \pm 2.93 $ &$ 16.21 \pm 13.09 $ &$ 10.10 \pm 0.35 $ \\
MATLAS-1174  & 38.0  &$ -13.47 \pm 0.02 $ &$ 8 \pm 2 $ &$ 1.89 \pm 2.83 $ &$ 2.24 \pm 3.36 $ &$ 9.13 \pm 13.71 $ &$ 9.89 \pm 0.65 $ \\
MATLAS-1216  & 39.0  &$ -13.25 \pm 0.03 $ &$ 5 \pm 2 $ &$ -2.67 \pm 2.24 $ &$ -3.23 \pm 2.71 $ & --  & --  \\
MATLAS-1225  & 19.0  &$ -13.21 \pm 0.01 $ &$ 8 \pm 2 $ &$ 0.54 \pm 2.93 $ &$ 0.55 \pm 2.95 $ &$ 2.85 \pm 15.42 $ &$ 9.27 \pm 2.35 $ \\
MATLAS-1262  & 32.0  &$ -14.10 \pm 0.02 $ &$ 7 \pm 2 $ &$ 1.72 \pm 2.85 $ &$ 1.87 \pm 3.12 $ &$ 4.29 \pm 7.13 $ &$ 9.81 \pm 0.72 $ \\
MATLAS-1302  & 37.0  &$ -12.33 \pm 0.03 $ &$ 0 \pm 0 $ &$ -1.41 \pm 0.29 $ &$ -1.65 \pm 0.33 $ & -- & -- \\
MATLAS-1321  & 37.0  &$ -14.67 \pm 0.01 $ &$ 11 \pm 3 $ &$ 8.03 \pm 3.33 $ &$ 9.40 \pm 3.90 $ &$ 12.70 \pm 5.27 $ &$ 10.51 \pm 0.18 $ \\
MATLAS-1332  & 35.8  &$ -15.52 \pm 0.01 $ &$ 30 \pm 5 $ &$ 25.03 \pm 5.48 $ &$ 28.74 \pm 6.29 $ &$ 17.81 \pm 3.9 $ &$ 11.00 \pm 0.10 $ \\
MATLAS-1400  & 17.4  &$ -14.38 \pm 0.01 $ &$ 8 \pm 2 $ &$ 2.79 \pm 2.84 $ &$ 2.80 \pm 2.85 $ &$ 4.94 \pm 5.02 $ &$ 9.98 \pm 0.44 $ \\
MATLAS-1408  & 14.0  &$ -14.57 \pm 0.01 $ &$ 1 \pm 1 $ &$ 0.38 \pm 1.00 $ &$ 0.38 \pm 1.00 $ &$ 0.56 \pm 1.49 $ &$ 9.11 \pm 1.16 $ \\
MATLAS-1412  & 17.0  &$ -13.32 \pm 0.003 $ &$ 7 \pm 2 $ &$ 3.80 \pm 2.70 $ &$ 3.81 \pm 2.71 $ &$ 17.93 \pm 12.76 $ &$ 10.12 \pm 0.31 $ \\
MATLAS-1413  & 41.0  &$ -14.68 \pm 0.02 $ &$ 37 \pm 6 $ &$ 25.70 \pm 6.11 $ &$ 32.19 \pm 7.65 $ &$ 43.30 \pm 10.32 $ &$ 11.05 \pm 0.10 $ \\
MATLAS-1437  & 17.0  &$ -13.44 \pm 0.01 $ &$ 4 \pm 2 $ &$ -0.35 \pm 2.00 $ &$ -0.35 \pm 2.01 $ & -- & -- \\
MATLAS-1470  & 17.0  &$ -14.42 \pm 0.02 $ &$ 9 \pm 3 $ &$ 6.19 \pm 3.03 $ &$ 6.21 \pm 3.04 $ &$ 10.62 \pm 5.20 $ &$ 10.33 \pm 0.21 $ \\
MATLAS-1485  & 13.88  &$ -13.59 \pm 0.01 $ &$ 12 \pm 3 $ &$ 6.93 \pm 3.99 $ &$ 6.93 \pm 3.99 $ &$ 25.36 \pm 14.61 $ &$ 10.38 \pm 0.25 $ \\
MATLAS-1530  & 40.0  &$ -13.45 \pm 0.03 $ &$ 1 \pm 1 $ &$ -0.58 \pm 1.01 $ &$ -0.71 \pm 1.24 $ & --  & --  \\
MATLAS-1534  & 40.0  &$ -13.02 \pm 0.02 $ &$ 14 \pm 3 $ &$ 11.19 \pm 3.74 $ &$ 13.76 \pm 4.60 $ &$ 85.37 \pm 28.58 $ &$ 10.68 \pm 0.15 $ \\
MATLAS-1539  & 40.0  &$ -14.37 \pm 0.01 $ &$ 2 \pm 1 $ &$ -1.06 \pm 1.44 $ &$ -1.30 \pm 1.77 $ & --  & --  \\
MATLAS-1545  & 40.0  &$ -12.74 \pm 0.04 $ &$ 3 \pm 1 $ &$ 1.14 \pm 1.74 $ &$ 1.40 \pm 2.13 $ &$ 11.26 \pm 17.14 $ &$ 9.68 \pm 0.66 $ \\
MATLAS-1550  & 32.0  &$ -13.27 \pm 0.01 $ &$ 7 \pm 2 $ &$ 1.93 \pm 2.68 $ &$ 2.11 \pm 2.93 $ &$ 10.35 \pm 14.40 $ &$ 9.86 \pm 0.60 $ \\
MATLAS-1558  & 32.0  &$ -13.28 \pm 0.01 $ &$ 14 \pm 3 $ &$ 2.58 \pm 3.78 $ &$ 2.82 \pm 4.13 $ &$ 13.73 \pm 20.09 $ &$ 9.99 \pm 0.64 $ \\
MATLAS-1577  & 32.0  &$ -12.46 \pm 0.05 $ &$3 \pm 1 $ &$ 0.38 \pm 1.73 $ &$ 0.41 \pm 1.89 $ &$ 4.28 \pm 19.61 $ &$ 9.15 \pm 1.99 $ \\
MATLAS-1589  & 30.0  &$ -13.56 \pm 0.01 $ &$ 10 \pm 3 $ &$ 5.62 \pm 3.25 $ &$ 6.01 \pm 3.48 $ &$ 22.70 \pm 13.13 $ &$ 10.32 \pm 0.25 $ \\
MATLAS-1616  & 30.0  &$ -14.01 \pm 0.02 $ &$ 18 \pm 4 $ &$ 10.52 \pm 4.30 $ &$ 11.25 \pm 4.60 $ &$ 28.10 \pm 11.49 $ &$ 10.59 \pm 0.18 $ \\
MATLAS-1618  & 30.0  &$ -12.25 \pm 0.10 $ &$ 4 \pm 2 $ &$ 0.24 \pm 2.07 $ &$ 0.26 \pm 2.21 $ &$ 3.27 \pm 27.94 $ &$ 8.95 \pm 3.71 $ \\
MATLAS-1630  & 30.0  &$ -13.48 \pm 0.02 $ &$ 8 \pm 2 $ &$ 2.25 \pm 2.83 $ &$ 2.40 \pm 3.02 $ &$ 9.77 \pm 12.30 $ &$ 9.92 \pm 0.55 $ \\
MATLAS-1647  & 37.0  &$ -12.40 \pm 0.02 $ &$ 2 \pm 1 $ &$ -1.94 \pm 1.42 $ &$ -2.27 \pm 1.66 $ & --  & --  \\
MATLAS-1662  & 37.0  &$ -13.39 \pm 0.01 $ &$ 3 \pm 1 $ &$ 1.11 \pm 1.73 $ &$ 1.29 \pm 2.03 $ &$ 5.71 \pm 8.95 $ &$ 9.65 \pm 0.68 $ \\
MATLAS-1667  & 30.0  &$ -14.07 \pm 0.01 $ &$ 5 \pm 2 $ &$ 2.07 \pm 2.24 $ &$ 2.22 \pm 2.39 $ &$ 5.25 \pm 5.66 $ &$ 9.88 \pm 0.47 $ \\
MATLAS-1740  & 26.0  &$ -12.36 \pm 0.02 $ &$ 5 \pm 2 $ &$ 1.24 \pm 2.24 $ &$ 1.28 \pm 2.31 $ &$ 14.59 \pm 26.38 $ &$ 9.64 \pm 0.79 $ \\
MATLAS-1779  & 39.0  &$ -13.97 \pm 0.02 $ &$ 20 \pm 4 $ &$ 9.43 \pm 4.48 $ &$ 11.39 \pm 5.41 $ &$ 29.31 \pm 13.92 $ &$ 10.59 \pm 0.21 $ \\
MATLAS-1794  & 29.0  &$ -13.55 \pm 0.02 $ &$ 7 \pm 2 $ &$ 0.43 \pm 2.66 $ &$ 0.46 \pm 2.82 $ &$ 1.74 \pm 10.68 $ &$ 9.20 \pm 2.66 $ \\
MATLAS-1801  & 39.0  &$ -12.99 \pm 0.03 $ &$ 4 \pm 2 $ &$ 1.86 \pm 2.01 $ &$ 2.25 \pm 2.43 $ &$ 14.24 \pm 15.42 $ &$ 9.89 \pm 0.47 $ \\
MATLAS-1865  & 27.0  &$ -13.93 \pm 0.01 $ &$ 11 \pm 3 $ &$ 7.37 \pm 3.36 $ &$ 7.68 \pm 3.51 $ &$ 20.63 \pm 9.42 $ &$ 10.42 \pm 0.20 $ \\
MATLAS-1888  & 27.0  &$ -13.22 \pm 0.02 $ &$ 4 \pm 2 $ &$ -0.15 \pm 2.00 $ &$ -0.15 \pm 2.09 $ & --  & --  \\
MATLAS-1907  & 24.0  &$ -14.07 \pm 0.004 $ &$ 6 \pm 2 $ &$ 0.43 \pm 2.46 $ &$ 0.44 \pm 2.51 $ &$ 1.04 \pm 5.93 $ &$ 9.18 \pm 2.48 $ \\
MATLAS-1938  & 17.8  &$ -15.03 \pm 0.004 $ &$ 32 \pm 5 $ &$ 28.38 \pm 5.69 $ &$ 28.5 \pm 5.71 $ &$ 27.70 \pm 5.55 $ &$ 10.99 \pm 0.09 $ \\
MATLAS-1975  & 26.0  &$ -13.18 \pm 0.07 $ &$ 8 \pm 2 $ &$ 1.83 \pm 2.87 $ &$ 1.90 \pm 2.97 $ &$ 10.15 \pm 15.92 $ &$ 9.82 \pm 0.68 $ \\
MATLAS-1985  & 26.0  &$ -14.53 \pm 0.01 $ &$ 17 \pm 4 $ &$ 9.31 \pm 4.12 $ &$ 9.64 \pm 4.27 $ &$ 14.88 \pm 6.59 $ &$ 10.52 \pm 0.19 $ \\
MATLAS-2019  & 20.3  &$ -14.04 \pm 0.01 $ &$ 45 \pm 6 $ &$ 37.31 \pm 6.75 $ &$ 37.65 \pm 6.81 $ &$ 90.97 \pm 16.49 $ &$ 11.11 \pm 0.08 $ \\
MATLAS-2021  & 22.0  &$ -12.64 \pm 0.08 $ &$ 12 \pm 3 $ &$ 2.82 \pm 3.57 $ &$ 2.86 \pm 3.62 $ &$ 25.07 \pm 31.82 $ &$ 9.99 \pm 0.55 $ \\
MATLAS-2069  & 25.0  &$ -12.16 \pm 0.02 $ &$ 5 \pm 2 $ &$ 0.74 \pm 2.24 $ &$ 0.77 \pm 2.30 $ &$ 10.50 \pm 31.56 $ &$ 9.42 \pm 1.31 $ \\
MATLAS-2176  & 23.0  &$ -14.80 \pm 0.01 $ &$ 12 \pm 3 $ &$ 8.89 \pm 3.47 $ &$ 9.05 \pm 3.54 $ &$ 10.91 \pm 4.26 $ &$ 10.49 \pm 0.17 $ \\
MATLAS-2184  & 29.0  &$ -13.24 \pm 0.05 $ &$ 10 \pm 3 $ &$ 6.39 \pm 3.20 $ &$ 6.77 \pm 3.39 $ &$ 34.11 \pm 17.14 $ &$ 10.37 \pm 0.22 $ \\
\hline
\end{tabular}
\label{tab:sample}
\end{table}

\newpage

\section{Globular cluster density maps}
\label{AppendixB}
\begin{figure*}[ht!]
\centerline{
\includegraphics[width=0.45\linewidth]{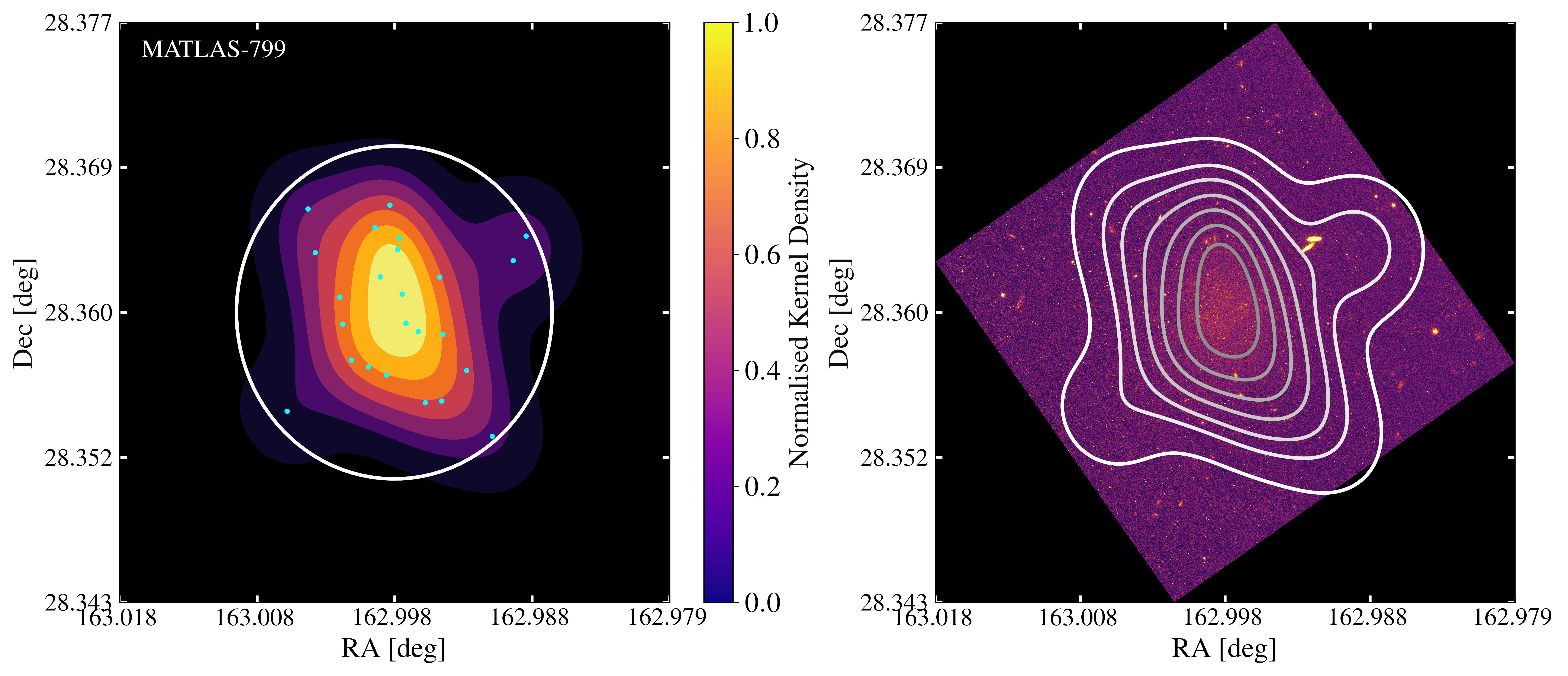}
\includegraphics[width=0.45\linewidth]{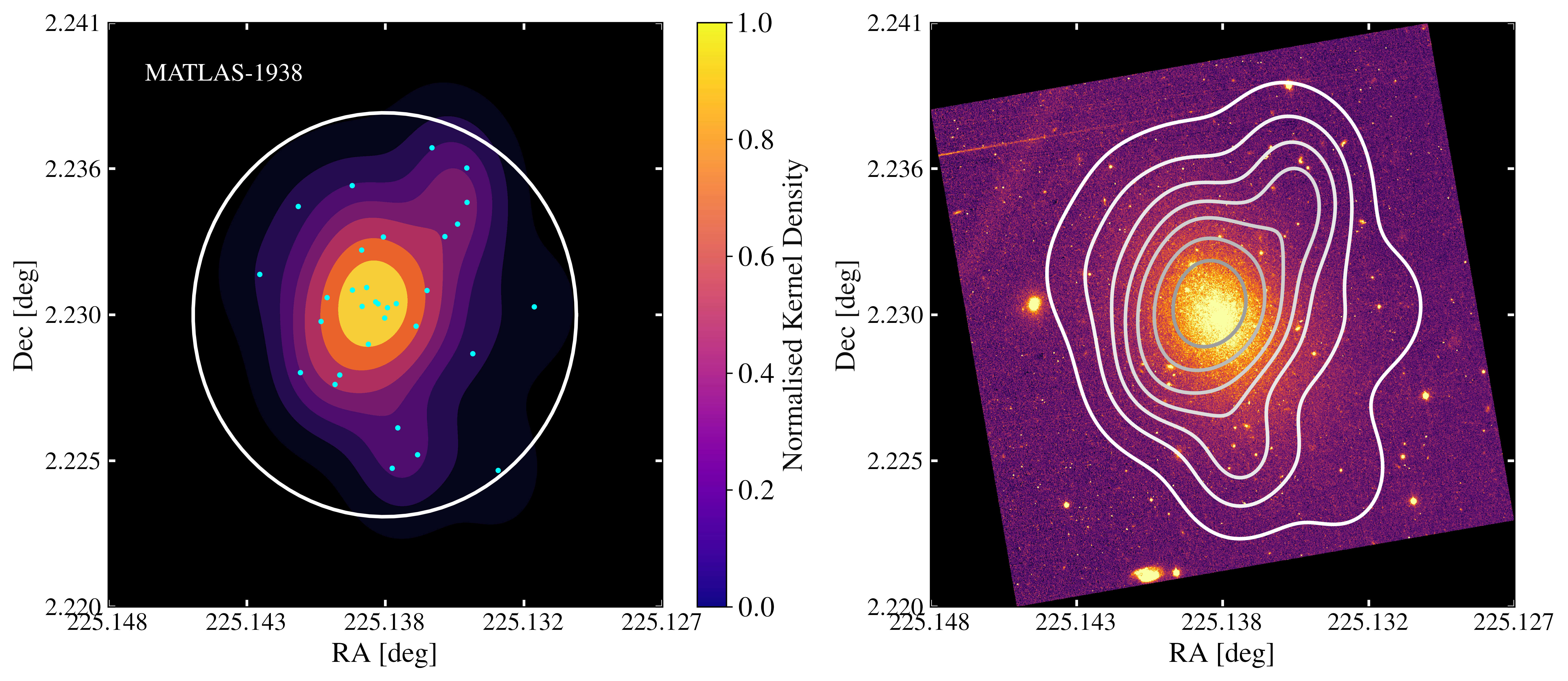}}
\centerline{
\includegraphics[width=0.45\linewidth]{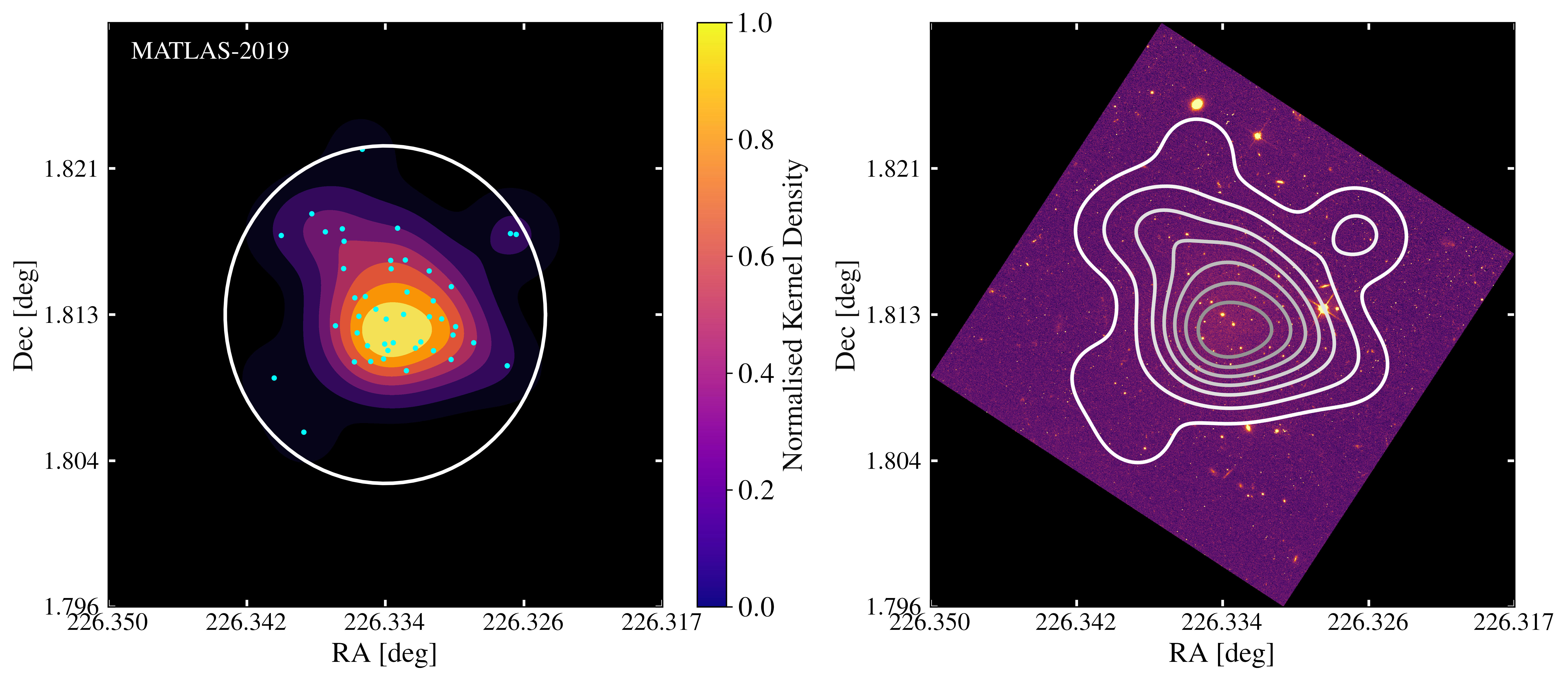}
\includegraphics[width=0.45\linewidth]{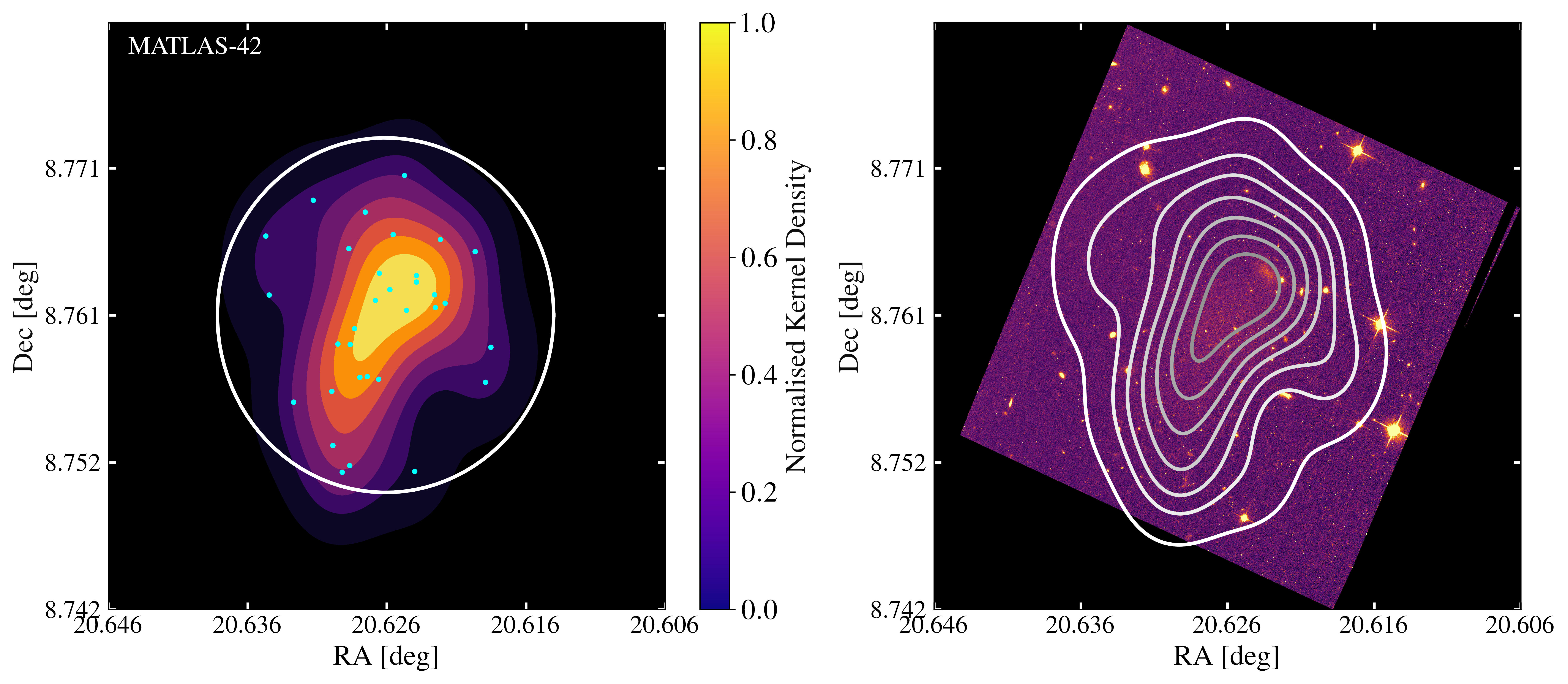}}
\centerline{
\includegraphics[width=0.45\linewidth]{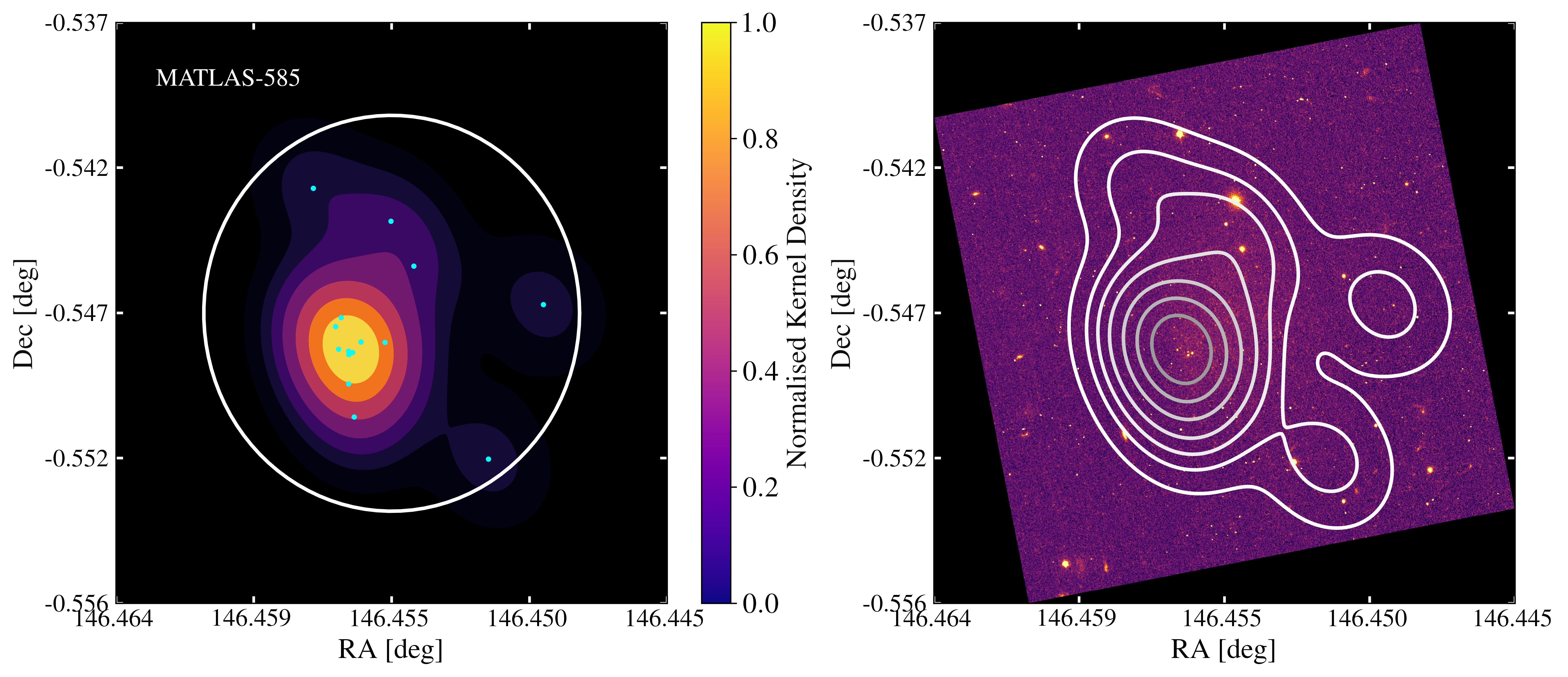}
\includegraphics[width=0.45\linewidth]{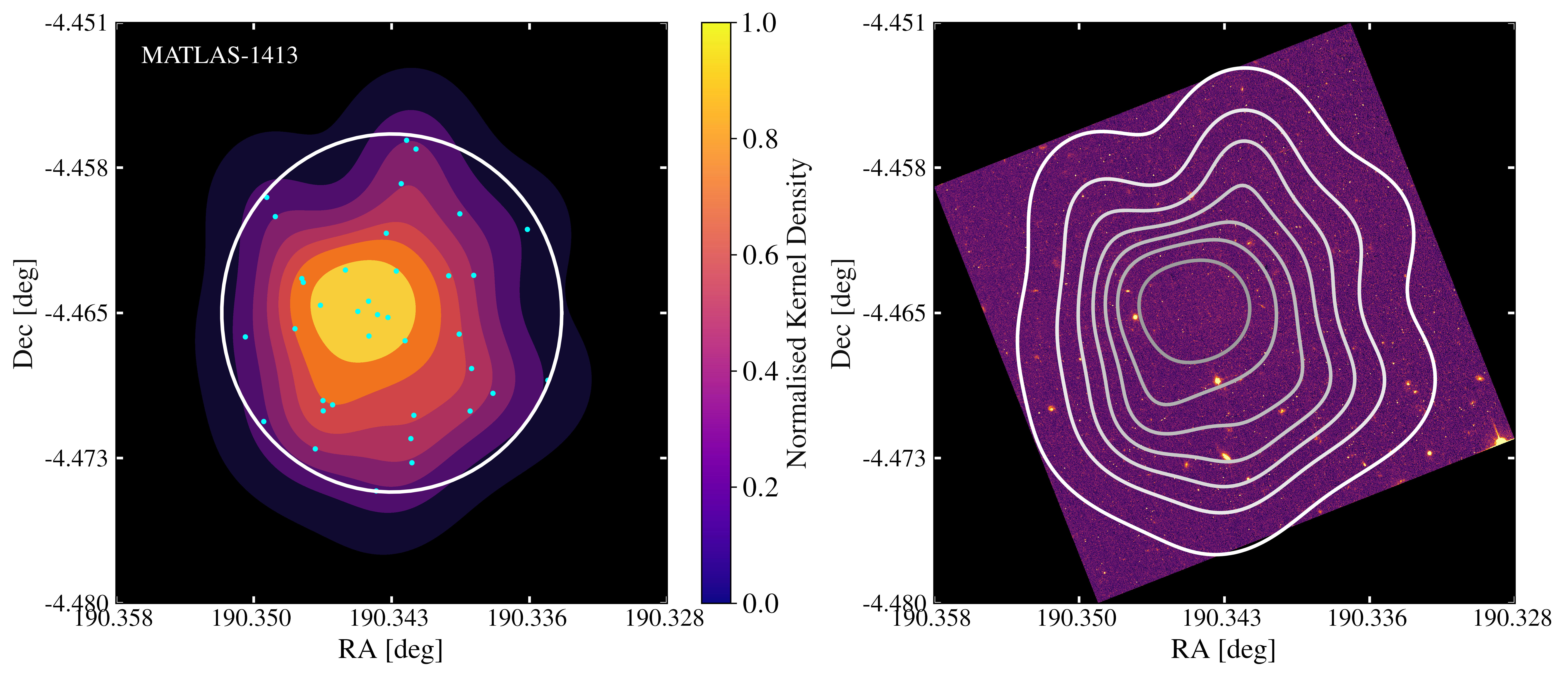}}
\centerline{
\includegraphics[width=0.45\linewidth]{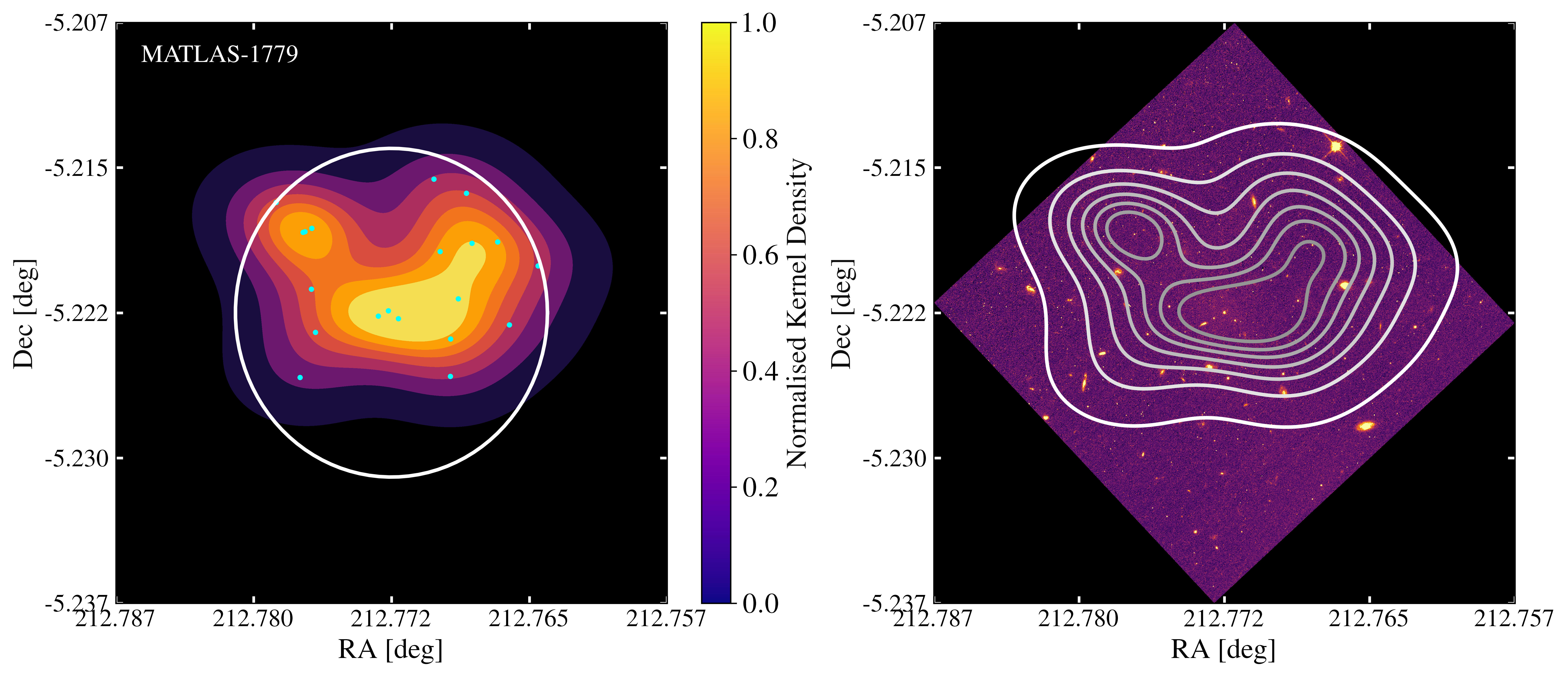}
\includegraphics[width=0.45\linewidth]{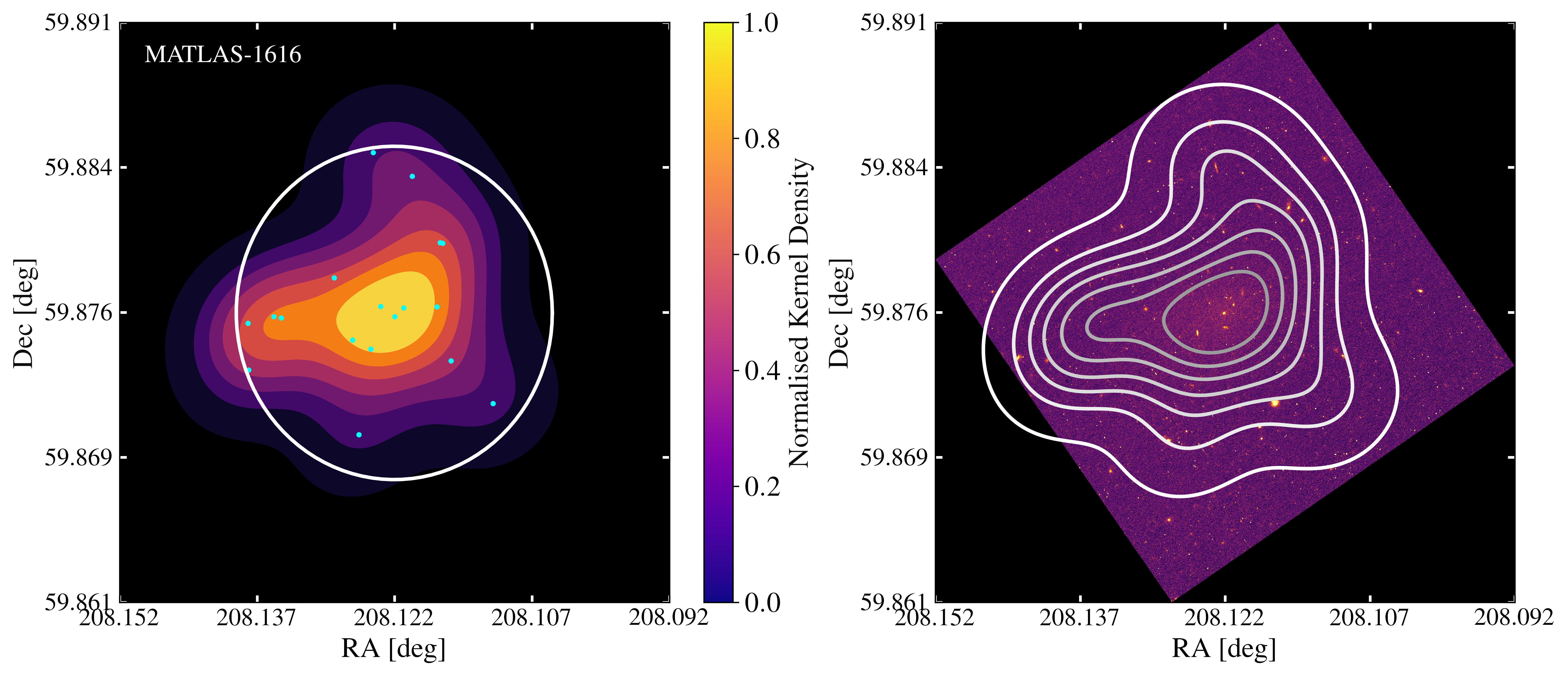}}
\centerline{
\includegraphics[width=0.45\linewidth]{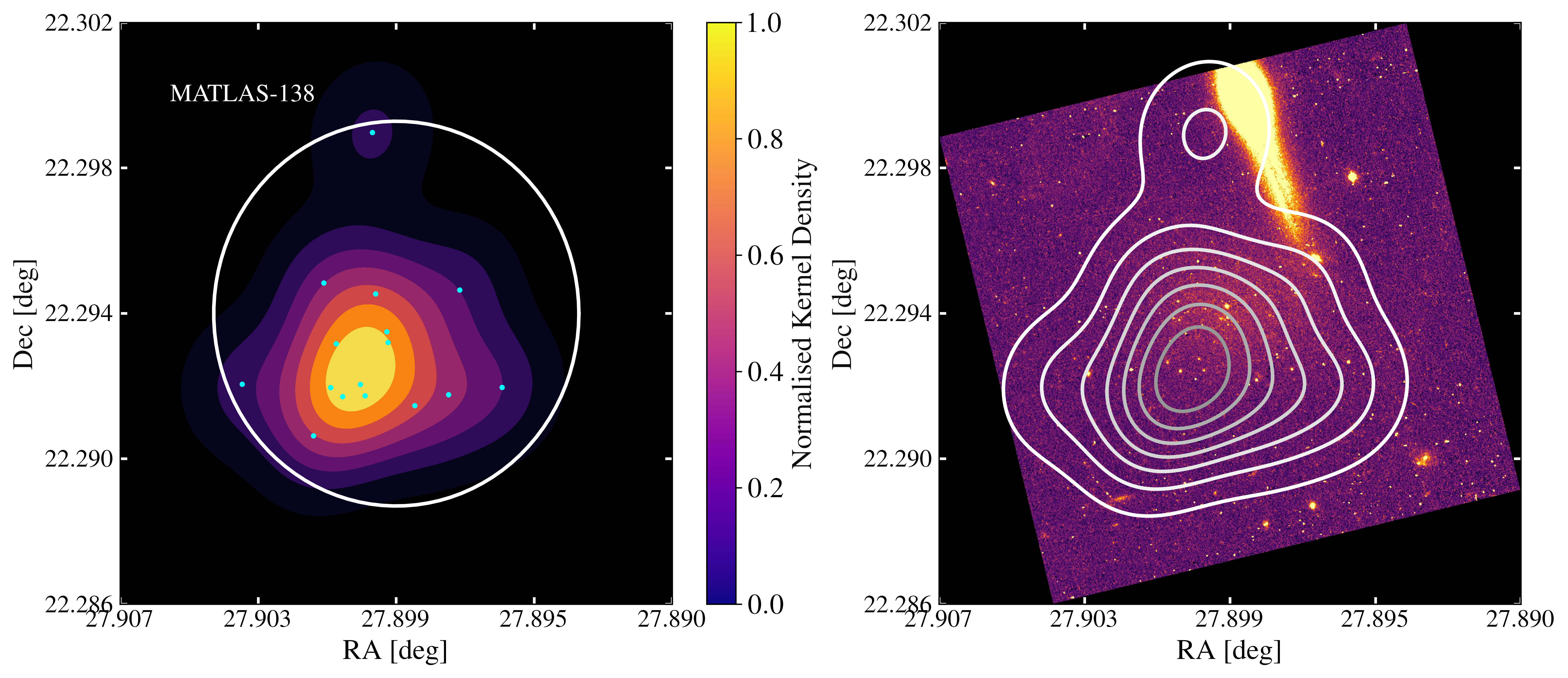}
\includegraphics[width=0.45\linewidth]{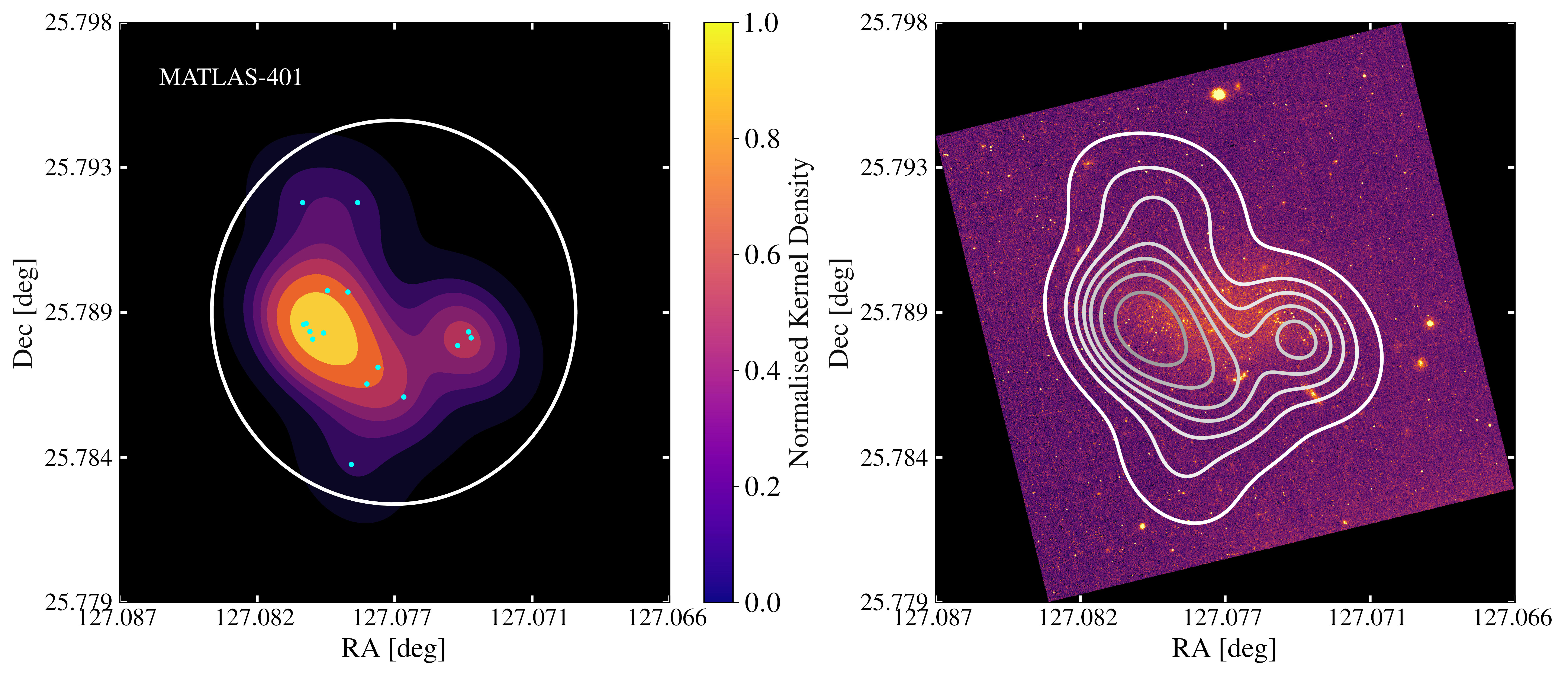}}
\centerline{
\includegraphics[width=0.45\linewidth]{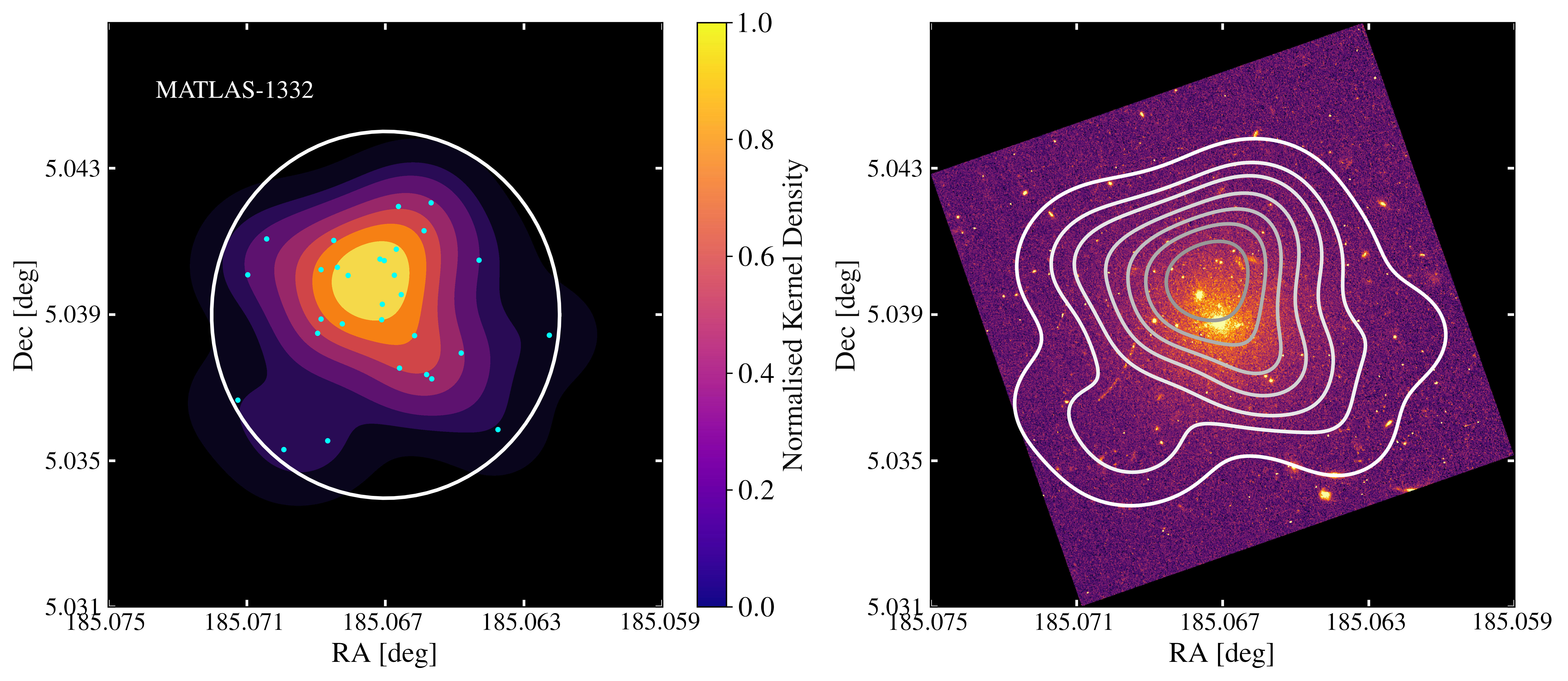}
\includegraphics[width=0.45\linewidth]{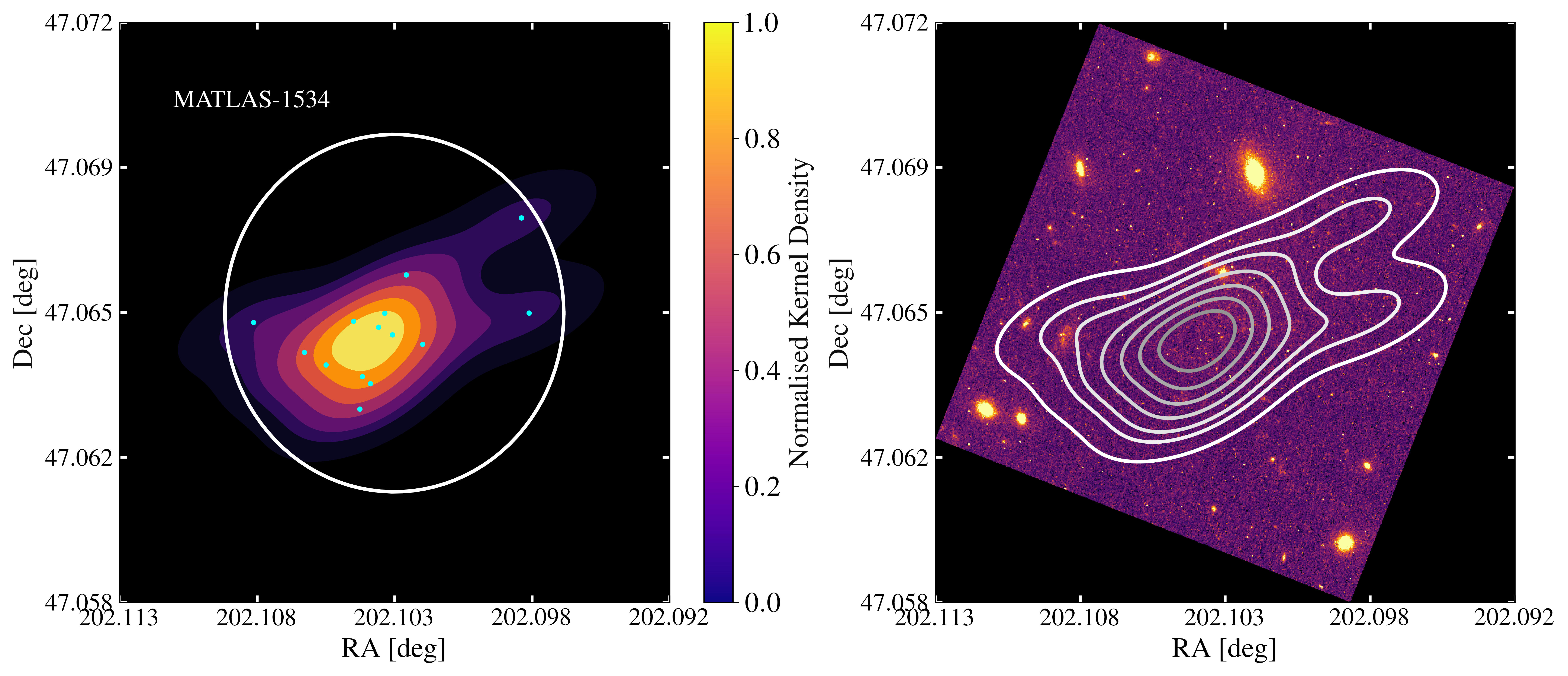}
}
\caption{{\it Left:} The projected distribution ({\it cyan dots}) and density map ({\it colored map}) of the GC candidates detected within 2$R_e$ ({\it large white circle}) for the 12 UDGs with $N_{GC}>10$. {\it Right:} The density map ({\it white contours}) of the GC candidates is overlayed on the HST $F814W$ image of the UDG. The images are 2.5$R_e$ on a side.
\label{fig:densitymapall}}
\end{figure*}

\newpage

\section{Globular cluster radial profiles}
\label{AppendixC}
\begin{figure*}[h!]
\centerline{
\includegraphics[width=0.33\linewidth]{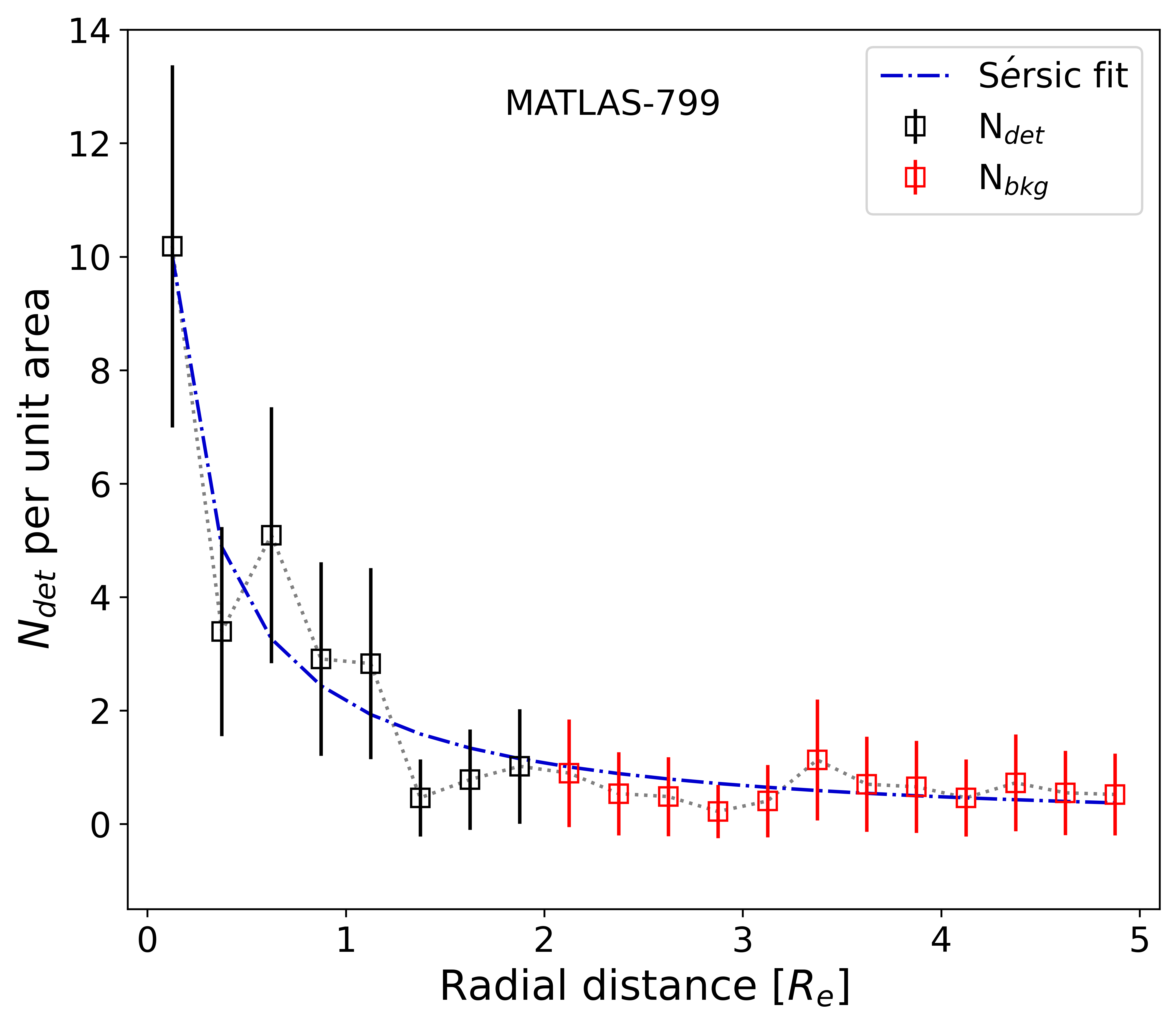}
\includegraphics[width=0.33\linewidth]{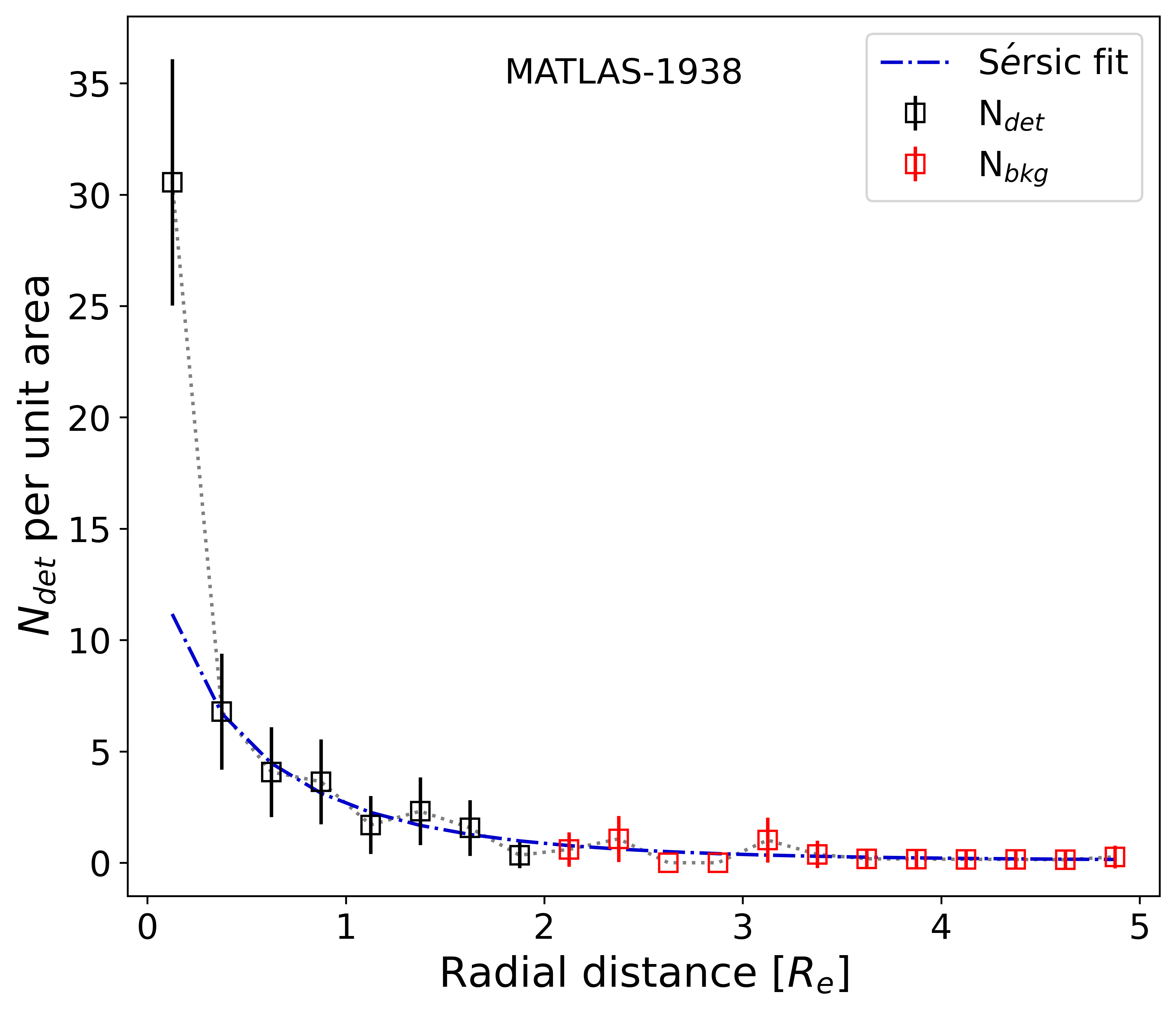}
\includegraphics[width=0.33\linewidth]{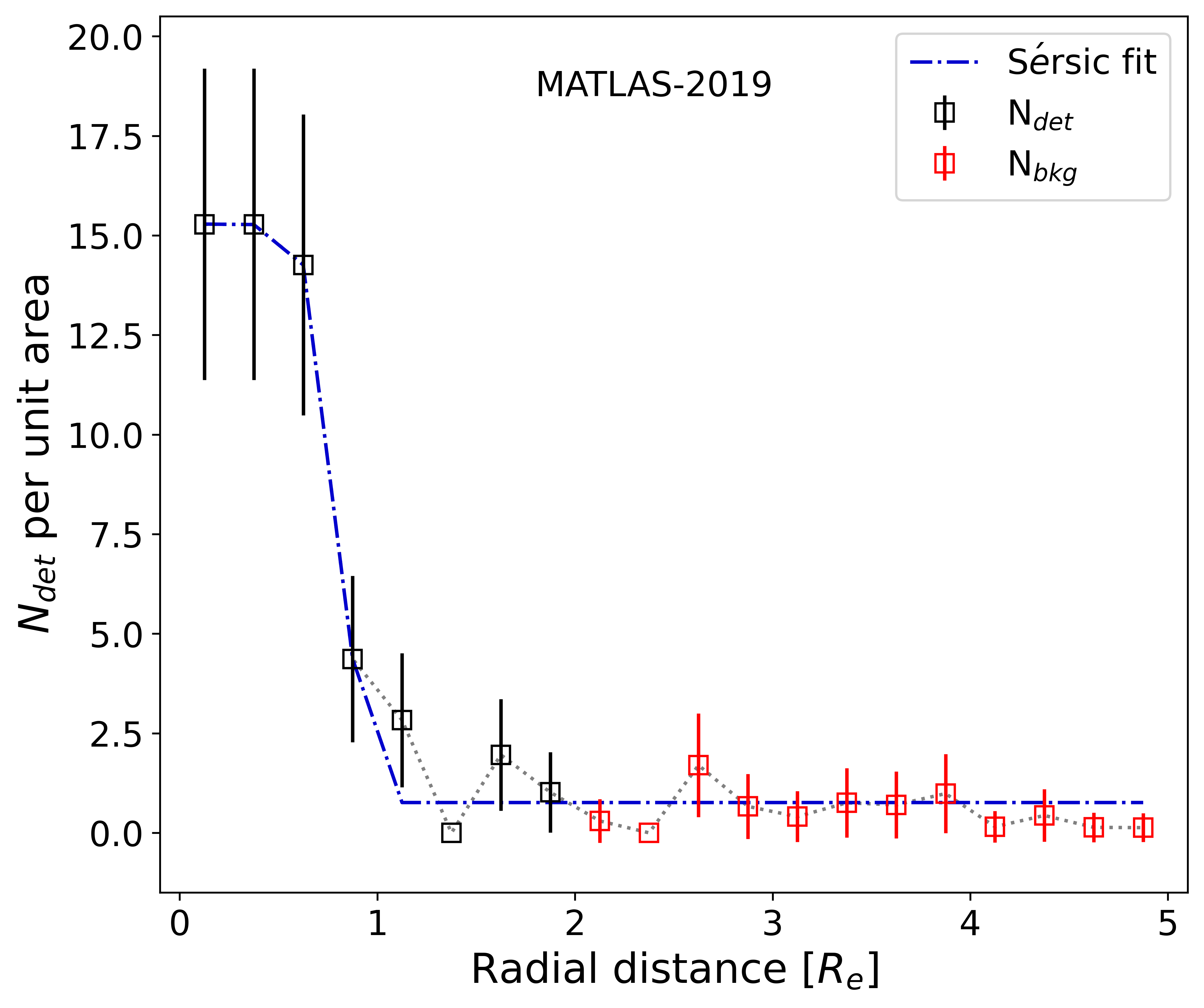}}
\centerline{
\includegraphics[width=0.33\linewidth]{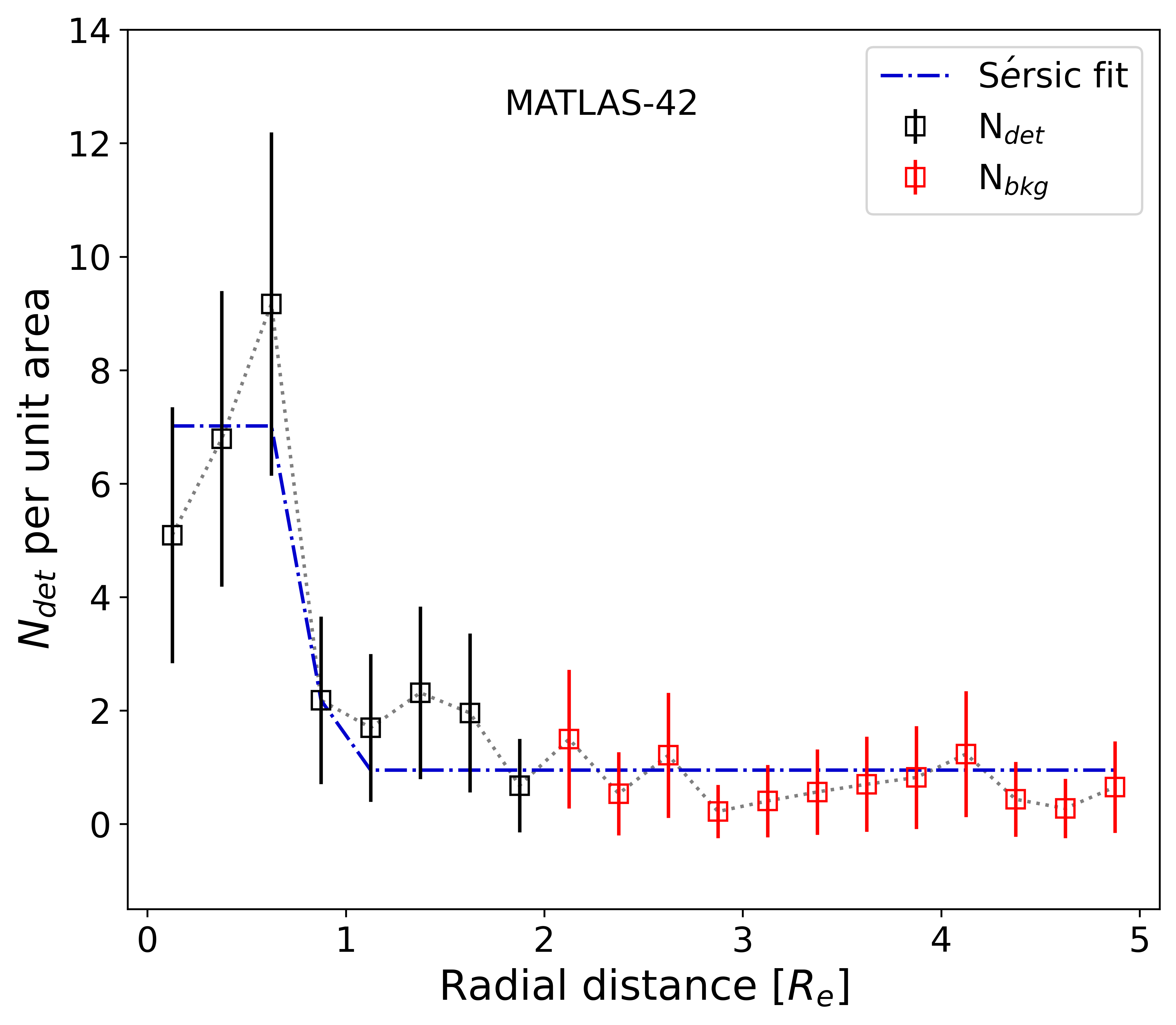}
\includegraphics[width=0.33\linewidth]{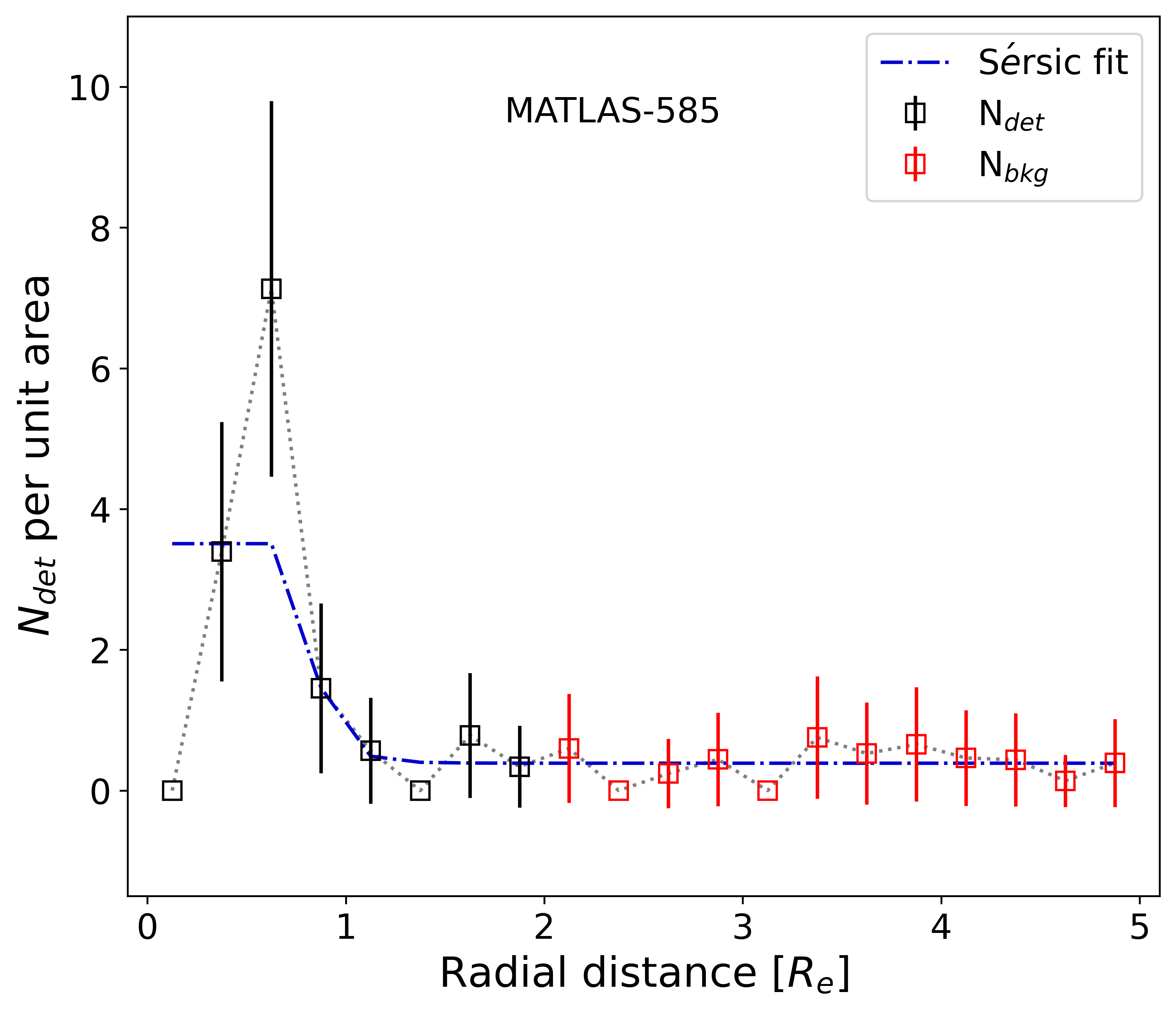}
\includegraphics[width=0.33\linewidth]{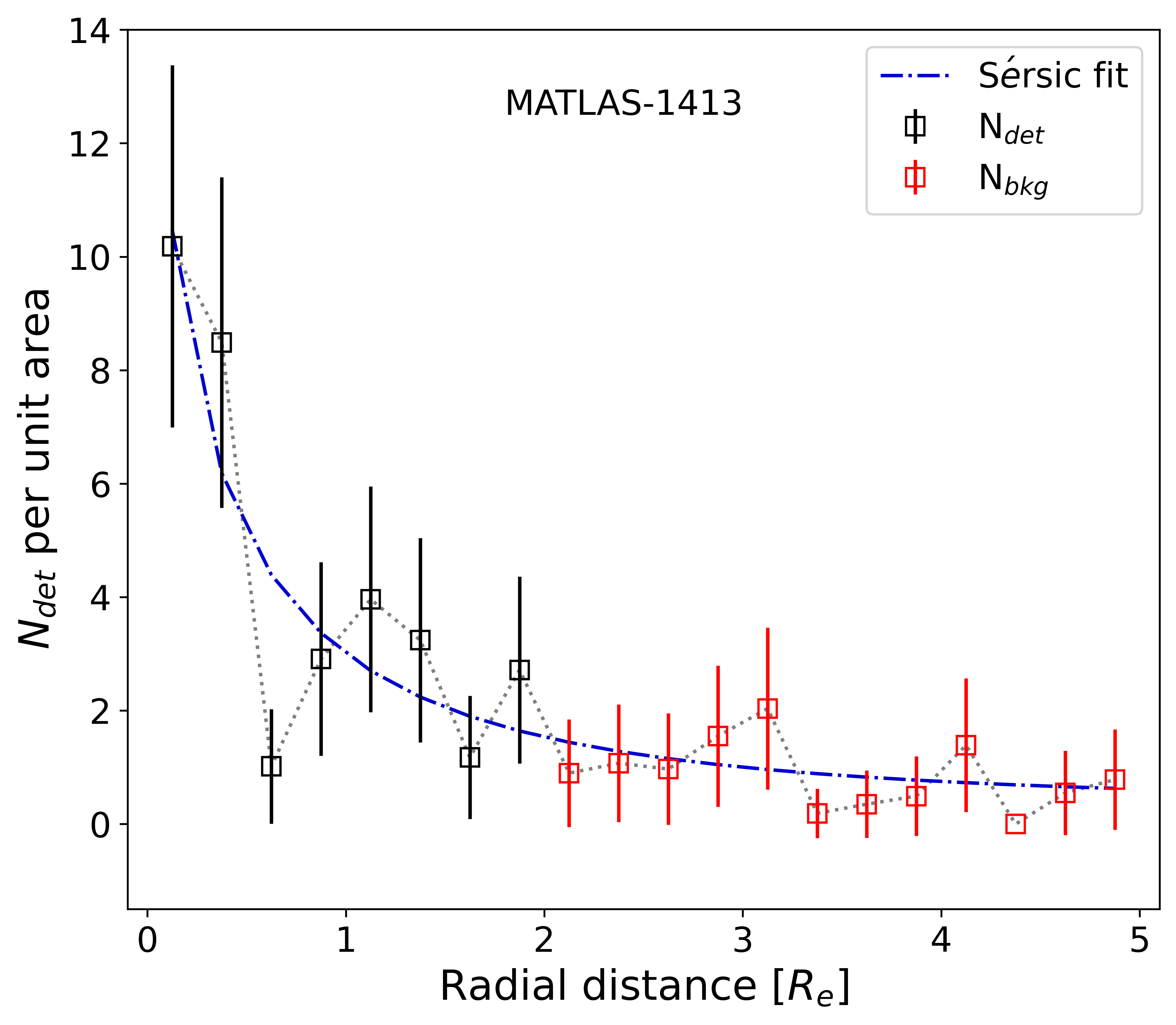}}
\centerline{
\includegraphics[width=0.33\linewidth]{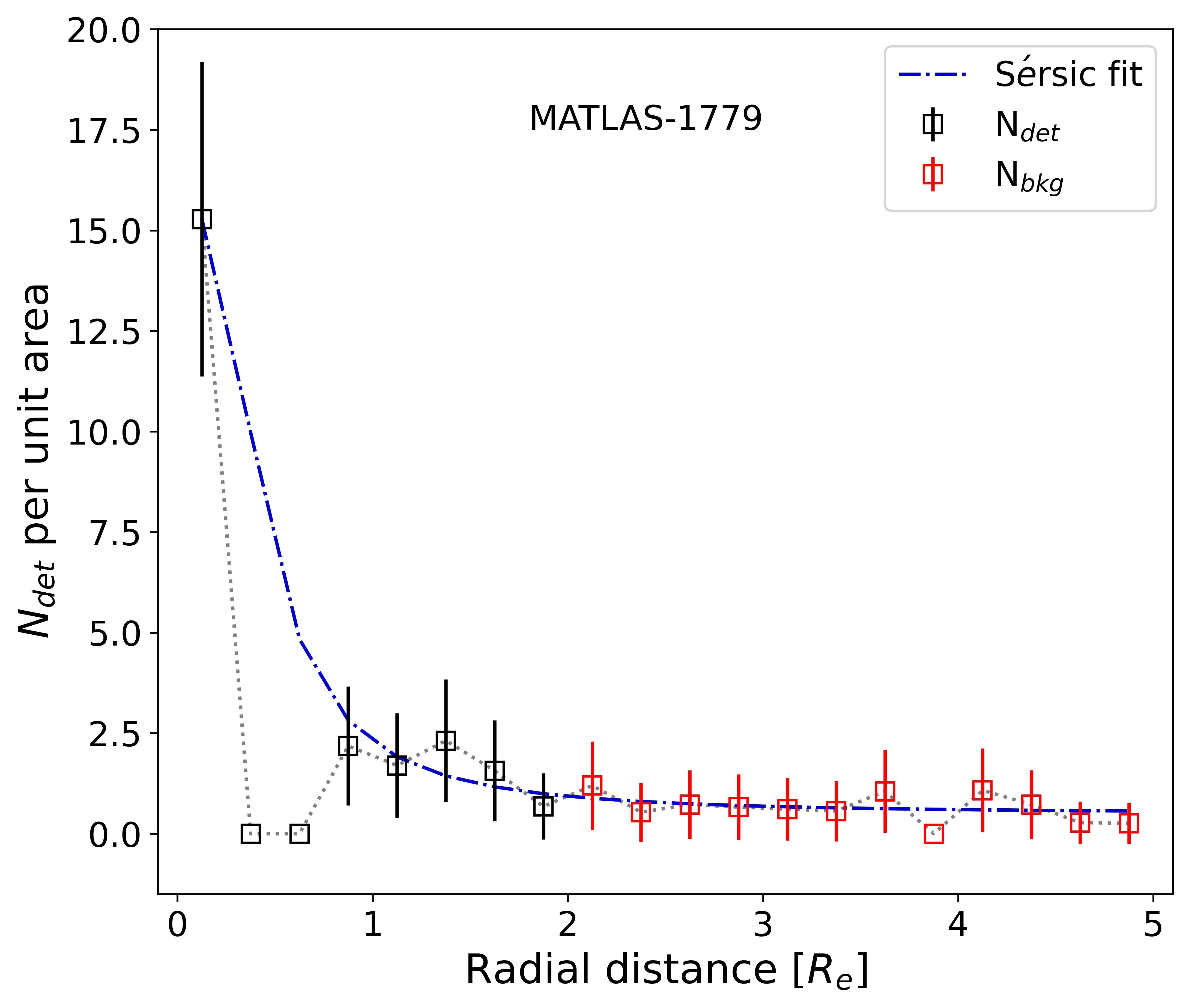}
\includegraphics[width=0.33\linewidth]{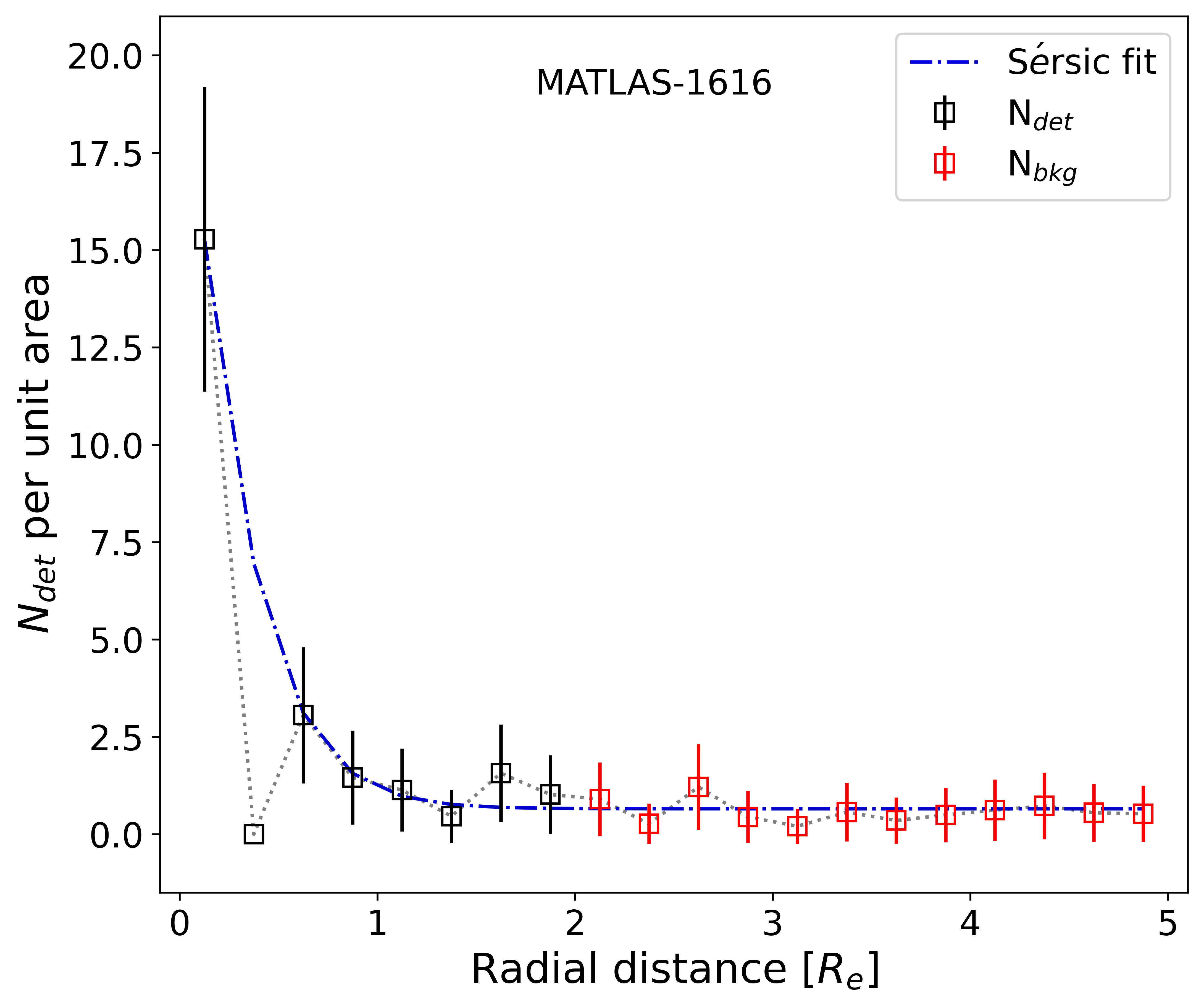}
\includegraphics[width=0.33\linewidth]{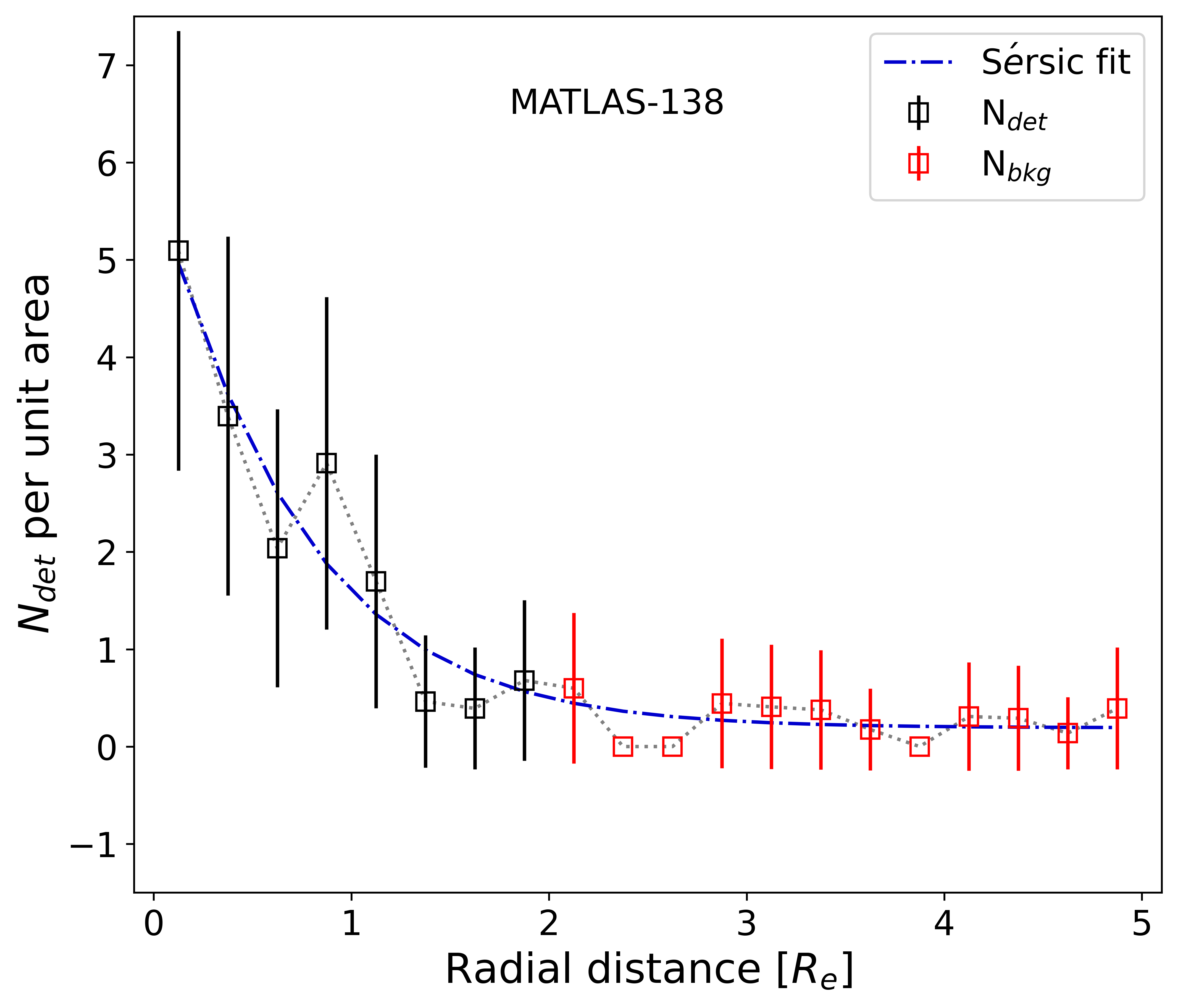}}
\centerline{
\includegraphics[width=0.33\linewidth]{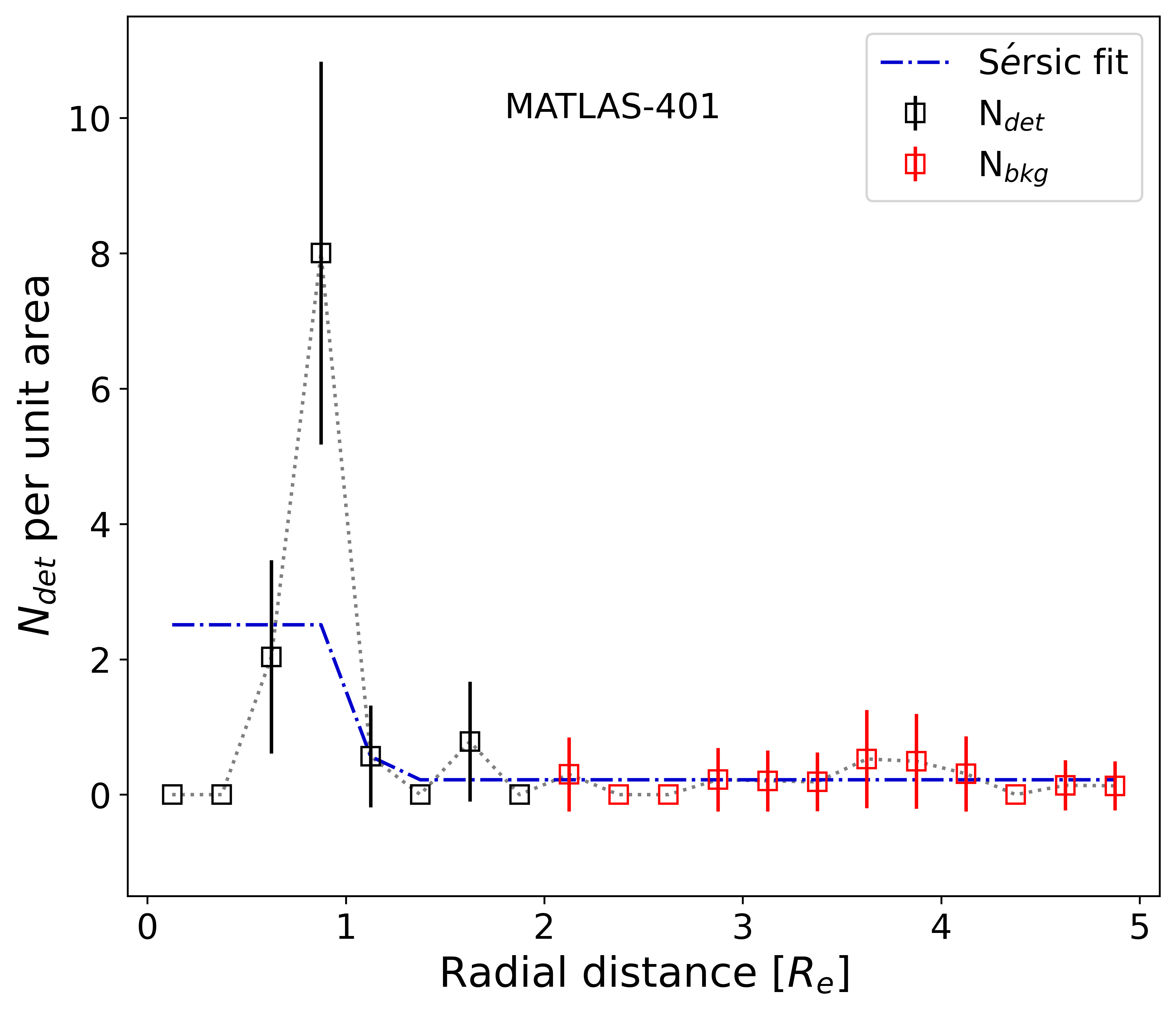}
\includegraphics[width=0.33\linewidth]{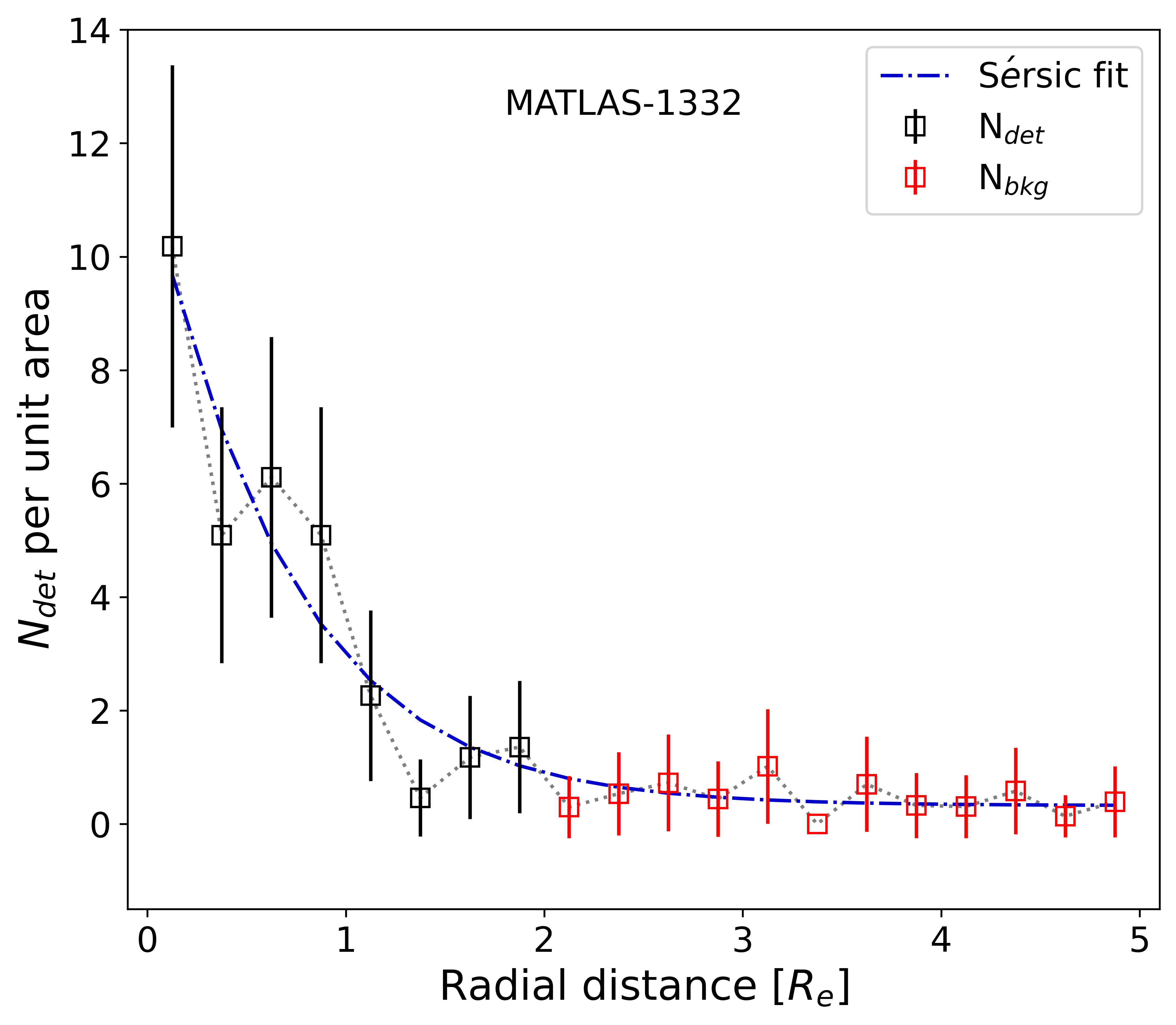}
\includegraphics[width=0.33\linewidth]{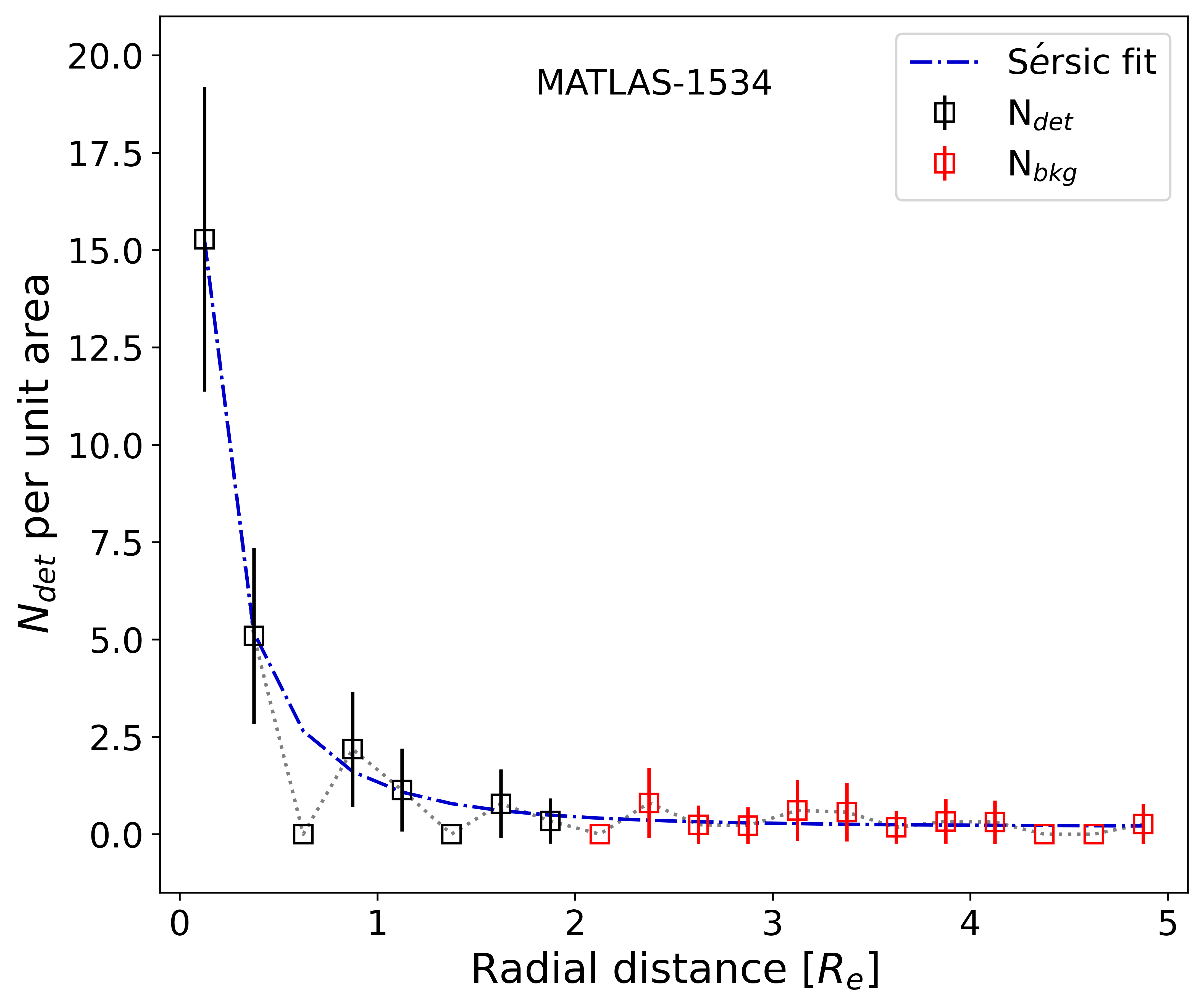}
}
\caption{{\it From top left to bottom right}: Radial distribution of all detected GCs for the 12 UDGs with $N_{GC} > 10$. The {\it black points} are for the GC counts within 2$R_e$ while the {\it red points} are the counts outside (background). The S\'ersic fit is shown with the {\it dash-dot blue line}. 
\label{fig:raddistap}}
\end{figure*}

\end{appendix}

\end{document}